\documentclass[journal]{IEEEtran}
\usepackage[dvipsnames]{xcolor}
\usepackage{color}
\usepackage{framed}

\definecolor{shadecolor}{rgb}{0.9,0.9,0.9}
\definecolor{mylightgray}{RGB}{220,220,220}

\definecolor{myblue}{RGB}{0, 68, 116}
\definecolor{mycyan}{RGB}{0, 97, 91}
\definecolor{mygreen}{RGB}{2, 102, 1}
\definecolor{myorange}{RGB}{240, 102, 0}
\definecolor{myred}{RGB}{172, 23, 0}
\definecolor{mymagenta}{RGB}{140,16,73}

\usepackage{mathtools}
\usepackage{amsmath}
\usepackage{amssymb}
\usepackage[shortlabels]{enumitem}
\usepackage{graphicx,epic,eepic,epsfig,latexsym,verbatim}
\usepackage{dsfont}
\usepackage{mathrsfs}

\usepackage{subcaption}
\usepackage{caption}
\usepackage{float}
\usepackage{tikz}
\usepackage{relsize}
\usepackage{pgfplots}

\usepackage{amsthm}
\usepackage{tcolorbox}
\tcbuselibrary{skins}
\tcbuselibrary{breakable}

\makeatletter
\newcommand{\DefineColoredTheoremStyle}[2]{%
  \newtheoremstyle{#1}
    {3pt}{3pt}
    {\itshape}
    {}
    {\bfseries\color{#2}}
    {\;}
    { }
    {\thmname{##1}\thmnumber{ ##2}\textnormal{\thmnote{ (##3)}}}
}
\makeatother

\DefineColoredTheoremStyle{lemmstyle}{black}
\DefineColoredTheoremStyle{propstyle}{black}
\DefineColoredTheoremStyle{thmstyle}{black}
\DefineColoredTheoremStyle{corostyle}{black}
\DefineColoredTheoremStyle{conjstyle}{black}
\DefineColoredTheoremStyle{problstyle}{black}
\DefineColoredTheoremStyle{plainstyle}{black}

\theoremstyle{thmstyle}\newtheorem{theorem}{Theorem}
\theoremstyle{lemmstyle}\newtheorem{lemma}[theorem]{Lemma}
\theoremstyle{propstyle}\newtheorem{prop}[theorem]{Proposition}
\theoremstyle{corostyle}\newtheorem{corollary}[theorem]{Corollary}
\theoremstyle{conjstyle}\newtheorem{conjecture}[theorem]{Conjecture}
\theoremstyle{problstyle}\newtheorem{problem}[theorem]{Problem}
\theoremstyle{plainstyle}\newtheorem{definition}[theorem]{Definition}
\theoremstyle{plainstyle}\newtheorem{assumption}[theorem]{Assumption}
\theoremstyle{plainstyle}\newtheorem{example}[theorem]{Example}
\theoremstyle{plainstyle}\newtheorem{remark}[theorem]{Remark}

\tcbset{
    commonstyle/.style n args={1}{
        oversize=0.5mm,
        colback={#1!8},
        colframe={#1!10},
        boxrule=1pt,
        boxsep=2mm,
        left=1mm,
        right=1mm,
        top=2mm,
        bottom=2mm,
        breakable
    }
}

\usepackage{colortbl}
\usepackage{pifont} 
\usepackage{booktabs}
\usepackage{makecell} 
\usepackage{diagbox}
\usepackage{multirow}

\newcommand{\nc}{\newcommand}
\nc{\rnc}{\renewcommand}

\nc{\<}{\langle}
\rnc{\>}{\rangle}
\nc{\bra}[1]{\langle#1|}
\nc{\ket}[1]{|#1\rangle}
\nc{\ketbra}[2]{|#1\rangle\!\langle#2|}
\nc{\braket}[2]{\langle#1|#2\rangle}
\nc{\braandket}[3]{\langle #1|#2|#3\rangle}
\nc{\proj}[1]{| #1\rangle\!\langle #1 |}
\nc{\avg}[1]{\langle#1\rangle}

\nc{\rank}{\operatorname{Rank}}
\nc{\id}{{\operatorname{id}}}
\nc{\supp}{{\operatorname{supp}}}
\nc{\smfrac}[2]{\mbox{$\frac{#1}{#2}$}}
\nc{\tr}{\operatorname{Tr}}
\nc{\ox}{\otimes}
\nc{\floor}[1]{\lfloor #1 \rfloor}
\nc{\trans}{\mathsf T}
\nc{\img}{\mathbf{i}}
\def\ve{\varepsilon}

\nc{\cA}{{\cal A}}
\nc{\cB}{{\cal B}}
\nc{\cC}{{\cal C}}
\nc{\cD}{{\cal D}}
\nc{\cE}{{\cal E}}
\nc{\cF}{{\cal F}}
\nc{\cG}{{\cal G}}
\nc{\cH}{{\cal H}}
\nc{\cI}{{\cal I}}
\nc{\cJ}{{\cal J}}
\nc{\cK}{{\cal K}}
\nc{\cL}{{\cal L}}
\nc{\cM}{{\cal M}}
\nc{\cN}{{\cal N}}
\nc{\cO}{{\cal O}}
\nc{\cP}{{\cal P}}
\nc{\cQ}{{\cal Q}}
\nc{\cR}{{\cal R}}
\nc{\cS}{{\cal S}}
\nc{\cT}{{\cal T}}
\nc{\cV}{{\cal V}}
\nc{\cU}{{\cal U}}
\nc{\cX}{{\cal X}}
\nc{\cY}{{\cal Y}}
\nc{\cZ}{{\cal Z}}
\nc{\cW}{{\cal W}}

\nc{\RR}{{{\mathbb R}}}
\nc{\CC}{{{\mathbb C}}}
\nc{\FF}{{{\mathbb F}}}
\nc{\NN}{{{\mathbb N}}}
\nc{\ZZ}{{{\mathbb Z}}}
\nc{\QQ}{{{\mathbb Q}}}
\nc{\UU}{{{\mathbb U}}}
\nc{\EE}{{{\mathbb E}}}

\nc{\bH}{{\mathfrak{H}}}

\nc{\sK}{{{\mathscr{K}}}}
\nc{\sS}{{{\mathscr{S}}}}
\nc{\sT}{{{\mathscr{T}}}}
\nc{\sA}{{{\mathscr{A}}}}
\nc{\sB}{{{\mathscr{B}}}}
\nc{\sC}{{{\mathscr{C}}}}
\nc{\sE}{{{\mathscr{E}}}}
\nc{\sL}{{{\mathscr{L}}}}
\nc{\sG}{{{\mathscr{G}}}}
\nc{\sF}{{{\mathscr{F}}}}
\nc{\sI}{{{\mathscr{I}}}}
\nc{\sN}{{{\mathscr{N}}}}
\nc{\sM}{{{\mathscr{M}}}}

\nc{\Choi}{Choi-Jamio\l{}kowski }
\nc{\reg}{\infty}
\nc{\amo}{\text{\rm amo}}
\nc{\Renyi}{R\'{e}nyi }

\nc{\conv}{\operatorname{conv}}
\nc{\cvxset}{\mathscr{C}}

\nc{\RM}{{{\mathscr{R}}}}

\nc{\END}{\operatorname{End}}
\nc{\PERM}{\mathfrak{\sigma}}

\nc{\Cone}{\text{\rm Cone}}
\nc{\sep}{{\SEP}}

\nc{\DD}{{{\mathbb D}}}
\nc{\BS}{{\scriptscriptstyle \rm {BS}}}
\nc{\Sand}{{\scriptscriptstyle  \rm S}}
\nc{\Petz}{{\scriptscriptstyle  \rm P}}
\nc{\Hypo}{{\scriptscriptstyle  \rm H}}
\nc{\Meas}{{\scriptscriptstyle \rm M}}
\nc{\Proj}{{{\scriptscriptstyle \rm P}}}

\nc{\suchthat}{\text{\rm s.t.}}

\nc{\pl}{{\scalebox{0.7}{+}}}
\nc{\HERM}{\mathscr{H}}
\nc{\PSD}{\HERM_{\pl}}
\nc{\PD}{\HERM_{\pl\pl}}
\nc{\density}{\mathscr{D}}
\nc{\subdensity}{\mathscr{D}_\bullet}

\nc{\polarPSD}[1]{{#1}_{\pl}^{\circ}}
\nc{\polarPSDre}[1]{{#1}_{\pl}^{\star}}
\nc{\polarPD}[1]{{#1}_{\pl\pl}^{\circ}}

\nc{\PPT}{\text{\rm PPT}}
\nc{\Rains}{\text{\rm Rains}}
\nc{\WD}{\text{\rm WD}}
\nc{\SEP}{\text{\rm SEP}}
\nc{\PSEP}{\text{\rm PSEP}}
\nc{\CPTP}{\text{\rm CPTP}}
\nc{\POVM}{\text{\rm POVM}}
\nc{\PVM}{\text{\rm PVM}}
\nc{\CP}{\text{\rm CP}}
\nc{\adv}{\text{\rm adv}}
\nc{\spec}{\text{\rm spec}}
\nc{\poly}{\text{\rm poly}}
\nc{\End}{\operatorname{End}}
\nc{\Par}{\operatorname{Par}}
\nc{\RNG}{\operatorname{RNG}}
\nc{\STAB}{\text{\rm STAB}}
\nc{\epi}{\boldsymbol{\operatorname{epi}}}
\nc{\op}{\boldsymbol{\operatorname{op}}}

\makeatletter
\newcommand*\rel@kern[1]{\kern#1\dimexpr\macc@kerna}
\newcommand*\widebar[1]{%
  \begingroup
  \def\mathaccent##1##2{%
    \rel@kern{0.8}%
    \overline{\rel@kern{-0.8}\macc@nucleus\rel@kern{0.2}}%
    \rel@kern{-0.2}%
  }%
  \macc@depth\@ne
  \let\math@bgroup\@empty \let\math@egroup\macc@set@skewchar
  \mathsurround\z@ \frozen@everymath{\mathgroup\macc@group\relax}%
  \macc@set@skewchar\relax
  \let\mathaccentV\macc@nested@a
  \macc@nested@a\relax111{#1}%
  \endgroup
}
\makeatother

\usepackage{changepage}

\ifCLASSINFOpdf
\else
\fi

\begin{document}
%
\title{Error Exponents of Quantum State Discrimination\\ with Composite Correlated Hypotheses}
%
%
%

\author{Kun Fang,~\IEEEmembership{Member,~IEEE,}
       and Masahito Hayashi,~\IEEEmembership{Fellow,~IEEE,}
\thanks{\quad K.F. is supported in part by the National Natural Science Foundation of China (Grant No. 12404569 and 92470113), the Shenzhen Science and Technology Program (Grant No. QNXMB20250701091826036 and JCYJ20240813113519025), the Shenzhen International Quantum Academy (Grant No. SIQA2025KFKT03), the Shenzhen Fundamental Research Program (Grant No. JCYJ20241202124023031), the General R\&D Projects of 1+1+1 CUHK-CUHK(SZ)-GDST Joint Collaboration Fund (Grant No. GRDP2025-022), and the University Development Fund (Grant No. UDF01003565). M.H. is supported in part by the General R\&D Projects of 1+1+1 CUHK-CUHK(SZ)-GDST Joint Collaboration Fund (Grant No. GRDP2025-022), the Guangdong Provincial Quantum Science Strategic Initiative (Grant No.
GDZX2505003),
the Shenzhen International Quantum Academy (Grant No. SIQA2025KFKT07),
and the National Natural Science Foundation of China under Grant 62171212. \textit{(Corresponding author: Kun Fang.)}}
\thanks{\quad Kun Fang is with the School of Data Science, The Chinese University of Hong Kong, Shenzhen, Guangdong 518172, China (e-mail: kunfang@cuhk.edu.cn).}
\thanks{\quad Masahito Hayashi is with the School of Data Science, The Chinese University of Hong Kong, Shenzhen, Guangdong 518172, China, and also with the International Quantum Academy, Futian District, Shenzhen 518048, China, and also with the Graduate School of Mathematics, Nagoya University, Chikusa-ku, Nagoya 464–8602, Japan (e-mail: hmasahito@cuhk.edu.cn).}
}

%
%

\markboth{}%
{Fang and Hayashi: Error Exponents of Quantum State Discrimination  with Composite Correlated Hypotheses}
%



\maketitle

\begin{abstract}
We study the error exponents in quantum hypothesis testing between two sets of quantum states, extending the analysis beyond the independent and identically distributed case to encompass composite correlated hypotheses. In particular, we introduce and compare two natural extensions of the quantum Hoeffding divergence and anti-divergence to sets of quantum states, establishing their equivalence or quantitative relations. In the error exponent regime, we generalize the quantum Hoeffding bound to stable sequences of convex, compact sets of quantum states, demonstrating that the optimal Type-I error exponent, under an exponential constraint on the Type-II error, is precisely characterized by the regularized quantum Hoeffding divergence between the sets. 
In the strong converse exponent regime, we establish a general lower bound on the exponent in terms of the regularized quantum Hoeffding anti-divergence, and we prove a matching upper bound when the null hypothesis is a singleton, {under additional assumptions}.
The generality of these results enables applications in various contexts, including (i) refining the generalized quantum Stein's lemma by [Fang, Fawzi \& Fawzi, 2024]; (ii) exhibiting counterexamples to the continuity of the regularized Petz \Renyi divergence and Hoeffding divergence; (iii) obtaining error exponents for adversarial channel discrimination and resource detection problems. 
\end{abstract}


%
\IEEEpeerreviewmaketitle

%
%
%
%



\section{Introduction}


\IEEEPARstart{D}istinguishability is a central topic in information theory from both theoretical and practical perspectives. A fundamental framework for studying distinguishability is \textit{asymmetric hypothesis testing}. In this setting, a source generates a sample $x$\ from one of two probability
distributions $p\equiv\{p(x)\}_{x\in\mathcal{X}}$ or $q\equiv\{q(x)\}_{x\in
\mathcal{X}}$. The objective of asymmetric hypothesis testing is to minimize the Type-II\ error (decides $p$ when the fact is $q$) while keeping the Type-I\ error (decides $q$ when the fact is $p$) within a certain threshold. The celebrated Chernoff-Stein’s Lemma~\cite{Chernoff1952} states that, for any constant bound on the Type-I error, the optimal Type-II error decays exponentially fast in the number of samples, and the decay rate is exactly the relative entropy,
\begin{equation}
D(p\Vert q)=\sum_{x\in\mathcal{X}}p(x)[\log p(x) - \log q(x)].
\end{equation}
In particular, this lemma also states the ``strong converse property'', a desirable mathematical property in information theory~\cite{Wolfowitz1978} that delineates a sharp boundary for the tradeoff between the Type-I and Type-II errors in the asymptotic regime: any possible scheme with Type-II error decaying to zero with an exponent larger than the relative entropy will result in the Type-I error converging to one in the asymptotic limit.  
Therefore, the Chernoff-Stein’s Lemma provides a rigorous operational interpretation of the relative entropy and establishes a crucial connection between hypothesis testing and information theory~\cite{Blahut1974}.

\begin{figure}[H]
\centering
\includegraphics[width=\linewidth]{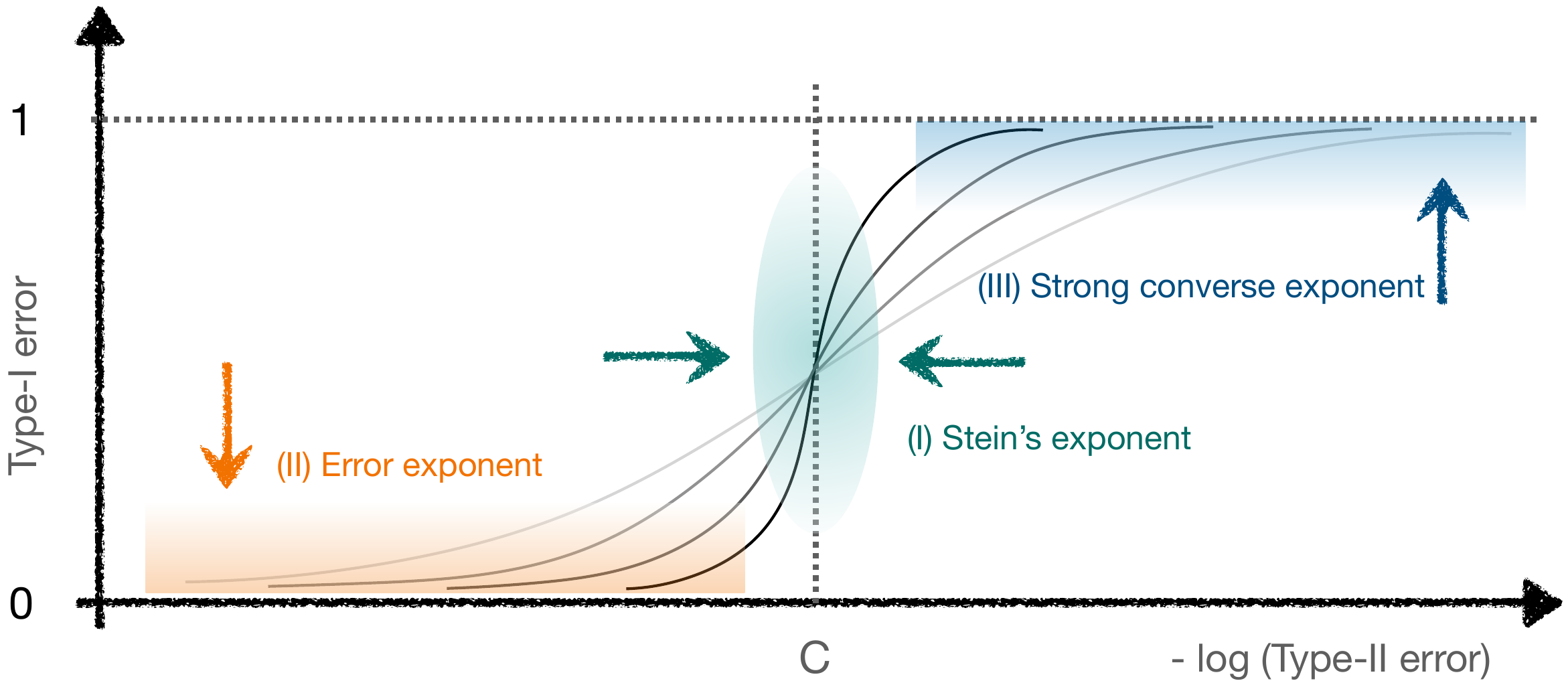}
\caption{Illustration of different study regimes of quantum hypothesis testing. Each curve represents the tradeoff between the Type-I and Type-II errors for varying block lengths $n$, with darker lines corresponding to longer block lengths. (I) represents the Stein exponent regime, (II) represents the error exponent regime, and (III) represents the strong converse exponent regime.}
\label{fig: regime error exponent}
\end{figure}

A natural question is whether the above result generalizes to the quantum case. Substantial efforts have been made to answer this fundamental question in quantum information community~(see, e.g., \cite{hiai1991proper,Ogawa2000,Hayashi2002,audenaert2008asymptotic,hayashi2007error,brandao2010generalization,cooney2016strong,mosonyi2015quantum,wang2019resource,WW2019,fang2025adversarial,fang2024generalized,hayashi2025entanglement}). Consider the problem of distinguishing between two quantum hypotheses: the system is prepared either in state $\rho_n$ (the null hypothesis) or in state $\sigma_n$ (the alternative hypothesis).  Operationally, the discrimination is carried out using a two-outcome positive operator-valued measure (POVM) $\{M_n, I-M_n\}$, with $0 \leq M_n \leq I$. Then, the Type-I and Type-II errors are, respectively, given by
\begin{align}
& \text{(Type-I)}\quad \alpha(\rho_n, M_n):= \tr[\rho_n (I-M_n)],\\
& \text{(Type-II)}\quad \beta(\sigma_n, M_n):= \tr[\sigma_n M_n].
\end{align}
It is generally impossible to find a quantum measurement that simultaneously makes both errors vanish; thus, one studies the asymptotic behavior of $\alpha$ and $\beta$ as $n \to \infty$, expecting a trade-off between minimizing $\alpha$ and minimizing $\beta$. The interplay between these errors can be analyzed in various operational regimes (see Figure~\ref{fig: regime error exponent}): (I) the \textbf{Stein exponent} regime, which focuses on the exponential decay rate of the Type-II error when the Type-I error is below a constant threshold; (II) the \textbf{error exponent} regime, which investigates the exponential rate at which the Type-I error vanishes when the Type-II error is required to decay exponentially at a prescribed rate; and (III) the \textbf{strong converse exponent} regime, which examines the exponential rate at which the Type-I error converges to one when the Type-II error decays exponentially at a given rate.

\paragraph{Stein exponent} In asymmetric hypothesis testing, one aims to minimize the Type-II error while keeping the Type-I error below a fixed threshold $\ve \in (0,1)$. The optimal Type-II error is given by
\begin{align}
    \beta_\ve(\rho_n\|\sigma_n):= \min_{0\leq M_n \leq I} \{\beta(\sigma_n,M_n): \alpha(\rho_n,M_n) \leq \ve\}.
\end{align}
The quantum version of the Chernoff-Stein's Lemma (also known as quantum Stein's lemma) states that
the optimal Type-II error decays exponentially with the number of copies $n$ of the states when the Type-I error is restricted below a constant threshold and the optimal exponent is given by the quantum relative entropy $D(\rho\|\sigma):= \tr[\rho(\log \rho - \log \sigma)]$~\cite{hiai1991proper,Ogawa2000},
\begin{align}
    \lim_{n\to \infty} -\frac{1}{n} \log \beta_\ve(\rho^{\ox n}\|\sigma^{\ox n}) = D(\rho\|\sigma),\quad \forall \ve \in (0,1).
\end{align}

\paragraph{Error exponent} As a refinement of the quantum Stein’s lemma, one can study the optimal Type-I error given that the Type-II error decays with a given exponential speed. One is then interested in the asymptotics of the Type-I error,
\begin{align}\label{eq: optimal Type-I error}
\alpha_{n, r}(\rho_n\|\sigma_n):= \min_{0\leq M_n \leq I} \{\alpha(\rho_n,M_n): \beta(\sigma_n,M_n) \leq 2^{-nr}\},
\end{align}
with a constant $r>0$. When $r < D(\rho\|\sigma)$, the optimal Type-I error $\alpha_{n, r}(\rho^{\ox n}\|\sigma^{\ox n})$, also decays with an exponential speed, as was shown in~\cite{ogawa2004error}. The exact decay rate is determined by the quantum Hoeffding bound~\cite{hayashi2007error,nagaoka2006converse,audenaert2008asymptotic} as
\begin{align}\label{eq: Hoeffding bound i.i.d.}
    \lim_{n\to \infty} -\frac{1}{n} \log \alpha_{n, r}(\rho^{\ox n}{\|} \sigma^{\ox n}) & = H_r(\rho\|\sigma),
\end{align}
where the quantum Hoeffding divergence is defined by
\begin{align}\label{eq: Hoeffding divergence}
    H_r(\rho\|\sigma) := \sup_{\alpha \in (0,1)} \frac{\alpha - 1}{\alpha} \left(r - D_{\Petz,\alpha}(\rho\|\sigma)\right),
\end{align}
and $D_{\Petz,\alpha}$ is the Petz \Renyi divergence. 

\paragraph{Strong converse exponent} On the other hand, when $r > D(\rho\|\sigma)$, the optimal Type-I error $\alpha_{n, r}(\rho^{\ox n}\|\sigma^{\ox n})$ goes to $1$ exponentially fast~\cite{Ogawa2000}. The rate of this convergence is~\cite{mosonyi2015quantum},
\begin{align}\label{eq: anti-Hoeffding bound i.i.d.}
    \lim_{n\to \infty} -\frac{1}{n} \log (1- \alpha_{n, r}(\rho^{\ox n}\|\sigma^{\ox n})) = H_r^*(\rho\|\sigma),
\end{align}
where the Hoeffding anti-divergence is defined by 
\begin{align}\label{eq: anti-Hoeffding divergence}
    H_r^*(\rho\|\sigma):= \sup_{\alpha > 1} \frac{\alpha - 1}{\alpha} \left(r - D_{\Sand,\alpha}(\rho\|\sigma)\right),    
\end{align} 
and $D_{\Sand,\alpha}$ is the sandwiched \Renyi divergence.

The results in~\eqref{eq: Hoeffding bound i.i.d.} and~\eqref{eq: anti-Hoeffding bound i.i.d.} provide a comprehensive characterization of the asymptotic trade-off between the Type-I and Type-II error probabilities. In particular, the quantum Stein’s lemma emerges as a special case in the limit $r \to D(\rho\|\sigma)$.

\begin{figure}[H]
    \centering
    \includegraphics[width=\linewidth]{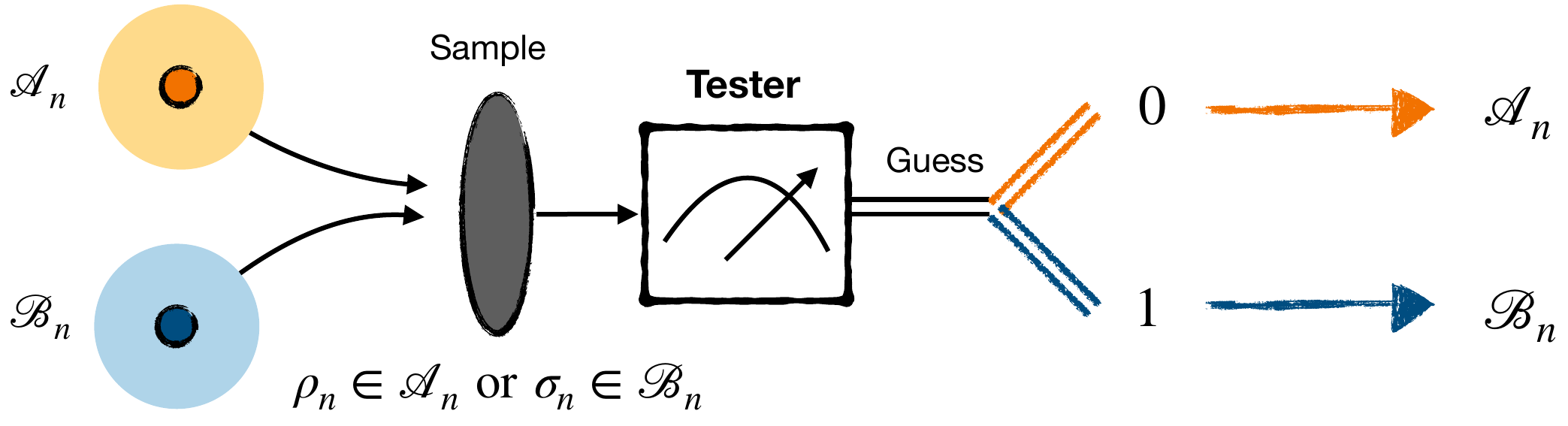}
    \caption{Quantum hypothesis testing between two sets of quantum states $\sA_n$ and $\sB_n$.}
    \label{fig: hypothesis testing demo}
\end{figure}

While much of the existing literature has focused on i.i.d.\ sources, practical scenarios often involve quantum states that are not fully specified (i.e., \emph{composite hypotheses}~\cite{brandao2010generalization,berta2021composite,Mosonyi_2022,hayashi2025entanglement,hayashi2016correlation,DWH25}) such as in adversarial or black-box settings~\cite{fang2025adversarial,watanabe2024black} and exhibit correlations that preclude a simple tensor product structure (i.e., \emph{correlated hypotheses}~\cite{hiai2007large,hiai2008error,mosonyi2015two,hayashi2025entanglement}). In this context, the task is to discriminate between two \emph{sets of correlated quantum states} (see Figure~\ref{fig: hypothesis testing demo}). That is, a tester receives samples prepared according to either the set $\sA_n$ or $\sB_n$, and determines, via a quantum measurement $\{M_n, I-M_n\}$, from which set the samples originate.

As in standard hypothesis testing, two types of errors can occur: a Type-I error, where a sample from $\sA_n$ is incorrectly classified as coming from $\sB_n$, and a Type-II error, where a sample from $\sB_n$ is incorrectly classified as coming from $\sA_n$.
Since we aim to control the errors for any state within the given sets, regardless of which one is drawn, the (worst-case) Type-I error is defined by
\begin{align}
    \alpha(\sA_n, M_n): = \sup_{\rho_n \in \sA_n} \tr[\rho_n (I-M_n)],
\end{align}
and the (worst-case) Type-II error is defined by
\begin{align}
    \beta(\sB_n, M_n): = \sup_{\sigma_n \in \sB_n} \tr[\sigma_n M_n].
\end{align} 

A few variants of quantum Stein's lemma has been shown in this worst-case setting under some structural assumptions on the sets $\sA_n$ and $\sB_n$~\cite{fang2024generalized,hayashi2024generalized,lami2024solutiongeneralisedquantumsteins}. Particularly, the Stein's lemma in~\cite{fang2024generalized} shows that
\begin{align}
    \lim_{n\to \infty} -\frac{1}{n} \log \beta_\ve(\sA_n\|\sB_n) = D^{\reg}(\sA\|\sB), \quad \forall \ve \in (0,1),
\end{align}
where 
\begin{align} 
    \beta_\ve(\sA_n\|\sB_n):=\min_{0\leq M_n\leq I}\{\beta(\sB_n,M_n): \alpha(\sA_n, M_n)\leq \ve\},
\end{align} 
and $D(\sA_n\|\sB_n):= \inf_{\rho_n \in \sA_n, \sigma_n \in \sB_n} D(\rho_n\|\sigma_n)$ is the quantum relative entropy between two sets of states and $D^{\reg}(\sA\|\sB):= \lim_{n\to \infty} \frac{1}{n} D(\sA_n\|\sB_n)$ is the regularization.


In this work, we refine the analysis of asymmetric hypothesis testing for composite correlated hypotheses, as developed in~\cite{hayashi2024generalized,lami2024solutiongeneralisedquantumsteins,fang2024generalized}, by extending it to the error exponent and strong converse exponent regimes. Specifically, we consider the optimal Type-I error~\footnote{Such version of definition is similarly used in~\cite{cooney2016strong,hayashi2016correlation,wilde2020amortized}.}:
\begin{align}\label{eq: optimal Type-I error sets}
    \alpha_{n, r}&(\sA_n\|\sB_n)\notag\\
    &:= \min_{0\leq M_n \leq I} \left\{\alpha(\sA_n, M_n): \beta(\sB_n, M_n) \leq 2^{-nr}\right\},
\end{align}
and seek to determine the error exponent
\begin{align}
\lim_{n\to \infty} -\frac{1}{n} \log \alpha_{n, r}(\sA_n\|\sB_n) = \quad ?
\end{align} 
and the strong converse exponent
\begin{align}
\lim_{n\to \infty} -\frac{1}{n} \log (1-\alpha_{n, r}(\sA_n\|\sB_n)) = \quad ? \label{eq: definition of sc}
\end{align}

Due to the uncertainty and correlation inherent in composite correlated hypotheses, several significant challenges arise when obtaining precise characterizations. First, discrimination strategies must be state-agnostic, ensuring that error probabilities are universally controlled for every state within the respective sets, regardless of which specific state is encountered. Second, the optimization problem exhibits a minimax structure, forming a competing scenario that requires simultaneous maximization over all possible states in the sets and minimization over all possible quantum measurements. Third, the non-i.i.d. structure significantly complicates the asymptotic analysis, as standard techniques relying on tensor product structures no longer directly apply. Fourth, it becomes necessary to define suitable extensions of the quantum Hoeffding divergence and anti-divergence to sets of quantum states that both recover existing results for i.i.d. sources and precisely capture the essential features of the general composite correlated setting.

\begin{figure}[H]
\centering
\includegraphics[width=\linewidth]{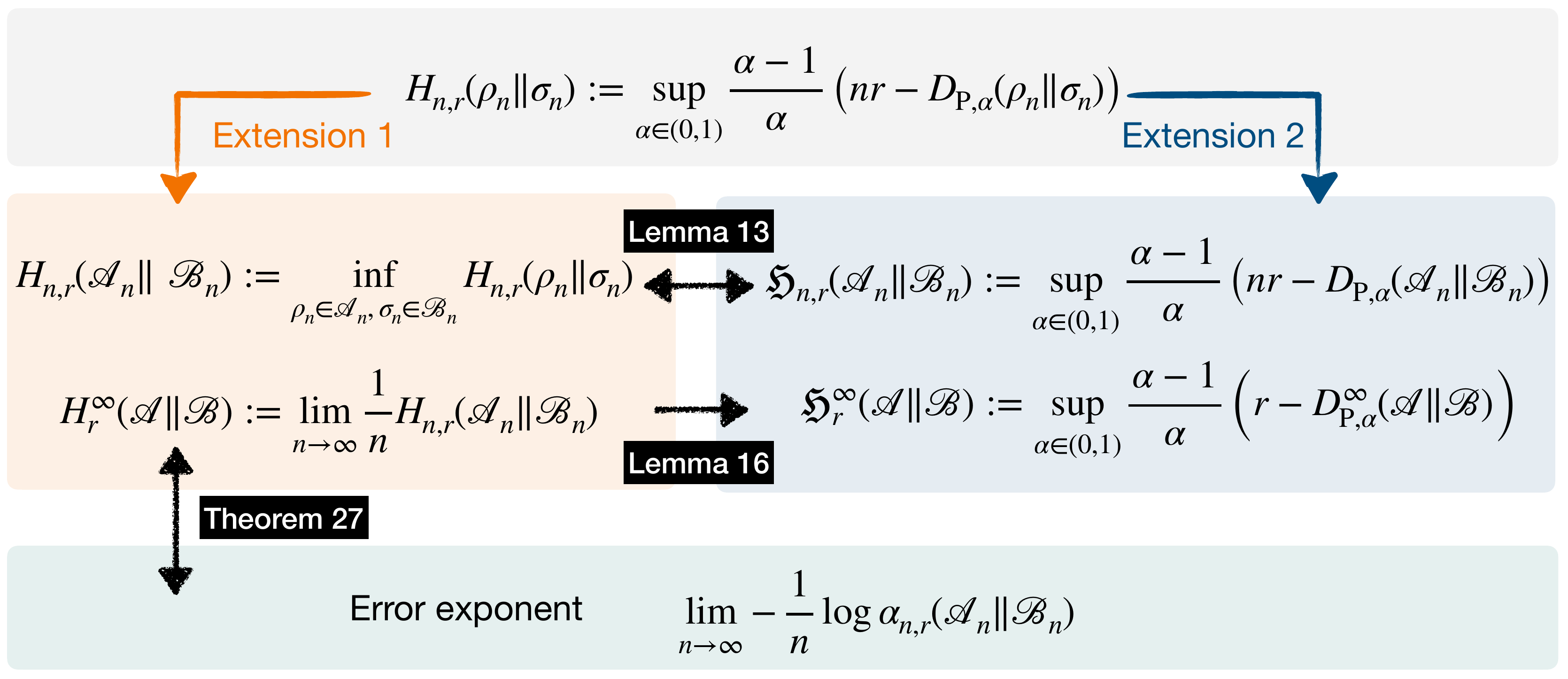}
\caption{Summary of results in the error exponent regime. Quantitative relationships between the various quantities are indicated by black arrows: the quantity at the tail of an arrow is always greater than or equal to the one at the head. A double arrow indicates an equality.}
\label{fig: outline error exponent}
\end{figure}

For the error exponent regime, we explore two natural approaches to extending the quantum Hoeffding divergence (see Figure~\ref{fig: outline error exponent}). The first treats the Hoeffding divergence as a quantum divergence and considers the minimal divergence between the sets, denoted as $H_{n, r}(\sA_n\|\sB_n)$. The second approach uses the explicit formula for the Hoeffding divergence in terms of the Petz \Renyi divergences: we first extend the \Renyi divergences to sets of quantum states, $D_{\Petz, \alpha}(\sA_n\|\sB_n)$, and then define the Hoeffding divergence accordingly, denoted as $\bH_{n, r}(\sA_n\|\sB_n)$. We show that these two extensions are equivalent for finite $n$ in general, and further show the inequality $H_{r}^\infty(\sA\|\sB) \geq \bH_{r}^\infty(\sA\|\sB)$ for their regularizations. As a main result, we prove that the error exponent is completely characterized by the regularized quantum Hoeffding divergence defined via the first approach.

\begin{figure}[H]
\centering
\includegraphics[width=\linewidth]{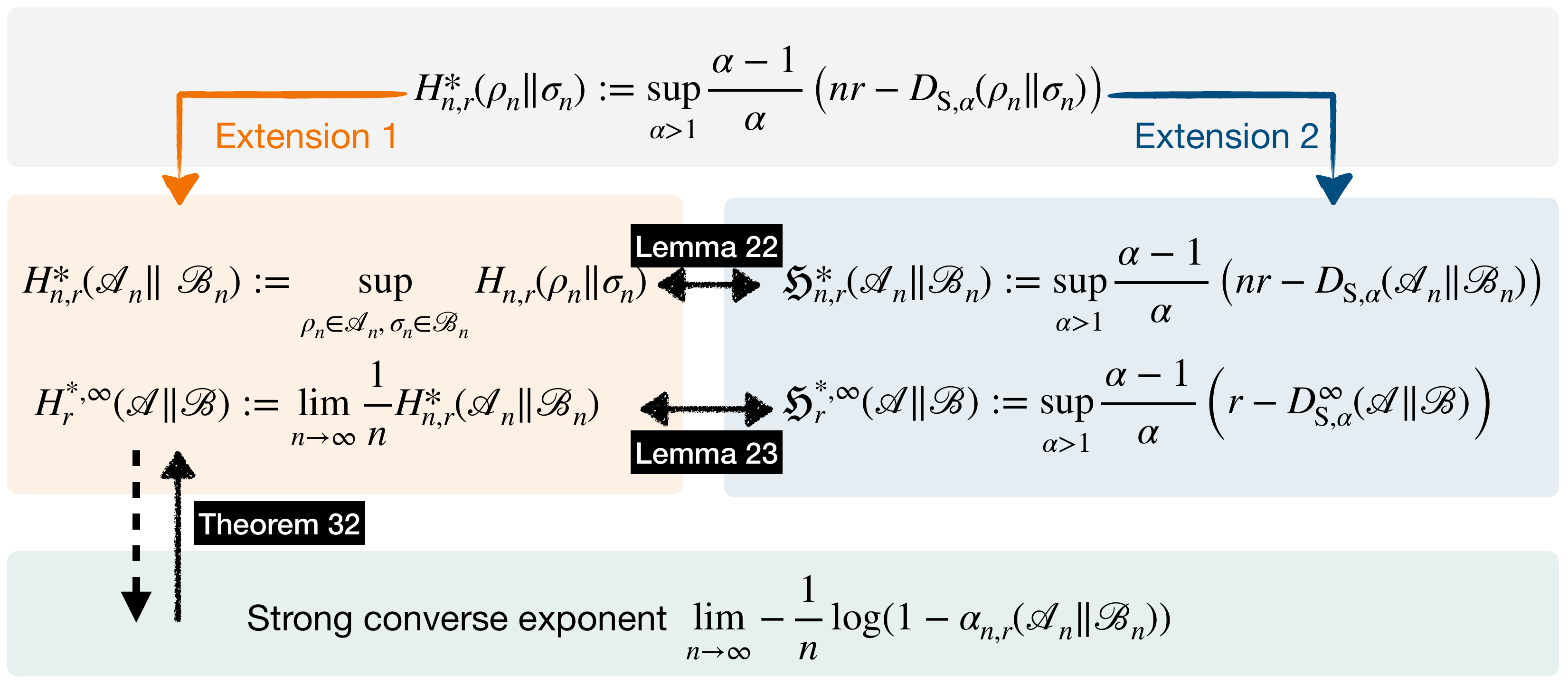}
\caption{Summary of results in the strong converse exponent regime. Quantitative relationships between the various quantities are indicated by black arrows: the quantity at the tail of an arrow is always greater than or equal to the one at the head. A double arrow indicates an equality. A dashed arrow indicates partial progress when $\sA_n$ is a singleton.}
\label{fig: outline strong converse exponent}
\end{figure}

In the strong converse regime, we likewise investigate two natural extensions of the quantum Hoeffding anti-divergence (see Figure~\ref{fig: outline strong converse exponent}). The first approach considers the maximal anti-divergence between the sets, denoted $H_{n, r}^*(\sA_n\|\sB_n)$. The second approach leverages the explicit formula for the anti-divergence in terms of the sandwiched \Renyi divergences: we first extend the sandwiched \Renyi divergences to sets, $D_{\Sand, \alpha}(\sA_n\|\sB_n)$, and then define the anti-divergence as $\bH_{n, r}^*(\sA_n\|\sB_n)$. We establish that these two approaches are equivalent for both finite $n$ and in the asymptotic setting. 
As a main result, we prove that the strong converse exponent is lower bounded by the regularized quantum Hoeffding anti-divergence in general, and we provide a matching upper bound when the null hypothesis is a singleton, {under additional assumptions}.

As the error exponent and strong converse exponent regimes provide a finer characterization of the trade-off between Type-I and Type-II errors than the Stein regime, we apply the results above to refine and extend several existing studies. In particular, we recover the quantum Stein's lemma for composite correlated hypotheses as established in~\cite{fang2024generalized}, and further strengthen it by providing explicit convergence rates for the Type-I error. Note that the Stein exponent only indicates whether the Type-I error decays, but does not quantify how fast it decays. Specifically, under the same assumptions as in~\cite{fang2024generalized}, we establish the following refined asymptotic trade-off between Type-I and Type-II errors:
\begin{align}
    \liminf_{n\to \infty} -\frac{1}{n} \log \alpha_{n, r}(\sA_n\| \sB_n) \geq \bH_{r}^\infty(\sA\|\sB)  > 0,
\end{align} 
for any $0 < r < D^\infty(\sA\|\sB)$, and 
\begin{align} 
    \liminf_{n\to \infty} -\frac{1}{n} \log (1-\alpha_{n, r}(\sA_n\| \sB_n)) \geq \bH_{r}^{*,\infty}(\sA\|\sB) > 0,
\end{align}  
for any $r > D^\infty(\sA\|\sB)$,
where $D^\infty(\sA\|\sB)$ is the regularized relative entropy between the sets.
These indicate that any Type-II error exponent $r$ below $D^\reg(\sA\|\sB)$ is achievable, with the corresponding Type-I error decaying exponentially at a rate at least $\bH_{r}^\infty(\sA\|\sB)$. Conversely, if the Type-II error exponent $r$ exceeds $D^\reg(\sA\|\sB)$, the Type-I error inevitably converges to one exponentially fast, with a rate at least $\bH_{r}^{*,\infty}(\sA\|\sB)$. Thus, the regularized quantum relative entropy delineates a {sharp threshold} for the asymptotic trade-off in hypothesis testing between two sets of quantum states. 

Besides this, we also provide a few more explicit applications, including: (i) finding counterexamples to the continuity of the regularized Petz \Renyi divergence and Hoeffding divergence; (ii) deriving error exponents for adversarial quantum channel discrimination; and (iii) obtaining error exponents for resource detection problems in coherence theory and entanglement theory.


The rest of the paper is organized as follows. In Section~\ref{sec: preliminaries} we introduce the notation and review a few quantum divergences, together with their extensions to sets of states. In Section~\ref{sec: Hoeffding divergence and anti-divergence for sets of states} we introduce two natural extensions of the Hoeffding divergence and anti‑divergence to sets and prove their equivalence or quantitative relation. In Section~\ref{sec: Optimal Type-I error probability} we formulate asymmetric hypothesis testing for composite correlated hypotheses and show how the worst‑case problem reduces to pairwise optimizations. Section~\ref{sec: Quantum Hoeffding bound for composite correlated hypotheses} establishes the quantum Hoeffding bound for composite correlated hypotheses, while Section~\ref{sec: Strong converse exponent for composite correlated hypotheses} studies the strong converse exponent. Section~\ref{sec: applications} discusses applications and implications of our results. In Section~\ref{sec: discussion} we conclude the paper and discuss a few open questions for future exploration. Some technical lemmas are delegated to the appendices.

\section{Preliminaries}
\label{sec: preliminaries}

\subsection{Notation}

Throughout this work, we adopt the following notational conventions. Finite-dimensional Hilbert spaces are denoted by $\cH$, with $|\cH|$ indicating their dimension. The set of all linear operators on $\cH$ is denoted by $\sL(\cH)$, while $\HERM(\cH)$ and $\HERM_{\pl}(\cH)$ denote the sets of Hermitian and positive semidefinite operators on $\cH$, respectively. The set of density operators (i.e., positive semidefinite operators with unit trace) on $\cH$ is denoted by $\density(\cH)$. Calligraphic letters such as $\sA$, $\sB$, and $\sC$ are used to represent sets of linear operators. Unless otherwise specified, all logarithms are taken to base two and denoted by $\log(x)$.
The positive semidefinite ordering is written as $X \geq Y$ if and only if $X - Y \geq 0$. The absolute value of an operator $X$ is defined by $|X|:= (X^\dagger X)^{1/2}$.
For a Hermitian operator $X$ with spectral decomposition $X = \sum_i x_i E_i$, the projection onto the
non-negative eigenspaces is denoted by $\{X \geq 0\} := \sum_{x_i \geq 0} E_i$. Similarly, $\{X > 0\} := \sum_{x_i > 0} E_i$.

\subsection{Quantum divergences}

A functional $\DD: \density \times \PSD \to \RR$ is called a \emph{quantum divergence} if it satisfies the data-processing inequality: for any completely positive and trace-preserving (CPTP) map $\cE$ and any $(\rho,\sigma) \in \density \times \PSD$, it holds that $\DD(\cE(\rho)\|\cE(\sigma)) \leq \DD(\rho\|\sigma)$. In the following, we introduce several quantum divergences that will be used throughout this work. We also define quantum divergences between two sets of quantum states.

\begin{definition}[Umegaki relative entropy~\cite{umegaki1954conditional}.]
For any $\rho\in \density$ and $\sigma \in \PSD$, the Umegaki relative entropy is defined by
\begin{align}\label{eq: Umegaki}
    D(\rho\|\sigma):= \tr[\rho(\log \rho - \log \sigma)],
\end{align}
if $\supp(\rho) \subseteq \supp(\sigma)$ and $\infty$ otherwise.
\end{definition}

\begin{definition}[Petz \Renyi divergence~\cite{petz1986quasi}.]
Let $\alpha \in (0,1) \cup (1,\infty)$. For any $\rho\in \density$ and $\sigma \in \PSD$, the Petz \Renyi divergence is defined by
\begin{align}\label{eq: Petz}
    D_{\Petz,\alpha}(\rho\|\sigma) := \frac{1}{\alpha-1}\log Q_{\alpha}(\rho \| \sigma),
\end{align}
with $Q_\alpha(\rho \| \sigma) := \tr\left[\rho^\alpha\sigma^{1-\alpha}\right]$,
if $\alpha \in (0,1)$ and $\rho \not\perp \sigma$, or if $\alpha \in (1,\infty)$ and $\supp(\rho) \subseteq \supp(\sigma)$; it is defined by $\infty$ otherwise.
\end{definition}

\begin{definition}[Sandwiched \Renyi divergence~\cite{muller2013quantum,wilde2014strong}.]
Let $\alpha \in (0,1) \cup (1,\infty)$. For any $\rho\in \density$ and $\sigma \in \PSD$, the sandwiched \Renyi divergence is defined by
\begin{align}\label{eq: Sandwiched}
    D_{\Sand,\alpha}(\rho\|\sigma) := \frac{1}{\alpha-1}\log\tr\left[\sigma^{\frac{1-\alpha}{2\alpha}}\rho\sigma^{\frac{1-\alpha}{2\alpha}}\right]^\alpha,
\end{align}
{if $\alpha \in (0,1)$ and $\rho \not\perp \sigma$, or if $\alpha \in (1,\infty)$ and $\supp(\rho) \subseteq \supp(\sigma)$; it is defined by $\infty$ otherwise.}
\end{definition}

\begin{definition}[Quantum divergence between two sets of states.]\label{def: divergence between two sets}
    Let $\DD$ be a quantum divergence between two quantum states. Let $\cH$ be a finite-dimensional Hilbert space. Then for any sets $\sA,\sB\subseteq \density(\cH)$, the quantum divergence between these two sets is defined by
    \begin{align}
        \DD(\sA\|\sB):= \inf_{\substack{\rho \in \sA\\ \sigma \in \sB}} \DD(\rho\|\sigma). 
    \end{align}
   Let $\sA = \{\sA_n\}_{n\in \NN}$ and $\sB = \{\sB_n\}_{n\in \NN}$ be two sequences of sets of quantum states\footnote{We abuse the notation $\sA,\sB$ to refer both to sets of states and to sequences of such sets, depending on the context.}, where each $\sA_n, \sB_n \subseteq \density(\cH^{\ox n})$. the regularized divergence between these sequences is defined by 
\begin{align}
    \underline{\DD}^{\reg}(\sA \| \sB) & := \liminf_{n \to \infty}\, \frac{1}{n} \DD(\sA_{n} \| \sB_{n}),\\
    \overline{\DD}^{\reg}(\sA \| \sB) & := \limsup_{n \to \infty} \frac{1}{n} \DD(\sA_{n} \| \sB_{n}).
\end{align}
If the limit exists, we define the regularized divergence as
\begin{align}
    \DD^{\reg}(\sA \| \sB) := \lim_{n \to \infty} \frac{1}{n} \DD(\sA_{n} \| \sB_{n}).
\end{align}
\end{definition}

\begin{remark}\label{rem: infimum divergence achievability}
    Note that if $\DD$ is lower semicontinuous (which is true for most quantum divergences of interest), and $\sA$ and $\sB$ are compact sets, the infimum in the above expression is always attained and can thus be replaced by a minimization~{(see, e.g.,~\cite[Lemma 2.8]{ben2023new})}.
\end{remark}

In many practical scenarios, the sequences of sets of interest are not arbitrary but possess a structure that is compatible with tensor products. This property, known as \emph{stability} (or closeness) under tensor product, is formalized as follows.

\begin{definition}[Stable sequence.]\label{def: closed under tensor product}
Let $\sA \subseteq \PSD(\cH_1)$, $\sB \subseteq \PSD(\cH_2)$, and $\sC \subseteq \PSD(\cH_1 \otimes \cH_2)$. We say that $\{\sA, \sB, \sC\}$ is stable under tensor product if, for any $X_1 \in \sA$ and $X_2 \in \sB$, it holds that $X_1 \otimes X_2 \in \sC$. In short, we write $\sA \otimes \sB \subseteq \sC$. A sequence of sets $\{\sC_n\}_{n \in \NN}$ with $\sC_n \subseteq \PSD(\cH^{\ox n})$ is stable under tensor product if $\sC_n \otimes \sC_m \subseteq \sC_{n+m}$ for all $n, m \in \NN$.
\end{definition}

\begin{remark}\label{rem: regularized divergence existence}
If the divergence $\DD$ is subadditive under tensor product states, its extension to sets of states is also subadditive for stable sequences of sets~\cite[Lemma 26]{fang2024generalized}. This implies the existence of the regularized divergence by Fekete's lemma,
\begin{align}
\DD^{\reg}(\sA \| \sB) & = \underline{\DD}^{\reg}(\sA \| \sB)\\
& = \overline{\DD}^{\reg}(\sA \| \sB) = \inf_{n \in \NN} \frac{1}{n} \DD(\sA_n \| \sB_n).
\end{align}
\end{remark}

\section{Hoeffding divergence and anti-divergence for sets of states}
\label{sec: Hoeffding divergence and anti-divergence for sets of states}

In this section, we develop extensions of the quantum Hoeffding divergence and anti-divergence to sets of quantum states. Two natural approaches arise for this purpose. The first approach treats the Hoeffding divergence as a quantum divergence and extends it to sets via Definition~\ref{def: divergence between two sets}. The second approach leverages the explicit formula for the Hoeffding divergence in terms of the Petz \Renyi divergences. Analogous constructions apply to the Hoeffding anti-divergence.

\subsection{Quantum Hoeffding divergence}

\begin{definition}
    Let $\cH$ be a finite-dimensional Hilbert space, $r > 0$ be a real number, and $n \in \NN$.
    Let $\rho_n, \sigma_n \in \density(\cH^{\ox n})$. The quantum Hoeffding divergence is defined by
    \begin{align}\label{eq: definition of Hoeffding for states}
        H_{n, r}(\rho_n\|\sigma_n) := \sup_{\alpha \in (0,1)} \frac{\alpha - 1}{\alpha} \left(nr - D_{\Petz,\alpha}(\rho_n\|\sigma_n)\right).
    \end{align}
\end{definition}

\begin{lemma}[Subadditivity.]\label{lem: subadditivity of Hoeffding divergence states}
For any $\rho_m, \sigma_m \in \density(\cH^{\ox m})$ and $\rho_n, \sigma_n \in \density(\cH^{\ox n})$, it holds that
\begin{align}\label{eq: subadditivity of Hoeffding divergence}
    H_{(m+n), r}&(\rho_m \otimes \rho_n\|\sigma_m \otimes \sigma_n)\notag \\
    & \leq H_{m, r}(\rho_m\|\sigma_m) + H_{n, r}(\rho_n\|\sigma_n).
\end{align}
\end{lemma}

\begin{proof}
This can be seen as follows:
\begin{align}
   & H_{(m+n), r}(\rho_m \otimes \rho_n\|\sigma_m \otimes \sigma_n) \notag\\
     & = \sup_{\alpha \in (0,1)} \frac{\alpha - 1}{\alpha} \Big\{(m+n)r\notag \\
     &\hspace{3cm} - D_{\Petz,\alpha}(\rho_m \otimes \rho_n\|\sigma_m \otimes \sigma_n)\Big\} \\
   & = \sup_{\alpha \in (0,1)} \frac{\alpha - 1}{\alpha} \Big\{(m+n)r \notag \\
   & \hspace{2.5cm}- D_{\Petz,\alpha}(\rho_m\|\sigma_m) - D_{\Petz,\alpha}(\rho_n\|\sigma_n)\Big\} \\
   & \leq \sup_{\alpha \in (0,1)} \frac{\alpha - 1}{\alpha} \Big\{mr - D_{\Petz,\alpha}(\rho_m\|\sigma_m)\Big\} \notag \\
   &\hspace{2cm} + \sup_{\alpha \in (0,1)} \frac{\alpha - 1}{\alpha} \Big\{nr - D_{\Petz,\alpha}(\rho_n\|\sigma_n)\Big\} \\
    & = H_{m, r}(\rho_m\|\sigma_m) + H_{n, r}(\rho_n\|\sigma_n),\label{eq: subadditivity of Hoeffding divergence tmp1}
\end{align}
where the second equality uses the additivity of $D_{\Petz,\alpha}$ under tensor product states, and the inequality follows from splitting the supremum over $\alpha$ for each term. 
\end{proof}

\begin{definition}[Quantum Hoeffding divergence between sets of states.]\label{def: Hoeffding divergence}
Let $\cH$ be a finite-dimensional Hilbert space, and $r>0$ be a real number, $n\in \NN$. Let $\sA_n, \sB_n \subseteq \density(\cH^{\ox n})$ be two sets of quantum states. Two variants of the quantum Hoeffding divergence between these sets are defined by
\begin{align}
    H_{n, r}(\sA_n\|\sB_n) & := \inf_{\substack{\rho_n \in \sA_n\\ \sigma_n \in \sB_n}} H_{n, r}(\rho_n\|\sigma_n),\\
    \bH_{n, r}(\sA_n\|\sB_n) & := \sup_{\alpha \in (0,1)} \frac{\alpha - 1}{\alpha} \left(nr - D_{\Petz,\alpha}(\sA_n\|\sB_n)\right),
\end{align}
where $D_{\Petz,\alpha}(\sA_n\|\sB_n)$ is defined by in Definition~\ref{def: divergence between two sets}.
Moreover, let $\sA = \{\sA_n\}_{n\in \NN}$ and $\sB = \{\sB_n\}_{n\in \NN}$ be two sequences of sets of quantum states, where each $\sA_n, \sB_n \subseteq \density(\cH^{\ox n})$. The regularized quantum Hoeffding divergences between these sequences are defined by
\begin{align}
    \underline{H}_r^\reg(\sA\|\sB) & := \liminf_{n\to \infty} \frac{1}{n} H_{n, r}(\sA_n\|\sB_n),\\
    \overline{H}_r^\reg(\sA\|\sB) & := \limsup_{n\to \infty} \frac{1}{n} H_{n, r}(\sA_n\|\sB_n),\\
    \underline{\bH}_r^\reg(\sA\|\sB) & := \sup_{\alpha \in (0,1)} \frac{\alpha - 1}{\alpha} \left(r - \underline{D}^\infty_{\Petz,\alpha}(\sA\|\sB)\right),\\
    \overline{\bH}_r^\reg(\sA\|\sB) & := \sup_{\alpha \in (0,1)} \frac{\alpha - 1}{\alpha} \left(r - \overline{D}^\infty_{\Petz,\alpha}(\sA\|\sB)\right),
\end{align}
where $\underline{D}^\infty_{\Petz,\alpha}(\sA\|\sB)$ and $\overline{D}^\infty_{\Petz,\alpha}(\sA\|\sB)$ are defined by in Definition~\ref{def: divergence between two sets}.
If the limits exist, we define the regularized quantum Hoeffding divergence as
\begin{align}
    H_r^\reg(\sA\|\sB) & := \lim_{n\to \infty} \frac{1}{n} H_{n, r}(\sA_n\|\sB_n),\\
    {\bH}_r^\reg(\sA\|\sB) & := \sup_{\alpha \in (0,1)} \frac{\alpha - 1}{\alpha} \left(r - {D}^\infty_{\Petz,\alpha}(\sA\|\sB)\right),
\end{align}
where ${D}^\infty_{\Petz,\alpha}(\sA\|\sB)$ is defined by in Definition~\ref{def: divergence between two sets}.
\end{definition}

\begin{remark}[Attainment.]\label{rem: hoeffding inf min}
 Since $D_{\Petz,\alpha}(\rho_n\|\sigma_n)$ is lower semicontinuous in $(\rho_n,\sigma_n)$ for any fixed $\alpha$~\cite[Proposition III.11]{mosonyi2023some}, it follows from Lemma~\ref{lem: inf usc} that $H_{n, r}(\rho_n\|\sigma_n)$ is also lower semicontinuous in $(\rho_n,\sigma_n)$. Consequently, if $\sA_n$ and $\sB_n$ are compact sets, we know from Lemma~\ref{lem: compact lsc} that the infimum in the definition of $H_{n, r}(\sA_n\|\sB_n)$ is attained.
\end{remark}

\begin{remark}[Limit existence.]\label{rem: subadditivity of Hoeffding divergence}
The quantum Hoeffding divergence is subadditive under tensor product by Lemma~\ref{lem: subadditivity of Hoeffding divergence states}. So this property extends to stable sequences of sets by Remark~\ref{rem: regularized divergence existence},
\begin{align}\label{eq: subadditivity of Hoeffding divergence}
    H_{(m+n), r}&(\sA_{m+n}\|\sB_{m+n})\notag \\
    & \leq H_{m, r}(\sA_m\|\sB_m) + H_{n, r}(\sA_n\|\sB_n).
\end{align}
As a consequence, the regularized quantum Hoeffding divergence exists and satisfies
\begin{align}
    H_r^\reg(\sA\|\sB) & = \overline{H}_r^\reg(\sA\|\sB)\\
    & = \underline{H}_r^\reg(\sA\|\sB) = \inf_{n \geq 1} \frac{1}{n} H_{n, r}(\sA_n\|\sB_n).\label{eq: multiplicativity of Hoeffding divergence}
\end{align}
Similarly, due to the additivity of $D_{\Petz,\alpha}$ under tensor product states, the regularized quantum Hoeffding divergence $\bH_r^\reg(\sA\|\sB)$ also exists for stable sequences and satisfies
\begin{align}
    \bH_r^\reg(\sA\|\sB) = \overline{\bH}_r^\reg(\sA\|\sB) = \underline{\bH}_r^\reg(\sA\|\sB).
\end{align}
\end{remark}

\bigskip
The following results establish the relationship between the two variants of the quantum Hoeffding divergence for sets and sequences of sets.

\begin{lemma}[Finite equivalence.]\label{lem: Hoeffding Petz sets}
    Let $\cH$ be a finite-dimensional Hilbert space, and $r>0$ be a real number, $n\in \NN$. Let $\sA_n, \sB_n \subseteq \density(\cH^{\ox n})$ be two convex compact sets of quantum states. Then 
\begin{align}
    H_{n, r}(\sA_n\|\sB_n) = \bH_{n, r}(\sA_n\|\sB_n).
\end{align}
\end{lemma}

\begin{proof}
    This was previously established in~\cite[Lemma II.8]{Mosonyi_2022}. 
\end{proof}

\begin{remark}[Computability.]\label{rem: computability of Hoeffding sets}
For any fixed $\alpha \in (0,1)$, the function $Q_\alpha(\rho_n\|\sigma_n)$ is jointly concave in $(\rho_n,\sigma_n)$. Consequently, the quasi-divergence $Q_{\alpha}(\sA_n\|\sB_n)$ can be efficiently computed using the QICS package~\cite{he2024qics}, provided that $\sA_n$ and $\sB_n$ admit semidefinite representations. If the sets $\sA_n$ and $\sB_n$ exhibit additional symmetries, the computational complexity can be further reduced. With this, $\bH_{n, r}(\sA_n\|\sB_n)$ can be efficiently evaluated by scanning over $\alpha\in (0,1)$. 
\end{remark}

\begin{remark}[Alternative expression.]\label{rem: alternative expression of Hoeffding divergence}
By the duality relation of the Petz \Renyi divergences, we have the following alternative expression for the Hoeffding divergence:
\begin{align}
H_{n, r}&
(\sA_n\|\sB_n) 
\notag \\ =
& \sup_{\alpha \in (0,1)} 
\frac{\alpha - 1}{\alpha} \left(nr - D_{\Petz,\alpha}
(\sA_n\|\sB_n) \right) \\
=& \sup_{\alpha \in (0,1)} 
\frac{\alpha - 1}{\alpha} \left(nr +
\frac{\alpha}{\alpha-1}
D_{\Petz,1-\alpha}(\sB_n\|\sA_n)
\right) \\
=& \sup_{\alpha \in (0,1)} 
\left(
\frac{\alpha - 1}{\alpha} nr +
D_{\Petz,1-\alpha}(\sB_n\|\sA_n)
\right) \\
=& \sup_{\alpha \in (0,1)} 
\left(
-\frac{\alpha}{1-\alpha} nr +
D_{\Petz,\alpha}(\sB_n\|\sA_n)
\right),
\end{align}
where we replace $\alpha$ by $1-\alpha$ in the {last} equality. This alternative formulation is particularly useful when analyzing the Stein's exponent by taking the limit $r \to 0$.
\end{remark}

\begin{lemma}[Asymptotic relation.]\label{lem: Hoeffding Petz sets regularized}
Let $\cH$ be a finite-dimensional Hilbert space, and $r > 0$ be a real number. Let $\sA = \{\sA_n\}_{n\in \NN}$ and $\sB = \{\sB_n\}_{n\in \NN}$ be two stable sequences of convex compact sets of quantum states, where each $\sA_n, \sB_n \subseteq \density(\cH^{\ox n})$. Then for any $n \in \NN$, 
    \begin{align}
      \frac{1}{n} \bH_{n, r}(\sA_n\|\sB_n) \geq  H_r^\infty(\sA\|\sB) \geq \bH_r^\infty(\sA\|\sB).
    \end{align}
\end{lemma}

\begin{proof}
The existence of the regularizations is ensured by Remark~\ref{rem: subadditivity of Hoeffding divergence}. Then we have the following chain of inequalities:
    \begin{align}
        H_r^\infty&(\sA\|\sB)\notag\\
         & = \inf_{n \geq 1} \frac{1}{n} H_{n, r}(\sA_n\|\sB_n)\\
        & = \inf_{n \geq 1} \sup_{\alpha \in (0,1)} \frac{\alpha - 1}{\alpha} \left(r - \frac{1}{n}D_{\Petz,\alpha}(\sA_n\|\sB_n)\right)\\
        & \geq \sup_{\alpha \in (0,1)} \inf_{n \geq 1}  \frac{\alpha - 1}{\alpha} \left(r - \frac{1}{n}D_{\Petz,\alpha}(\sA_n\|\sB_n)\right)\\
        & = \sup_{\alpha \in (0,1)}  \frac{\alpha - 1}{\alpha} \left(r - \inf_{n \geq 1} \frac{1}{n}D_{\Petz,\alpha}(\sA_n\|\sB_n)\right)\\
        & = \sup_{\alpha \in (0,1)} \frac{\alpha - 1}{\alpha} \left(r - D_{\Petz,\alpha}^\infty(\sA\|\sB)\right)\\
        & = \bH_r^\infty(\sA\|\sB),
    \end{align}
    where the first equality follows from Remark~\ref{rem: subadditivity of Hoeffding divergence}, the second equality follows from Lemma~\ref{lem: Hoeffding Petz sets}, the inequality follows by minimax inequality, the fourth equality follows from Remark~\ref{rem: subadditivity of Hoeffding divergence}, and the last equality follows by definition.
Moreover, we have 
\begin{align}
    H_r^\infty(\sA\|\sB) & = \inf_{n \geq 1} \frac{1}{n} H_{n, r}(\sA_n\|\sB_n)\\
    & \leq \frac{1}{n} H_{n, r}(\sA_n\|\sB_n) = \frac{1}{n} \bH_{n, r}(\sA_n\|\sB_n).
\end{align}
This completes the proof.
\end{proof}

\begin{remark}\label{rem: equality of Hoeffding Petz sets regularized}
If the minimax equality in the above proof can be established, then we would have $H_r^\infty(\sA\|\sB) = \bH_r^\infty(\sA\|\sB)$. However, this appears to be challenging, as existing minimax theorems typically require at least one of the spaces to be compact—a condition that is not directly satisfied here. Moreover, as we show later that the Petz \Renyi divergence and the Hoeffding divergence between two sets can be nonadditive in general (see Section~\ref{sec: applications to adversarial quantum channel discrimination}), which further complicates the situation. Nevertheless, in the special case where the Petz \Renyi divergence is additive for all $\alpha \in (0,1)$, i.e., $\frac{1}{n} D_{\Petz,\alpha}(\sA_n\|\sB_n) = D_{\Petz,\alpha}(\sA_1\|\sB_1)$, we have that
\begin{align}
    H_r^\infty(\sA\|\sB) & = \bH_r^\infty(\sA\|\sB)\\
    & = H_{1,r}(\sA_1\|\sB_1) = \bH_{1,r}(\sA_1\|\sB_1).
\end{align}
\end{remark}

\subsection{Quantum Hoeffding anti-divergence}

Analogous to the quantum Hoeffding divergence, we can also define the quantum Hoeffding anti-divergence for sets of quantum states.

\begin{definition}
   Let $\cH$ be a finite-dimensional Hilbert space, $r>0$ be a real number, $n\in \NN$. Let $\rho_n, \sigma_n \in \density(\cH^{\ox n})$. The quantum Hoeffding anti-divergence is defined by
    \begin{align}\label{eq: definition of Hoeffding anti-divergence}
        H^*_{n, r}(\rho_n\|\sigma_n) := \sup_{\alpha > 1} \frac{\alpha - 1}{\alpha} \left(nr - D_{\Sand,\alpha}(\rho_n\|\sigma_n)\right).
    \end{align}
\end{definition}

\begin{lemma}[Subadditivity.]
For any $\rho_m, \sigma_m \in \density(\cH^{\ox m})$ and $\rho_n, \sigma_n \in \density(\cH^{\ox n})$, it holds that
\begin{align}\label{eq: subadditivity of Hoeffding anti-divergence}
    H_{(m+n), r}^*& (\rho_m \otimes \rho_n\|\sigma_m \otimes \sigma_n)\notag \\
    & \leq H_{m, r}^*(\rho_m\|\sigma_m) + H^*_{n, r}(\rho_n\|\sigma_n).
\end{align}
\end{lemma}

\begin{proof}
This can be seen as follows:
\begin{align}
    &H_{(m+n), r}^*(\rho_m \otimes \rho_n\|\sigma_m \otimes \sigma_n) \notag\\
     & = \sup_{\alpha > 1} \frac{\alpha - 1}{\alpha} \Big\{(m+n)r\notag \\
     &\hspace{3cm} - D_{\Sand,\alpha}(\rho_m \otimes \rho_n\|\sigma_m \otimes \sigma_n)\Big\} \\
   & = \sup_{\alpha > 1} \frac{\alpha - 1}{\alpha} \Big\{(m+n)r\notag \\
   &\hspace{2.5cm} - D_{\Sand,\alpha}(\rho_m\|\sigma_m) - D_{\Sand,\alpha}(\rho_n\|\sigma_n)\Big\} \\
   & \leq \sup_{\alpha > 1} \frac{\alpha - 1}{\alpha} \Big\{mr - D_{\Sand,\alpha}(\rho_m\|\sigma_m)\Big\} \notag \\
   & \hspace{2cm}+ \sup_{\alpha > 1} \frac{\alpha - 1}{\alpha} \Big\{ nr - D_{\Sand,\alpha}(\rho_n\|\sigma_n)\Big\} \\
    & = H_{m, r}^*(\rho_m\|\sigma_m) + H^*_{n, r}(\rho_n\|\sigma_n),\label{eq: subadditivity of Hoeffding anti-divergence tmp1}
\end{align}
where the second equality uses the additivity of $D_{\Sand,\alpha}$ under tensor product states, and the inequality follows from splitting the supremum over $\alpha$ for each term.
\end{proof}

\begin{definition}[Quantum Hoeffding anti-divergence between sets of states.]\label{def: Hoeffding anti-divergence}
Let $\cH$ be a finite-dimensional Hilbert space, and $r>0$ be a real number, $n\in \NN$. Let $\sA_n, \sB_n \subseteq \density(\cH^{\ox n})$. Two variants of the quantum Hoeffding anti-divergence between these sets are defined by~\footnote{The anti-divergence is monotone non-decreasing under CPTP maps, so the extension to sets is based on the supremum rather than the infimum.}
\begin{align}
    H^*_{n, r}(\sA_n\|\sB_n) & := \sup_{\substack{\rho_n \in \sA_n\\ \sigma_n \in \sB_n}} H^*_{n, r}(\rho_n\|\sigma_n),\\
     \bH^*_{n, r}(\sA_n\|\sB_n) & := \sup_{\alpha > 1} \frac{\alpha - 1}{\alpha} \left(nr - D_{\Sand,\alpha}(\sA_n\|\sB_n)\right),
\end{align}
where $D_{\Sand,\alpha}(\sA_n\|\sB_n)$ is defined by in Definition~\ref{def: divergence between two sets}.
Moreover, let $\sA = \{\sA_n\}_{n\in \NN}$ and $\sB = \{\sB_n\}_{n\in \NN}$ be two sequences of sets of quantum states, where each $\sA_n, \sB_n \subseteq \density(\cH^{\ox n})$. The regularized quantum Hoeffding anti-divergences between these sequences are defined by
\begin{align}
    \underline{H}_r^{*,\reg}(\sA\|\sB) & := \liminf_{n\to \infty} \frac{1}{n} H^*_{n, r}(\sA_n\|\sB_n),\\
    \overline{H}_r^{*,\reg}(\sA\|\sB) & := \limsup_{n\to \infty} \frac{1}{n} H^*_{n, r}(\sA_n\|\sB_n),\\
    \underline{\bH}_r^{*,\reg}(\sA\|\sB) & := \sup_{\alpha > 1} \frac{\alpha - 1}{\alpha} \left(r - \underline{D}_{\Sand,\alpha}^\infty(\sA\|\sB)\right),\\
    \overline{\bH}_r^{*,\reg}(\sA\|\sB) & := \sup_{\alpha > 1} \frac{\alpha - 1}{\alpha} \left(r - \overline{D}_{\Sand,\alpha}^\infty(\sA\|\sB)\right).
\end{align}
where $\underline{D}^\infty_{\Sand,\alpha}(\sA\|\sB)$ and $\overline{D}^\infty_{\Sand,\alpha}(\sA\|\sB)$ are defined by in Definition~\ref{def: divergence between two sets}.
If the limits exist, we define the regularized Hoeffding divergence as
\begin{align}
    H_r^{*,\reg}(\sA\|\sB) & := \lim_{n\to \infty} \frac{1}{n} H^*_{n, r}(\sA_n\|\sB_n),\\
    \bH_r^{*,\reg}(\sA\|\sB) & := \sup_{\alpha > 1} \frac{\alpha - 1}{\alpha} \left(r - {D}_{\Sand,\alpha}^\infty(\sA\|\sB)\right),
\end{align}
where ${D}^\infty_{\Sand,\alpha}(\sA\|\sB)$ is defined by in Definition~\ref{def: divergence between two sets}.
\end{definition}

\begin{remark}[Attainment.]
    It is known that $H_{n,r}^*(\rho_n\|\sigma_n)$ is upper semicontinuous in $(\rho_n,\sigma_n)$ \cite[Corollary V.16]{Mosonyi_2022}. So the supremum in $H^*_{n, r}(\sA_n\|\sB_n)$ is achieved for any compact sets. 
\end{remark}

The following results aim to establish the relationship between the two variants of the quantum Hoeffding anti-divergence for sets and sequences of sets.

\begin{lemma}[Finite equivalence.]\label{lem: Hoeffding anti equivalence}
  Let $\cH$ be a finite-dimensional Hilbert space, $r>0$ be a real number, and $n\in \NN$. Let $\sA_n, \sB_n \subseteq \density(\cH^{\ox n})$. Then it holds that
\begin{align}
    H^*_{n, r}(\sA_n\|\sB_n) = \bH^*_{n, r}(\sA_n\|\sB_n).
\end{align}
\end{lemma}

\begin{proof}
    By definition, we have 
    \begin{align}
        H^*_{n, r}&(\sA_n\|\sB_n) \notag\\
        & = \sup_{\substack{\rho_n \in \sA_n\\ \sigma_n \in \sB_n}} \sup_{\alpha > 1} \frac{\alpha - 1}{\alpha} \left(nr - D_{\Sand,\alpha}(\rho_n\|\sigma_n)\right)\\
        & =  \sup_{\alpha > 1} \sup_{\substack{\rho_n \in \sA_n\\ \sigma_n \in \sB_n}}\frac{\alpha - 1}{\alpha} \left(nr - D_{\Sand,\alpha}(\rho_n\|\sigma_n)\right)\\
        & =  \sup_{\alpha > 1} \frac{\alpha - 1}{\alpha} \left(nr - \inf_{\substack{\rho_n \in \sA_n\\ \sigma_n \in \sB_n}} D_{\Sand,\alpha}(\rho_n\|\sigma_n)\right)\\
        & = \sup_{\alpha > 1} \frac{\alpha - 1}{\alpha} \big(nr - D_{\Sand,\alpha}(\sA_n\|\sB_n)\big)\\
        & = \bH^*_{n, r}(\sA_n\|\sB_n),
    \end{align}
    where in the {second} line we exchange the two suprema.
\end{proof}

It is important to note that $H^*_{n, r}(\sA_n\|\sB_n)$ is defined by a supremum over the feasible states, which makes its additivity property for stable sequences unclear—even though we know the Hoeffding anti-divergence for states is subadditive. As a result, we cannot directly apply Remark~\ref{rem: regularized divergence existence} as in previous discussions of Remark~\ref{rem: subadditivity of Hoeffding divergence}. Nevertheless, the following result shows that the regularization $H_r^{*,\infty}$ indeed exists for stable sequences and coincides with $\bH_r^{*,\infty}$ in general.

\begin{lemma}[Asymptotic equivalence.]\label{lem: Hoeffding anti Sand sets regularized}
Let $\cH$ be a finite-dimensional Hilbert space, and $r>0$ be a real number. Let $\sA = \{\sA_n\}_{n\in \NN}$ and $\sB = \{\sB_n\}_{n\in \NN}$ be two stable sequences of sets of quantum states, where each $\sA_n, \sB_n \subseteq \density(\cH^{\ox n})$. Then it holds that
\begin{align}
    {H}_r^{*,\reg}(\sA\|\sB) = {\bH}_r^{*,\reg}(\sA\|\sB).
\end{align}
\end{lemma}

\begin{proof}
We have the following chain of inequalities:
\begin{align}
    \underline{H}_r^{*,\reg}&(\sA\|\sB)\notag \\
     & = \liminf_{n\to \infty} \frac{1}{n} H^*_{n, r}(\sA_n\|\sB_n)\\
    & = \liminf_{n\to \infty} \frac{1}{n} \bH^*_{n, r}(\sA_n\|\sB_n)\\
    & = \liminf_{n\to \infty} \frac{1}{n} \sup_{\alpha > 1} \frac{\alpha - 1}{\alpha} \big(nr - D_{\Sand,\alpha}(\sA_n\|\sB_n)\big)\\
    & \geq \sup_{\alpha > 1} \liminf_{n\to \infty} \frac{1}{n} \frac{\alpha - 1}{\alpha} \big(nr - D_{\Sand,\alpha}(\sA_n\|\sB_n)\big)\\
    & = \sup_{\alpha > 1} \frac{\alpha - 1}{\alpha} \big(r - \limsup_{n\to \infty} \frac{1}{n} D_{\Sand,\alpha}(\sA_n\|\sB_n)\big)\\
    & = \sup_{\alpha > 1} \frac{\alpha - 1}{\alpha} \big(r - D^\infty_{\Sand,\alpha}(\sA_n\|\sB_n)\big)\\
    & = {\bH}_r^{*,\reg}(\sA\|\sB),\label{eq: anti-divergence regularization proof tmp1}
\end{align}
where the second line follows from Lemma~\ref{lem: Hoeffding anti equivalence}, the inequality follows by the fact that for any sequence of numbers $x_{\alpha,n}$, $\liminf_{n\to \infty} \sup_{\alpha > 1} x_{\alpha,n} \geq \sup_{\alpha > 1} \liminf_{n\to \infty}  x_{\alpha,n}$, the second last line follows from Remark~\ref{rem: regularized divergence existence} and the stability of the sequences.

In the other direction, we have 
\begin{align}
        \overline{H}_r^{*,\reg}&(\sA\|\sB)\notag \\
         & = \limsup_{n\to \infty} \frac{1}{n} H^*_{n, r}(\sA_n\|\sB_n)\\
        & = \limsup_{n\to \infty} \frac{1}{n} \bH^*_{n, r}(\sA_n\|\sB_n)\\
        & = \limsup_{n\to \infty} \frac{1}{n} \sup_{\alpha > 1} \frac{\alpha - 1}{\alpha} \big(nr - D_{\Sand,\alpha}(\sA_n\|\sB_n)\big),
\end{align}
where the second line follows from Lemma~\ref{lem: Hoeffding anti equivalence}.
Note that $D_{\Sand,\alpha}(\sA_n\|\sB_n)$ is subadditive for stable sequences, so we have $\frac{1}{n} D_{\Sand,\alpha}(\sA_n\|\sB_n) \geq D_{\Sand,\alpha}^\infty(\sA\|\sB)$ for any $n \in \NN$. This gives 
\begin{align}
    \frac{1}{n}\sup_{\alpha > 1}  \frac{\alpha - 1}{\alpha} &  \big(nr - D_{\Sand,\alpha}(\sA_n\|\sB_n)\big)\notag \\
    & \leq \sup_{\alpha > 1} \frac{\alpha - 1}{\alpha} \big(r - D_{\Sand,\alpha}^\infty(\sA\|\sB)\big).
\end{align}
Then taking the limit of $n$, we have 
\begin{align}\label{eq: anti-divergence regularization proof tmp2}
    \overline{H}_r^{*,\reg}(\sA\|\sB) & \leq \sup_{\alpha > 1} \frac{\alpha - 1}{\alpha} \big(r - D_{\Sand,\alpha}^\infty(\sA\|\sB)\big)\\
    & = {\bH}_r^{*,\reg}(\sA\|\sB).
\end{align}
Combining~\eqref{eq: anti-divergence regularization proof tmp1} and~\eqref{eq: anti-divergence regularization proof tmp2}, we have the asserted result.
\end{proof}

\section{Optimal Type-I error probability}
\label{sec: Optimal Type-I error probability}

The error exponent and strong converse exponent regimes study the optimal behavior of the Type-I error provided that the Type-II error exponentially decays. {Recall that} the optimal Type-I error for hypothesis testing between two sets of quantum states, $\sA_n$ and $\sB_n$, is defined by
\begin{align}\label{eq: optimal Type-I error sets 2}
    \alpha_{n, r}&(\sA_n\|\sB_n)\notag \\
    & := \min_{0\leq M_n \leq I} \left\{\alpha(\sA_n, M_n): \beta(\sB_n, M_n) \leq 2^{-nr}\right\},
\end{align}
where the measurement $M_n$ is chosen to minimize the worst-case Type-I error $\alpha(\sA_n, M_n)$, subject to the constraint that the Type-II error $\beta(\sB_n, M_n)$ decays exponentially at a rate $r$. In other words, the measurement must perform universally well for all states in $\sA_n$ and $\sB_n$. 

The following shows that the optimal Type-I error for hypothesis testing between two sets of quantum states is precisely determined by the most challenging pair of states from these sets. This implies a universal measurement for $\sA_n$ and $\sB_n$ whose performance matches that of the optimal measurement for the worst-case pair of states, providing a key simplification that enables our subsequent analysis.

\begin{lemma}\label{lem: optimal Type-I error minimax}
    Let $\cH$ be a finite-dimensional Hilbert space, $r >0$ be a real number, and $n\in \NN$. Let $\sA_n, \sB_n\subseteq \density(\cH^{\ox n})$ be two convex sets of quantum states. Then it holds that
\begin{align}
    \alpha_{n, r}(\sA_n\| \sB_n) = \sup_{\substack{\rho_n \in \sA_n\\ \sigma_n \in \sB_n}} \alpha_{n, r}(\rho_n\| \sigma_n).
\end{align}
\end{lemma}

\begin{proof}
We begin by noting the following symmetry role between Type-I and Type-II errors:
\begin{align}
    \alpha(\sA_n, M_n) &= \beta(\sA_n, I-M_n), \\
     \beta(\sB_n, M_n) & = \alpha(\sB_n, I-M_n). \label{eq: Type-I and II exchange role}
\end{align}
This allows us to rewrite the optimization in Eq.~\eqref{eq: optimal Type-I error sets 2} as
\begin{align}
    & \alpha_{n, r}(\sA_n\| \sB_n) \notag \\
    &= \min_{\substack{0\leq M_n \leq I\\\alpha(\sB_n, I-M_n) \leq 2^{-nr}}}  \beta(\sA_n, I-M_n) \\
    &= \min_{\substack{0\leq M_n \leq I\\\alpha(\sB_n, M_n) \leq 2^{-nr}}} \beta(\sA_n, M_n) \\
    &= \sup_{\substack{\rho_n \in \sA_n \\ \sigma_n \in \sB_n}} \ \min_{\substack{0\leq M_n \leq I\\\alpha(\sigma_n, M_n) \leq 2^{-nr}}} \beta(\rho_n, M_n) \\
    &= \sup_{\substack{\rho_n \in \sA_n \\ \sigma_n \in \sB_n}} \ \min_{\substack{0\leq M_n \leq I\\\alpha(\sigma_n, I-M_n) \leq 2^{-nr}}} \beta(\rho_n, I-M_n) \\
    &= \sup_{\substack{\rho_n \in \sA_n \\ \sigma_n \in \sB_n}} \ \min_{\substack{0\leq M_n \leq I\\\beta(\sigma_n, M_n) \leq 2^{-nr}}}  \alpha(\rho_n, M_n)  \\
    &= \sup_{\substack{\rho_n \in \sA_n \\ \sigma_n \in \sB_n}} \alpha_{n, r}(\rho_n\| \sigma_n),
\end{align}
where the first and {fifth} equalities use~\eqref{eq: Type-I and II exchange role}, the second and {fourth} equalities follow by substituting $M_n $ to $I - M_n$ in the optimization, the third equality uses~\cite[Lemma 31]{fang2024generalized} which allows us to pull out the optimization over $\sA_n, \sB_n$ for the optimal Type-II error probability when Type-I error is restricted to a constant threshold, and the last equality follows by definition.
\end{proof}

\begin{remark}\label{rem: type errors convexity}
Note that for general sets $\sA_n$ and $\sB_n$, i.e., not necessarily convex, we have 
\begin{align}
    \alpha(\sA_n, M_n) &= \alpha(\conv(\sA_n), M_n),\\
    \beta(\sB_n, M_n) & = \beta(\conv(\sB_n), M_n),
\end{align}
where $\conv(\sC)$ denotes the convex hull of set $\sC$.
Therefore, we have the relation that 
\begin{align}
    \alpha_{n, r}(\sA_n\| \sB_n) = \alpha_{n, r}(\conv(\sA_n)\| \conv(\sB_n)).
\end{align}
\end{remark}

\section{Quantum Hoeffding bound for composite correlated hypotheses} 
\label{sec: Quantum Hoeffding bound for composite correlated hypotheses}

In this section, we establish the quantum Hoeffding bound for composite correlated hypotheses. Notably, the result holds under \emph{minimal and standard} assumptions, which are satisfied by many frameworks of interest, such as those considered in~\cite{hayashi2024generalized,lami2024solutiongeneralisedquantumsteins,fang2024generalized}.

\begin{assumption}[Assumptions for sets of quantum states.]
We denote the following assumptions for a sequence of sets of quantum states $\sC = \{\sC_n\}_{n\in \NN}$ where each $\sC_n \subseteq \density(\cH^{\ox n})$.
\begin{itemize}
    \item[(C1)] Convexity: For any $n \in \NN$, the set $\sC_n$ is convex.
    \item[(C2)] Compactness: For any $n \in \NN$, the set $\sC_n$ is compact.
    \item[(C3)] Stability under tensor product: For any $m,n \in \NN$, it holds that $\sC_{m} \ox \sC_{n} \subseteq \sC_{m+n}$.
\end{itemize}
{For two sequences of sets of quantum states $\sA = \{\sA_n\}_{n\in \NN}$ and $\sB = \{\sB_n\}_{n\in \NN}$ where each $\sA_n, \sB_n \subseteq \density(\cH^{\ox n})$, we denote the additional assumption as follows:}
\begin{itemize}
    \item[(C4)] Finiteness: $D_{\Petz, \alpha}(\sA_1\|\sB_1) < \infty$ for $\alpha \in (0,1)$.~\footnotemark
\end{itemize}
\end{assumption}

\footnotetext{This is a mild technical assumption, requiring that there exist $\rho \in \sA_1$ and $\sigma \in \sB_1$ which are not orthogonal.}

\begin{theorem}[Quantum Hoeffding bound for composite correlated hypotheses.]\label{thm: Hoeffding bound sets}
    Let $\cH$ be a finite-dimensional Hilbert space. Let $\sA = \{\sA_n\}_{n\in \NN}$ and $\sB = \{\sB_n\}_{n\in \NN}$ be two sequences of sets of quantum states, where each $\sA_n, \sB_n \subseteq \density(\cH^{\ox n})$. Let $0< r < D^\reg(\sA\|\sB)$ be a real number.
    \begin{itemize}
    \item If $\sA$ and $\sB$ satisfy assumptions (C1) and (C2):
    \begin{align}
        \liminf_{n\to \infty} -\frac{1}{n} \log \alpha_{n, r}(\sA_n\| \sB_n) & \geq \underline{H}_r^\infty(\sA\|\sB).
    \end{align}
    \item If $\sA$ and $\sB$ satisfy assumptions (C3) and (C4):
    \begin{align}
        \limsup_{n\to \infty} -\frac{1}{n} \log \alpha_{n, r}(\sA_n\| \sB_n) & \leq H_r^\infty(\sA\|\sB).
    \end{align}
    \end{itemize}

    Consequently, if $\sA$ and $\sB$ satisfy (C1), (C2), (C3) and (C4), then the following limit exists and 
    \begin{align}
        \lim_{n\to \infty} -\frac{1}{n} \log \alpha_{n, r}(\sA_n\| \sB_n)  = H_r^\infty(\sA\|\sB).
    \end{align}
\end{theorem}

\begin{remark}
    By Remark~\ref{rem: type errors convexity}, we can remove the convexity on $\sA_n$ and $\sB_n$ in Theorem~\ref{thm: Hoeffding bound sets} and get
    \begin{align}
        \lim_{n\to \infty} -\frac{1}{n} \log \alpha_{n, r}(\sA_n\| \sB_n)  = H_r^\infty(\conv(\sA)\|\conv(\sB)),
    \end{align}
    where $\conv(\sC) := \{\conv(\sC_n)\}_{n\in \NN}$ represents the sequences of the convex hulls.  This result strengthens~\cite[Eq.~(II.65)]{Mosonyi_2022} by establishing the tightness of the regularized Hoeffding bound. 
    Another related analysis of the Hoeffding bound for composite correlated hypotheses was undertaken in~\cite[Theorem 2]{hayashi2025entanglement} for the specific case where $\sA_n$ is a singleton and $\sB_n$ is the set of separable states. However, the lower bound provided therein is not tight in general.
\end{remark}

\subsection{Proof of the lower bound}

Recall that for any $V,W \in \PSD$ and $\alpha \in (0,1)$, it
 holds that~\cite{audenaert2007discriminating},
\begin{align}
    \tr[V^\alpha W^{1-\alpha}] \geq \tr W\{W \leq V\} + \tr V\{W > V\}.
\end{align}
Let $\rho_n \in \sA_n$ and $\sigma_n \in \sB_n$. 
Applying the inequality with the choice $V =  \rho_n$ and $W = 2^{nR}\sigma_n$ with an arbitrary real number $R$. 
Then we have 
 \begin{align}
    \tr 2^{nR}\sigma_n \{2^{nR}\sigma_n \leq \rho_n\} & + \tr \rho_n \{2^{nR}\sigma_n > \rho_n\}\notag \\ 
    & \leq 2^{n(1-\alpha)R} Q_{\alpha}(\rho_n\|\sigma_n).
 \end{align}
 This implies that 
 \begin{align}
    \tr \rho_n \{2^{nR}\sigma_n > \rho_n\} & \leq 2^{n(1-\alpha)R} Q_{\alpha}(\rho_n\|\sigma_n),\\
    \tr \sigma_n \{2^{nR}\sigma_n \leq \rho_n\} & \leq 2^{-n\alpha R} Q_{\alpha}(\rho_n\|\sigma_n).
 \end{align}
Now, letting the constant
\begin{align}
    R = \frac{nr + \log Q_{\alpha}(\rho_n\|\sigma_n)}{n\alpha}.
\end{align}
we get
\begin{align}
   \tr \rho_n \{2^{nR}\sigma_n > \rho_n\} & \leq 2^{\frac{1-\alpha}{\alpha}n\left(r-\frac{1}{n}D_{\Petz,\alpha}(\rho_n\|\sigma_n)\right)},\\
    \tr \sigma_n \{2^{nR}\sigma_n \leq \rho_n\} & \leq 2^{-nr}. \label{eq: feasible test tmp2}
\end{align}
Let $M_n = \{2^{nR}\sigma_n \leq \rho_n\}$, which is a valid measurement operator. Then Eq.~\eqref{eq: feasible test tmp2} implies that it is a feasible solution to the optimization problem in Eq.~\eqref{eq: optimal Type-I error}. Therefore, we have
\begin{align}
    \alpha_{n,r}(\rho_n\|\sigma_n) \leq 2^{\frac{1-\alpha}{\alpha}n\left(r-\frac{1}{n}D_{\Petz,\alpha}(\rho_n\|\sigma_n)\right)}.
\end{align}
This gives 
\begin{align}
   - \frac{1}{n}\log \alpha_{n, r}(\rho_n{\|}\sigma_n) \geq  \frac{\alpha-1}{\alpha}\left(r-\frac{1}{n}D_{\Petz,\alpha}(\rho_n\|\sigma_n)\right).
\end{align}
Taking infimum over $\rho_n \in \sA_n$ and $\sigma_n \in \sB_n$ on both sides, we have 
\begin{align}
    - \frac{1}{n}\log \alpha_{n, r}(\sA_n{\|}\sB_n) \geq  \frac{\alpha-1}{\alpha}\left(r-\frac{1}{n}D_{\Petz,\alpha}(\sA_n\|\sB_n)\right),
\end{align}
where we use Lemma~\ref{lem: optimal Type-I error minimax} and the assumption (C1).
Taking supremum over $\alpha \in (0,1)$, we have 
\begin{align}
   - \frac{1}{n}\log \alpha_{n, r}(\sA_n{\|}\sB_n) & \geq  \frac{1}{n} \bH_{n,r}(\sA_n\|\sB_n)\\
   & = \frac{1}{n}H_{n, r}(\sA_n\|\sB_n),
\end{align}
where the equality follows from Lemma~\ref{lem: Hoeffding Petz sets} and the assumptions (C1) and (C2).

Taking limit of $n$, we have
\begin{align}
    \liminf_{n\to \infty} & -\frac{1}{n}\log \alpha_{n, r}(\sA_n{\|}\sB_n) \notag \\
    & \geq \liminf_{n\to \infty}  \frac{1}{n}H_{n, r}(\sA_n\|\sB_n) = \underline{H}_{r}^\infty(\sA\|\sB).
\end{align}

\subsection{Proof of the upper bound}

We can easily prove the upper bound for limit inferior as follows. For any fixed $m \in \NN$ any $\rho_m \in \sA_m, \sigma_m \in \sB_m$, then 
 \begin{align}
   & \liminf_{n\to \infty}  -\frac{1}{n} \log \alpha_{n, r}(\sA_n\| \sB_n) \notag \\
    & \leq \liminf_{n\to \infty} -\frac{1}{mn} \log \alpha_{mn, r}(\sA_{mn}{\|} \sB_{mn})\\
    & \leq \liminf_{n\to \infty} -\frac{1}{mn} \log \sup_{\substack{\rho_{mn} \in \sA_{mn}\\ \sigma_{mn} \in \sB_{mn}}}\alpha_{mn, r}(\rho_{mn}{\|} \sigma_{mn})\\
    & \leq \liminf_{n\to \infty} -\frac{1}{mn} \log \alpha_{mn, r}(\rho_{m}^{\ox n}{\|} \sigma_{m}^{\ox n})\\
    & = \frac{1}{m} H_{m, r}(\rho_m\|\sigma_m),
 \end{align}
where the first inequality follows as the lower limit of a subsequence is no smaller than the lower limit of the sequence, the second inequality holds trivially as $\alpha_{n,r}(\sA_n\|\sB_n) \geq \alpha_{n,r}(\rho_n\|\sigma_n)$ for any $\rho_n \in \sA_n, \sigma_n \in \sB_n$ by definition, the third inequality follows by taking a particular feasible solution and the assumption (C3), the equality follows from the quantum Hoeffding bound between two quantum states (see Eq.~\eqref{eq: Hoeffding bound i.i.d.}). As this holds for any $\rho_m \in \sA_m, \sigma_m \in \sB_m$, 
\begin{align}
    \liminf_{n\to \infty} -\frac{1}{n} \log \alpha_{n, r}(\sA_n\| \sB_n) \leq \frac{1}{m} H_{m, r}(\sA_m\|\sB_m).
\end{align}
    Taking limit of $m$, we get
    \begin{align}
         \liminf_{n\to \infty} &-\frac{1}{n} \log \alpha_{n, r}(\sA_n\| \sB_n)\notag \\
         & \leq \liminf_{m\to \infty}\frac{1}{m} H_{m, r}(\sA_m\|\sB_m) = {H}_r^\infty(\sA\|\sB),
    \end{align}
    where the equality follows from the stability assumption (C3) and Eq.~\eqref{eq: multiplicativity of Hoeffding divergence}. 

\bigskip
The above arguement only gives the upper bound for limit inferior by choosing a suitable subsequence of i.i.d.~states. However, we show that the upper bound can be strengthened further to limit superior by carefully designing a sequence of states whose limit superior is also upper bounded by the regularized Hoeffding divergence. However, as this sequence is not i.i.d. states anymore, its analysis is more challenging and requires the Nussbaum-Szko\l{}a distributions~\cite{Nussbaum2009} and the G\"{a}rtner-Ellis theorem~\cite{Dembo2010}.

Let the spectral decompositions of $\rho$ and $\sigma$ be given by 
\begin{align}
    \rho = \sum_{i=1}^d \lambda_i \ketbra{u_i}{u_i}\quad \text{and} \quad \sigma = \sum_{j=1}^d \mu_j \ketbra{v_j}{v_j},
\end{align}
where $\ket{u_i}$ and $\ket{v_j}$ are two orthonormal bases and $\lambda_i$ and $\mu_j$ are the corresponding eigenvalues, respectively. Then the Nussbaum-Szko\l{}a distributions of 
$\rho,\sigma$ are defined by
\begin{align}
(P_{\rho,\sigma})(i,j) &= \lambda_i|\<u_i|v_j\>|^2, \\ (Q_{\rho,\sigma})(i,j) & = \mu_j |\<u_i|v_j\>|^2,
\end{align}
where $i,j \in \{1,\cdots, d\}$. 

In the remaining discussion of this section, let $\log$ be a logarithm with natural base $e$ for simplicity.
Given a sequence of random variables $\{X_n\}_{n\in \NN}$, the asymptotic cumulant generating function is defined by
\begin{align}
    \Lambda_X(t):= \lim_{n\to \infty} \frac{1}{n} \log \mathbb{E}\left[\exp({nt X_n})\right],
\end{align}
provided that the limit exists. 

{We will use the G\"{a}rtner--Ellis theorem~\cite[Theorem~2.3.6]{Dembo2010} in the following slightly tailored form, which suffices for our later arguments. Compared with the version employed in~\cite[Proposition~17]{hayashi2016correlation}, we state the bound for the open set \((x,\infty)\). Since \(\Pr\{X_n > x\} \leq \Pr\{X_n \geq x\}\), the same upper bound also applies to the closed set \([x,\infty)\).}

{
\begin{lemma}\label{lem: Gartner-Ellis}
Assume that the asymptotic cumulant generating function
\(
t \mapsto \Lambda_X(t)
\)
exists for all \(t\in\RR\), and that \(\Lambda_X\) is strictly convex and continuously differentiable on \(\RR\) (i.e., \(\Lambda_X\in C^1(\RR)\)).
Fix an open interval \((a,b)\subseteq \RR\).
Then, for any
\(
x \in \bigl(\Lambda_X'(a),\, \Lambda_X'(b)\bigr),
\)
    \begin{align}
        \limsup_{n\to \infty} -\frac{1}{n} \log \Pr\{X_n > x\} \leq \sup_{t\in (a,b)} \{tx - \Lambda_X(t)\}.
    \end{align}
\end{lemma}
\begin{proof}
    This is a direct consequence of the G\"{a}rtner--Ellis theorem (see, e.g.,~\cite[Theorem~2.3.6]{Dembo2010}), with minor adaptations.
    By assumption, $\Lambda_X$ exists on $\RR$ and is $C^{1}(\RR)$ and strictly convex; in particular, it is lower semicontinuous and essentially smooth (see, e.g.,~\cite[Page~251]{rockafellar}).
    Hence, by~\cite[Theorem~2.3.6(c)]{Dembo2010}, the sequence $\{X_n\}$ satisfies the large deviation principle with good rate function
    \begin{align}
        \Lambda_X^*(w):=\sup_{t\in\RR}\{tw-\Lambda_X(t)\}.
    \end{align}
    Therefore, for the open set $(x,\infty)$,
    \begin{align}
        \limsup_{n\to\infty}-\frac{1}{n}\log\Pr\{X_n>x\}
        \le \inf_{w>x}\Lambda_X^*(w).
    \end{align}

    Since $\Lambda_X$ is strictly convex and differentiable, the supremum defining $\Lambda_X^*(w)$
    is attained at the unique maximizer $t_w$ satisfying the first-order condition
    $\Lambda_X'(t_w)=w$, and
    \begin{align}
        \Lambda_X^*(w)= t_w\, w-\Lambda_X(t_w).
    \end{align}
    Moreover, $\Lambda_X'$ is strictly increasing and continuous, so $w\mapsto t_w=(\Lambda_X')^{-1}(w)$ is continuous, and consequently $\Lambda_X^*$ is continuous on the corresponding interval of $w$'s.
    Thus, 
    \begin{align}
        \inf_{w>x}\Lambda_X^*(w)
        \le \lim_{k\to\infty}\Lambda_X^*\left(x+\frac1k\right)
        = \Lambda_X^*(x),
    \end{align}
    where the inequality follows as $x + 1/k > x$ and the equality follows from the continuity of $\Lambda_X^*$.

    Finally, the assumption
    $
    x\in\bigl(\Lambda_X'(a),\,\Lambda_X'(b)\bigr)
    $
    ensures that the unique maximizer $t_x$ satisfying $\Lambda_X'(t_x)=x$ lies in $(a,b)$.
    Hence the supremum defining $\Lambda_X^*(x)$ can be restricted to the interval $(a,b)$:
    \begin{align}
        \Lambda_X^*(x)=\sup_{t\in(a,b)}\{tx-\Lambda_X(t)\}.
    \end{align}
    Combining the above relations yields the desired bound.
\end{proof}}

\begin{lemma}\label{LH1}
Let $m \in \NN$ be any integer. Let 
$\rho_1 \in \sA_1$, $\sigma_1 \in \sB_1$ 
and
$\rho_m \in \sA_m$, $\sigma_m \in \sB_m$ be quantum states such that $D_{\Petz,\alpha}(\rho_1\|\sigma_1)<\infty $ and $D_{\Petz,\alpha}(\rho_m\|\sigma_m)<\infty $ for any $\alpha \in (0,1)$. We set $k:=\lfloor n/m\rfloor$ and construct quantum states
\begin{align}
    \rho^{(n)} &:=
\rho_{1}^{\otimes n-km}\otimes \rho_m^{\otimes k},\\
\sigma^{(n)}&:=
\sigma_{1}^{\otimes n-km}\otimes \sigma_m^{\otimes k}.
\end{align} 
Let $\phi(s):= \frac{1}{m}\log {\tr}\rho_m^{1-s} \sigma_m^{s}$. If 
\begin{align} 
    R \in (-\frac{1}{m}D(\sigma_m \|\rho_m ), \frac{1}{m} D(\rho_m \|\sigma_m )),
\end{align} 
then for any $0 \leq T_n \leq I$,
\begin{align}
\limsup_{n\to \infty} -\frac{1}{n}\log 
&\left[\tr
e^{-nR}\rho^{(n)}
(I-T_n)
+\tr\sigma^{(n)}T_n \right] \notag \\
& \leq \max_{s\in (0,1)} (1-s)R -\phi(s).
\end{align}
\end{lemma}

\begin{proof}
Let ${P^{(n)}}$ and ${Q^{(n)}}$ be the Nussbaum-Szko\l{}a distributions of $\rho^{(n)}$ and $\sigma^{(n)}$. Let 
\begin{align} 
    S_n = \left\{e^{-nR } {P^{(n)}} > {Q^{(n)}}\right\},
\end{align} 
be a likelihood ratio test. Consider the random variable
\begin{align}
    X_n(x) := \frac{1}{n} \left(\log {Q^{(n)}}(x) - \log {P^{(n)}}(x)\right),
\end{align}
where $x$ is drawn from the distribution ${P^{(n)}}$.
Then we have the asymptotic cumulant generating function of the random variable $X_n$ as,
\begin{align}
&\lim_{n\to \infty} \frac{1}{n}\log \sum_{x} {P^{(n)}}(x) \exp \left(s n X_n(x) \right)\notag\\
&  = \lim_{n\to \infty}\frac{1}{n}\log \tr {(Q^{(n)})}^s {(P^{(n)})}^{1-s}\\
& = \lim_{n\to \infty} \frac{1}{n} \log \tr (\sigma^{(n)})^s (\rho^{(n)})^{1-s}\\
& =
\lim_{n\to \infty}\frac{1}{n}
\left((n-km) \log \tr \rho_1^{1-s} \sigma_1^{s}
+k \log \tr \rho_m^{1-s} \sigma_m^{s}\right)\\
& =\phi(s),
\end{align}
where the second equality is a simple fact of the Nussbaum-Szko\l{}a distributions (c.f.~\cite[Proposition 1]{audenaert2008asymptotic}). Note that $\phi(s)$ is {$C^1$ continuous and strictly convex}, with $\phi'(0) = - \frac{1}{m}D(\rho_m\|\sigma_m)$ and $\phi'(1) = \frac{1}{m}D(\sigma_m\|\rho_m)$ (see Appendix~\ref{sec: math properties of phi}).  Applying the G\"{a}rtner-Ellis theorem in Lemma~\ref{lem: Gartner-Ellis} for the random variable $X_n$, interval $(0,1)$ and $x=-R$, we have
\begin{align}
    \limsup_{n\to \infty} - \frac{1}{n} \log \Pr\{X_n \geq -R\} \leq \sup_{s\in (0,1)} -s R -\phi(s). \label{eq: NS relation tmp1}
\end{align}
Similarly, consider the random variable 
\begin{align} 
    Y_n(x):= \frac{1}{n} \left(\log {P^{(n)}}(x) - \log {Q^{(n)}}(x)\right),
\end{align} 
where $x$ is drawn from the distribution ${Q^{(n)}}$. Then we have the asymptotic cumulant generating function of the random variable $Y_n$ as,
\begin{align}
&\lim_{n\to \infty} \frac{1}{n}\log \sum_{x} {Q^{(n)}}(x) \exp \left(t n Y_n(x)\right)\notag\\
&  = \lim_{n\to \infty}\frac{1}{n}\log \tr {(Q^{(n)})}^{1-t} {(P^{(n)})}^{t}\\
& = \lim_{n\to \infty} \frac{1}{n} \log \tr (\sigma^{(n)})^{1-t} (\rho^{(n)})^{t}\\
& =
\lim_{n\to \infty}\frac{1}{n}
\left((n-km) \log \tr \rho_1^{t} \sigma_1^{1-t}
+k \log \tr \rho_m^{t} \sigma_m^{1-t}\right)\\
& =\phi(1-t).
\end{align}
Applying again the G\"{a}rtner-Ellis theorem in Lemma~\ref{lem: Gartner-Ellis} for the random variable $Y_n$, interval $(0,1)$ and $x=R$, we have
\begin{align}
\limsup_{n\to \infty} -\frac{1}{n} \log \Pr\{Y_n {\ > \ } R\} & \leq \sup_{t \in (0,1)} t R -\phi(1-t).\label{eq: NS relation tmp2}
\end{align}
By direct calculation, we have the relations
\begin{align}
    \limsup_{n\to \infty} & - \frac{1}{n} \log \Pr\{X_n \geq -R\}\notag \\
     & = \limsup_{n\to \infty}  -\frac{1}{n}\log \tr P^{(n)} (I-S_n),\label{eq: NS relation tmp3}\\
\limsup_{n\to \infty} &- \frac{1}{n} \log \Pr\{Y_n {\ > \ }  R\}\notag \\
 & = \limsup_{n\to \infty}  -\frac{1}{n}\log \tr Q^{(n)} S_n. \label{eq: NS relation tmp4}
\end{align}
Moreover, the Nussbaum-Szko\l{}a theorem (c.f.~\cite[Lemma 3.4]{hayashi2017quantum}) implies that for any test $T_n$,
\begin{align}\label{eq: NS relation tmp5}
\tr &  e^{-nR} \rho^{(n)} (I-T_n)  + \tr \sigma^{(n)} T_n \notag \\
& \geq 
\frac{1}{2} \left(\tr  e^{-nR } {P^{(n)}} (I- S_n) + \tr {Q^{(n)}} S_n\right).
\end{align}
Combining Eqs.~\eqref{eq: NS relation tmp1},~\eqref{eq: NS relation tmp2},~\eqref{eq: NS relation tmp3},~\eqref{eq: NS relation tmp4} and~\eqref{eq: NS relation tmp5},
we have
\begin{align}
&\limsup_{n\to \infty}  -\frac{1}{n}\log 
\left[\tr e^{-nR}\rho^{(n)} (I-T_n) +\tr\sigma^{(n)}T_n \right]\notag \\
& \leq \limsup_{n\to \infty} - \frac{1}{n}\log \frac{1}{2} \left[\tr e^{-nR} P^{(n)} (I-S_n) +\tr {Q^{(n)}} S_n\right] \\
& \leq \min \left\{ R+ \sup_{s\in (0,1)} -s R -\phi(s),
\sup_{t \in (0,1)} t R -\phi(1-t)\right\}\\
& =\sup_{s\in (0,1)} (1-s)R -\phi(s),\label{eq: lemma tmp1}
\end{align}
where the last equality follows by replacing $t$ with $1-s$. Finally, as the objective function in Eq.~\eqref{eq: lemma tmp1} is concave and its first derivative is given by $-R - \phi'(s)$, there is a unique critical point that achieves the maximum if $R \in (- \phi'(1), -\phi'(0))$. Therefore, the supremum is attained.
\end{proof}

With the above Lemma~\ref{LH1}, we are ready to show the upper bound for limit superior. That is, we aim to 
show that
    \begin{align}
        \limsup_{n\to \infty} -\frac{1}{n} \log \alpha_{n, r}(\sA_n\| \sB_n)  \le H_r^\infty(\sA\|\sB).
    \end{align}
For this, we plan to show for any fixed $m \in \NN$,
\begin{align}\label{eq: upper bound proof tmp2} \limsup_{n\to \infty} - \frac{1}{n}\log 
\alpha_{n, r}(\sA_n\| \sB_n)
& \leq \frac{1}{m}H_{m,r} (\sA_m\|\sB_m).
\end{align}

If $H_{m,r}(\sA_m\|\sB_m) = \infty$, the upper bound holds trivially. It remains to show Eq.~\eqref{eq: upper bound proof tmp2} when $H_{m,r}(\sA_m\|\sB_m) < \infty$. So for any $\delta > 0$, there exist $\rho_m \in \sA_m$ and $\sigma_m \in \sB_m$ such that
\begin{align}\label{eq: choice of rhom and sigmam}
    H_{m,r}(\rho_m\|\sigma_m) \leq H_{m,r}(\sA_m\|\sB_m) + \delta < \infty.
\end{align}
This implies that $D_{\Petz, \alpha}(\rho_m\|\sigma_m) < \infty$ for any $\alpha \in (0,1)$. Otherwise, we will have a contradiction to the finiteness of $H_{m,r}(\rho_m\|\sigma_m)$ by definition in Eq.~\eqref{eq: definition of Hoeffding for states}.
Using this choice of $\rho_m$ and $\sigma_m$, we construct the sequence of states $\rho^{(n)}$ and $\sigma^{(n)}$ as in Lemma~\ref{LH1}.

Combining Lemma~\ref{LH1} and Lemma~\ref{lem: log rate min}, we have for any test with $\tr \sigma^{(n)} T_n \le e^{-nr}$, that
\begin{align}
    & \limsup_{n\to \infty} -\frac{1}{n}\log 
 \left[\tr
e^{-nR}\rho^{(n)}
(I-T_n)
+\tr\sigma^{(n)}T_n \right]\notag \\
 & \geq  \limsup_{n\to \infty} -\frac{1}{n}\log 
\left[\tr
e^{-nR}\rho^{(n)}
(I-T_n)
+ e^{-nr} \right]\\
& = \min \left\{r, R +
\limsup_{n\to \infty} -\frac{1}{n}\log 
\tr \rho^{(n)} (I-T_n) \right\},
\end{align}
where the equality follows from Lemma~\ref{lem: log rate min}. Therefore, by Lemma~\ref{LH1}, we have
\begin{align}\label{eq: limsup proof tmp1}
\min & \left\{r, R +
\limsup_{n\to \infty} -\frac{1}{n}\log 
\tr \rho^{(n)} (I-T_n) \right\}\notag \\
& \hspace{2cm}\le\max_{s\in (0,1)} (1-s)R -\phi(s).
\end{align}
Let $r'<r$. From Lemma~\ref{lem: exponent optimization}, we know that the optimization 
\begin{align}
    \max_{s\in (0,1)} \frac{-\phi(s)-sr'}{1-s}
\end{align} 
has a unique maximizer $s_{r'} \in (0,1)$ such that $r' = (s_{r'}-1) \phi'(s_{r'}) - \phi(s_{r'})$.
We set 
\begin{align}\label{eq: limsup proof tmp2}
    R_{r'}:=\frac{\phi(s_{r'})+r'}{1-s_{r'}} = -\phi'(s_{r'}).
\end{align}
It is clear that 
\begin{align} 
    R_{r'} \in & (-\phi'(1), -\phi'(0))\\
    & = \left(-\frac{1}{m}D(\sigma_m\|\rho_m), \frac{1}{m}D(\rho_m\|\sigma_m)\right).
\end{align} 
By the proof of Lemma~\ref{LH1}, $\max_{s\in (0,1)} (1-s)R -\phi(s)$ is uniquely achieved at point $s$ such that $R = -\phi'(s)$. For $R = R_{r'}$, Eq.~\eqref{eq: limsup proof tmp2} implies that the maximum is uniquely achieved at $s = s_{r'}$. That is,
\begin{align}
\max_{s\in (0,1)} (1-s)R_{r'} -\phi(s) = (1-s_{r'})R_{r'} -\phi(s_{r'}) = r'.
\end{align}
Then, by Eq.~\eqref{eq: limsup proof tmp1} we have 
\begin{align}
\min \left\{ r, R_{r'} +
\limsup_{n\to \infty} -\frac{1}{n}\log 
\tr \rho^{(n)} (I-T_n) \right\}
\leq r'.
\end{align}
Then, we have
\begin{align}
R_{r'} +
\limsup_{n\to \infty} -\frac{1}{n}\log 
\tr \rho^{(n)} (I-T_n)
\le r.
\end{align}
Otherwise, it will contradict to the assumption that $r' < r$.
Thus, we have
\begin{align}
\limsup_{n\to \infty} -\frac{1}{n}\log 
\tr \rho^{(n)} (I-T_n)
\le r - R_{r'} .
\end{align}

Since $s_r$ is continuous function in $r$ (see Lemma~\ref{lem: sr differentiable}) and $\phi(s)$ is a continuous function in $s$ (see Lemma~\ref{lem: first and second derivative of phi}), $R_{r'}$ is a continuous function in $r'$. Therefore, 
\begin{align}
    \lim_{r'\to r^-} R_{r'} = R_r:=\frac{\phi(s_{r})+r}{1-s_{r}}.
\end{align} 
This gives
\begin{align}
\limsup_{n\to \infty} - \frac{1}{n}\log 
\tr \rho^{(n)} (I-T_n)
& \le r - R_{r}.
\end{align}
By direct calculation, the right-hand side gives
\begin{align}
r - R_{r} & = \frac{-\phi(s_r)-s_r r}{1-s_r}\\
& = \max_{s\in (0,1)} \frac{-\phi(s)-s r}{1-s}\\
& = \frac{1}{m}H_{m,r}(\rho_m\|\sigma_m),
\end{align}
where the second equality follows from the optimality of $s_r$. By the choice of $\rho_m,\sigma_m$ in Eq.~\eqref{eq: choice of rhom and sigmam}, we have 
\begin{align} 
\limsup_{n\to \infty} & - \frac{1}{n}\log 
\tr \rho^{(n)} (I-T_n)\notag \\
& \leq \frac{1}{m}H_{m,r} (\sA_m\|\sB_m) + \frac{\delta}{m}.  
\end{align} 
This implies that 
\begin{align}
\limsup_{n\to \infty} & - \frac{1}{n}\log 
\alpha_{n, r}(\rho^{(n)}\| \sigma^{(n)})\notag \\
& \leq \frac{1}{m}H_{m,r} (\sA_m\|\sB_m) + \frac{\delta}{m}.
\end{align}
As we have the trivial relation that $\alpha_{n,r}(\sA_n\|\sB_n) \geq \alpha_{n,r}(\rho_n\|\sigma_n)$ for any $\rho_n \in \sA_n, \sigma_n \in \sB_n$ by definition, this implies that 
\begin{align}
\limsup_{n\to \infty} & - \frac{1}{n}\log 
\alpha_{n, r}(\sA_n\| \sB_n)\notag \\
& \leq \frac{1}{m}H_{m,r} (\sA_m\|\sB_m) + \frac{\delta}{m}.
\end{align}
As this holds for any $\delta > 0$, we have 
\begin{align}\limsup_{n\to \infty} - \frac{1}{n}\log 
\alpha_{n, r}(\sA_n\| \sB_n)
& \leq \frac{1}{m}H_{m,r} (\sA_m\|\sB_m).
\end{align}
Finally, taking the limit $m\to \infty$, we have 
    \begin{align}
        \limsup_{n\to \infty} -\frac{1}{n} \log \alpha_{n, r}(\sA_n\| \sB_n)  \le H_r^\infty(\sA\|\sB),
    \end{align}
 where the existence of the limit on the right hand side follows from the stability assumption (C3) and Eq.~\eqref{eq: multiplicativity of Hoeffding divergence}.
This finishes the proof of the upper bound.

\section{Strong converse exponent for composite correlated hypotheses}
\label{sec: Strong converse exponent for composite correlated hypotheses}

In this section, we extend the strong converse exponent from the i.i.d. setting to the broader context of composite correlated hypotheses and show that the strong converse exponent is lower bounded by a regularized quantum Hoeffding anti-divergence between the sets in general. We also provide a matching upper bound when the null hypothesis is a singleton i.i.d. state {under additional assumptions}.

For the upper bound, we need to introduce the permutation invariance. Let $U_{A^n}$ denote the natural unitary representation of the permutation group $S_n$ that permutes the subsystems of $A^n := A_1 A_2 \cdots A_n$. An linear operator $X_{A^n} \in \sL(A^n)$ is called {permutation invariant} if it satisfies $X_{A^n} = U_{A^n}(\pi) X_{A^n} U_{A^n}(\pi)^\dagger$ for all $\pi \in S_n$. A set of quantum state $\sA_n$ is called permutation invariant if $\rho_n \in \sA_n$ implies that $U_{A^n}(\pi) \rho_{A^n} U_{A^n}(\pi)^\dagger \in \sA_n$ for all $\pi \in S_n$. Let $A$ be a system with $|A| = d$. For all $n\in\NN$ there exists a permutaion invariant state $\omega_{A^n}^n \in \density(A^n)$, which we call \emph{universal state}, such that for all permutation invariant states $\rho_{A^n} \in \density(A^n)$, we have 
\begin{align}\label{eq: universal state}
    \rho_{A^n} \leq g_{n,d}\, \omega_{A^n}^n,\quad \text{with} \quad g_{n,d} \leq (n+1)^{d^2-1}.
\end{align}
Note that such a universal state exists and has been explicitly constructed in~\cite{hayashi2016correlation}.

\begin{assumption}[Assumptions for sets of quantum states.]
We denote the following assumptions for a sequence of sets of quantum states $\sC = \{\sC_n\}_{n\in \NN}$ where each $\sC_n \subseteq \density(\cH^{\ox n})$.
\begin{itemize}
    \item[(C5)] Permutation invariance: For any $n \in \NN$, the set $\sC_n$ is permutation invariant.
    \item[(C6)] Universal state containment: For any $n \in \NN$, the set $\sC_n$ contains a universal state.
\end{itemize}
\end{assumption}

\begin{theorem}[Strong converse exponent for composite correlated hypotheses.]\label{thm: strong converse exponent for composite correlated hypotheses}  Let $\cH$ be a finite-dimensional Hilbert space. Let $\sA = \{\sA_n\}_{n\in \NN}$ and $\sB = \{\sB_n\}_{n\in \NN}$ \,be two sequences of sets of quantum states, where each $\sA_n, \sB_n \subseteq \density(\cH^{\ox n})$. Let $r > D^\reg(\sA\|\sB)$ be a real number. 
\begin{itemize}
    \item Without any additional assumptions on $\sA$ or $\sB$, 
        \begin{align}\label{eq: strong converse exponent lower bound}
            \liminf_{n\to \infty} -\frac{1}{n} \log \bigl(1-\alpha_{n, r}(\sA_n\| \sB_n)\bigr)
            \geq \underline{H}_{r}^{*,\reg}(\sA\|\sB).
        \end{align}
    \item Suppose that $\sA_n=\{\rho^{\ox n}\}$ is a singleton i.i.d.\ sequence, and that $\sB$ satisfies assumptions (C1), (C5), and (C6). Assume further that $D_{\Sand,\alpha}^\infty(\sA\|\sB)$ exists and is differentiable in $\alpha$ for all $\alpha\ge 1$. Then, for
    {$D^\reg(\sA\|\sB) < r < \sup_{\alpha > 1} D_{\Sand,\alpha}^\infty(\sA\|\sB)$,}
\begin{align}
    \limsup_{n\to \infty} -\frac{1}{n} \log (1-\alpha_{n, r}(\sA_n\| \sB_n)) \leq {\bH}_{r}^{*,\reg}(\sA\|\sB).
\end{align}
\end{itemize}
Consequently, with the same assumptions for the upper bound and additionally assuming that $\sB$ satisfies assumption (C3), then the limit exists and is given by
\begin{align}
    \lim_{n\to \infty} -\frac{1}{n} \log (1-\alpha_{n, r}(\sA_n\| \sB_n)) = {H}_{r}^{*,\reg}(\sA\|\sB).
\end{align}
\end{theorem}

\begin{remark}
    This result applies to a few interesting cases, such as the resource theory of coherence and entanglement, as discuss in Section~\ref{sec: applications to resource theories}. Moreover, by Remark~\ref{rem: type errors convexity},
    the convexity of $\sA_n$ and $\sB_n$ can be removed by replacing the right hand side with
    $H_{r}^{*,\infty}(\conv(\sA)\|\conv(\sB))$, where
    $\conv(\sC):=\{\conv(\sC_n)\}_{n\in\mathbb N}$ denotes the set of convex hulls.
\end{remark}

\subsection{Proof of the lower bound}

For any $\delta>0$, let $\rho_n \in \sA_n, \sigma_n\in\sB_n$ such that 
\begin{align} 
    D_{\Sand,\alpha}(\rho_n\|\sigma_n) \leq D_{\Sand,\alpha}(\sA_n\|\sB_n) + \delta.
\end{align} 
By standard arguments, e.g.~\cite[Lemma 5]{cooney2016strong}, we have for any $0\leq M_n \leq I$, that
\begin{align}
    \frac{1}{n} & \log \left(1- \tr[(I-M_n)\rho_n]\right)\\
    & \leq \frac{\alpha-1}{\alpha} \left(\frac{1}{n}D_{\Sand,\alpha}(\rho_n\|\sigma_n) + \frac{1}{n} \log \tr[M_n \sigma_n]\right).\notag
\end{align}
Since $\tr[(I-M_n)\rho_n] \leq \alpha(\sA_n, M_n)$ and $\tr[M_n \sigma_n] \leq \beta(\sB_n, M_n)$, it follows that
\begin{align}
     & \frac{1}{n}  \log \left(1- \alpha(\sA_n, M_n)\right) \\
     & \leq \frac{\alpha-1}{\alpha} \left(\frac{1}{n}D_{\Sand,\alpha}(\sA_n\|\sB_n) + \delta + \frac{1}{n} \log \beta(\sB_n, M_n)\right).\notag
\end{align}
As this holds for any $\delta > 0$, we have
\begin{align}
     \frac{1}{n}& \log \left(1- \alpha(\sA_n, M_n)\right) \\
     & \leq \frac{\alpha-1}{\alpha} \left(\frac{1}{n}D_{\Sand,\alpha}(\sA_n\|\sB_n) + \frac{1}{n} \log \beta(\sB_n, M_n)\right).\notag
\end{align}
For any $0 \leq M_n \leq I$ such that $\beta(\sB_n, M_n) \leq 2^{-nr}$, we obtain
\begin{align}
    \frac{1}{n} & \log \left(1- \alpha(\sA_n, M_n)\right) \\
    & \leq \frac{\alpha-1}{\alpha} \left(\frac{1}{n}D_{\Sand,\alpha}(\sA_n\|\sB_n) - r\right).\notag
\end{align}
Taking the supremum over all such $M_n$, we find
\begin{align}\label{eq: strong converse exponent for sets tmp3}
     - \frac{1}{n} & \log \left(1- \alpha_{n, r}(\sA_n\|\sB_n)\right) \\
     & \geq \frac{\alpha-1}{\alpha} \left(r- \frac{1}{n}D_{\Sand,\alpha}(\sA_n\|\sB_n)\right).\notag 
\end{align}
Since this holds for any $\alpha > 1$, we have
\begin{align}
     - \frac{1}{n} \log \left(1- \alpha_{n, r}(\sA_n\|\sB_n)\right) \geq H_{n,r}^*(\sA_n\|\sB_n),
\end{align}
where we use the finite equivalence in Lemma~\ref{lem: Hoeffding anti equivalence}.
Taking limit inferior of $n$ on both sides,
\begin{align}
    \liminf_{n\to \infty} -\frac{1}{n} \log \left(1-\alpha_{n, r}(\sA_n\|\sB_n)\right) 
    & \geq \underline{H}_{r}^{*,\reg}(\sA\|\sB).
\end{align}

\subsection{Proof of the upper bound}

Before the proof of the upper bound, we require the following lemma that simplifies the optimization over $\sB_n$ via the universal state. Let $X = \sum_{\lambda \in \spec(X)} \lambda E_\lambda$ be the spectral decomposition of the Hermitian operator $X$, where $E_\lambda$ are projectors and $\spec(X)$ is its discrete specturm. Then the pinching map for this spectral decomposition is defined by $\cP_X(\cdot):= \sum_i E_i (\cdot) E_i$.

\begin{lemma}
\label{lem: pinched renyi divergence}
Let $\sA_n = \{\rho_n\}$ be a permutation invariant state  and $\sB_n$ be a convex and permutation invariant set, which contains a universal state $\omega_n$ as defined in Eq.~\eqref{eq: universal state}. Then for any $\alpha \geq 1/2$, 
\begin{align}
    \frac{1}{n} D_{\Sand,\alpha}(\cP_{\omega_n}(\rho_n)\|\omega_n) = \frac{1}{n} D_{\Sand,\alpha}(\rho_n\|\sB_n) + O\left(\frac{\log n}{n}\right).
\end{align}
In particular,
\begin{align}
   \lim_{n\to \infty} \frac{1}{n} D_{\Sand,\alpha}(\cP_{\omega_n}(\rho_n)\|\omega_n) & = D_{\Sand,\alpha}^\infty(\sA\|\sB).
\end{align}
\end{lemma}

\begin{proof}
By the permutation invariance and the convexity of the sandwiched \Renyi divergence, we know that $D_{\Sand,\alpha}(\rho_n\|\sB_n)$ can be achieved at some permutation invariant state $\sigma_n \in \sB_n$.
    Note also that $\sigma_n\leq g_{n,d}\; \omega_n$ for any permutation invariant state $\sigma_n$. Therefore, we have
\begin{align}
    D_{\Sand,\alpha}(\cP_{\omega_n}(\rho_n)\|\omega_n) & \leq D_{\Sand,\alpha}(\rho_n\|\omega_n)\\
    & \leq D_{\Sand,\alpha}(\rho_n\|\sigma_n)+ \log g_{n,d},
\end{align}
where the first inequality uses the data processing inequality and the second inequality uses the monotonicity of the sandwiched Renyi divergence under scaling of the second argument. As this holds for any permutation invariant state $\sigma_n \in \sB_n$, we have
\begin{align}
    D_{\Sand,\alpha}(\cP_{\omega_n}(\rho_n)\|\omega_n) & \leq D_{\Sand,\alpha}(\rho_n\|\sB_n) + \log g_{n,d}.
\end{align}
On the other hand, we have~\cite[Lemma 3]{hayashi2016correlation}, 
\begin{align}
    D_{\Sand,\alpha}(\cP_{\omega_n}(\rho_n)\|\omega_n) & \geq D_{\Sand,\alpha}(\rho_n\|\omega_n) - 2 \log |\spec(\omega_n)|,
\end{align}
where $|\spec(\cdot)|$ denotes the number of distinct eigenvalues. Since $\omega_n \in \sB_n$, we have 
\begin{align}
     D_{\Sand,\alpha}(\cP_{\omega_n}(\rho_n)\|\omega_n) & \geq D_{\Sand,\alpha}(\rho_n\|\sB_n) - 2 \log |\spec(\omega_n)|.
\end{align}
As $\omega_n$ is permutation invariant, we have $|\spec(\omega_n)| \leq (n+1)^d$. This completes the proof.
\end{proof}

\begin{remark}\label{rem: pinch continuity and convexity}
In the above case, $D^\infty_{\Sand,\alpha}(\sA\|\sB)$ is a pointwise limit of $\frac{1}{n} D_{\Sand,\alpha}(\cP_{\omega_n}(\rho_n)\|\omega_n)$ and the convergence is uniform in $\alpha$. Therefore, $D^\infty_{\Sand,\alpha}(\sA\|\sB)$ is continuous and monotonically increasing in $\alpha$. Similarly, the function $t D^\infty_{\Sand,1+t}(\sA\|\sB)$ is continuous and convex in $t$.
\end{remark}

Now, we are ready to prove the upper bound. The proof follows the similar idea as in~\cite{hayashi2016correlation}. Let $\cP_{\omega_n}(\rho_n)$ be the pinching of $\rho_n$ with respect to $\omega_n$. Then $\cP_{\omega_n}(\rho_n)$ and $\omega_n$ commute. Let $\ket{v_{x_n}}$ be a common orthonormal eignbasis and we define the classical probability distributions
\begin{align}
    P_n(x)&:= \<v_{x_n}|\cP_{\omega_n}(\rho_n)|v_{x_n}\>,\\
     Q_n(x)& := \<v_{x_n}|\omega_n|v_{x_n}\>.
\end{align}
Let $X_n$ be a random variable distributed according to $P_n$ and $X_n'$ be a random variable distributed according to $Q_n$. Define 
\begin{align}
    S_n:= \{ \cP_{\omega_n}(\rho_n) \geq \exp(\lambda_n) \omega_n\}.
\end{align}
Then we have the relation that 
\begin{align}
    \alpha(\rho_n, S_n) & = \tr[\rho_n (I-S_n)]\\
    & = \tr[\cP_{\omega_n}(\rho_n) (I-S_n)]\\
    & = \Pr[P_n(X_n) < \exp(\lambda_n) Q_n(X_n)],
\end{align}
where the second equality follows as $\tr \cP_X(Y) X = \tr YX$ for all $Y$. Similarly, we have 
\begin{align}
        \beta(\sB_n, S_n) & = \sup_{\sigma_n \in \sB_n} \tr [\sigma_n S_n]\\
        & = \tr [\sigma_n' S_n]\\
        & \leq \tr [g_{n,d}\, \omega_n S_n]\\
        & \leq g_{n,d} \Pr[P_n(X_n') \geq \exp(\lambda_n) Q_n(X_n')].
\end{align}
Here we use the fact that $S_n$ is permutation invariant and therefore the optimal solution is achieved at some permutation invariant state $\sigma_n'$, which is then upper bounded by $g_{n,d}\, \omega_n$.

{Moreover, by direct calculation, for any $s > 1$, we have
\begin{align}
    \Pr&[P_n(X_n') \geq \exp(\lambda_n) Q_n(X_n')]\notag \\
    & = \sum_{x:P_n(x) \geq \exp(\lambda_n) Q_n(x)} Q_n(x)\\
    & \leq \sum_{x:P_n(x) \geq \exp(\lambda_n) Q_n(x)} Q_n(x)^{1-s} [\exp(-\lambda_n) P_n(x)]^s\\
    & \leq  \sum_x Q_n(x)^{1-s} [\exp(-\lambda_n) P_n(x)]^s \\
    & = \exp(-s\lambda_n) \tr[P_n^s Q_n^{1-s}],
\end{align}
where the first inequality uses the condition $P_n(x) \geq \exp(\lambda_n) Q_n(x)$, and the second inequality extends the sum to all $x$. Consequently, we obtain
\begin{align}
    \beta(\sB_n, S_n) \leq g_{n,d} \exp(-s\lambda_n) \tr[P_n^s Q_n^{1-s}].
\end{align}}

For arbitrary fixed $s \in (1,\infty)$, we choose $\lambda_n$ such that
\begin{align}
    \lambda_n := \frac{1}{s} \bigg[\phi(1-s|P_n\|Q_n) + \log g_{n,d} + n r\bigg].
\end{align}
Consider a random variable
\begin{align}
    Z_n := & \frac{1}{n} \left(\log P_n(X_n) - \log Q_n(X_n) - \lambda_n\right).
\end{align}
Hence, by the choice of $\lambda_n$, we have $\beta(\sB_n, S_n) \leq \exp({-n r})$. Morover,
\begin{align}
    1-\alpha(\rho_n, S_n) &= \Pr[P_n(X_n) \geq \exp(\lambda_n) Q_n(X_n)] \\ & = \Pr[Z_n \geq 0].
\end{align}
So the optimal Type-I error satisfies
\begin{align}
    1-\alpha_{n,r}(\rho_n\| \sB_n) & \geq 1-\alpha(\rho_n, S_n) = \Pr[Z_n \geq 0].
\end{align}
This implies that 
\begin{align}\label{eq: optimal type I and random variable}
    \limsup_{n\to \infty} & -\frac{1}{n} \log (1-\alpha_{n, r}(\rho_n\| \sB_n))\notag \\
     & \leq \limsup_{n\to \infty} -\frac{1}{n} \log \Pr[Z_n \geq 0].
\end{align}
In the following, we will use the G\"{a}rtner-Ellis theorem to upper bound the right-hand side. For this, we need to compute the asymptotic cumulant generating function of $Z_n$. 

For $t\geq -\frac{1}{2}$, we have 
\begin{align}
    & \Lambda_Z(t)  := \lim_{n\to \infty} \frac{1}{n} \log \EE[e^{n t Z_n}]\\
    & = \lim_{n\to \infty} \frac{1}{n} \log \Bigg\{\sum_x P_n(x)\cdot \notag \\
    &\hspace{2cm} \exp(t (\log P_n(x) - \log Q_n(x) - \lambda_n))\Bigg\}\\
    & = \lim_{n\to \infty} \frac{1}{n} \Bigg\{\log \sum_x P_n(x)^{1+t} Q_n(x)^{-t} \notag \\
    & \hspace{1.7cm}- \frac{t}{s} \left[\phi(1-s|P_n\|Q_n) + \log g_{n,d} + n r\right]\Bigg\}\\
    & = \lim_{n\to \infty} \frac{1}{n} \Bigg\{\log \tr P_n^{1+t} Q_n^{-t} \notag \\
    & \hspace{1.6cm}- \frac{t}{s} \big[\phi(1-s|P_n\|Q_n) + \log g_{n,d} + n r\big]\Bigg\}\\
    & = \lim_{n\to \infty} \frac{1}{n} \Bigg\{\phi(-t|P_n\|Q_n)\notag\\
    &\hspace{1.6cm} - \frac{t}{s} \left[\phi(1-s|P_n\|Q_n) + \log g_{n,d} + n r\right]\Bigg\}.
\end{align}
Denote $g(\alpha):=D^\infty_{\Sand,\alpha}(\sA\|\sB)$ for $\alpha > 1$ and $g(1) = D^\infty(\sA\|\sB)$. We have the relation that
\begin{align}
    \phi(s|P_n\|Q_n) & = \log \tr[P_n^{1-s} Q_n^s]= -s D_{1-s}(P_n\|Q_n).
\end{align}
This means that 
\begin{align}
    \lim_{n\to \infty}& \frac{1}{n} \phi(-t|P_n\|Q_n)\notag \\
     & = \lim_{n\to \infty} \frac{1}{n} t D_{1+t}(P_n\|Q_n)\\
    &  = \lim_{n\to \infty} \frac{1}{n} t D_{\Sand, 1+t}(\cP_{\omega_n}(\rho_n)\|\omega_n)\\
    & = t g(1+t),
\end{align}
where the second equality follows as $\cP_{\omega_n}(\rho_n)$ and $\omega_n$ commute and the last equality follows from Lemma~\ref{lem: pinched renyi divergence}. Similarly, we have 
\begin{align}
    \lim_{n\to \infty} & \frac{1}{n} \phi(1-s|P_n\|Q_n)\notag\\
     & = \lim_{n\to \infty} \frac{1}{n} (s-1) D_{s}(P_n\|Q_n)\\
    & = \lim_{n\to \infty} \frac{1}{n} (s-1) D_{\Sand, s}(\cP_{\omega_n}(\rho_n)\|\omega_n)\\
    & = (s-1) g(s).
\end{align} 
Define the function
\begin{align}
    f(s,t) := t \big(r - s g(1+t) + (s-1) g(s)\big).
\end{align}
Then we have 
\begin{align}
    \Lambda_Z(t) & = -\frac{f(s,t)}{s}.
\end{align}
As we assume that $g(\alpha)$ is differentiable in $\alpha$ for $\alpha \geq 1$. Then $f(s,t)$ is differentiable in both $s$ and $t$. Moreover, we have
\begin{align}
    \Lambda_Z'(t) & = -\frac{1}{s} \left(r - s g(1+t) + (s-1) g(s)- ts g'(1+t))\right).
\end{align}
Since $g(\alpha)$ is monotonically increasing and differentiable for $\alpha \geq 1$, we have for any $s > 1$,
\begin{align}
    \lim_{t \to 0^+} \Lambda_Z'(t) & = -\frac{1}{s} \left(r - s g(1) + (s-1) g(s)\right)\\
    & = g(1) - \frac{s-1}{s} g(s) - \frac{r}{s}\\
    & \leq g(1) - \frac{s-1}{s} g(1) - \frac{r}{s}\\
    & = \frac{1}{s} (g(1) - r)\\
    & < 0,
\end{align}
where we use the assumption that $r > g(1) = D^\infty(\sA\|\sB)$ in the last inequality.
On the other hand, using the convexity of $t\mapsto \psi(t):= t g(1+t)$ and $\psi(0) = 0$ (see Remark~\ref{rem: pinch continuity and convexity}), we have $\psi(\lambda t) \leq \lambda \psi(t)$ for all $\lambda \in (0,1)$. This implies that
\begin{align}
    \psi'(t) = \lim_{\lambda \to 1^-} \frac{\psi(t) - \psi(\lambda t)}{t(1-\lambda)} \geq \frac{\psi(t)}{t},
\end{align}
and therefore we have $tg'(1+t) \geq 0$.
Let $t_0$ be such that $r < g(t_0+1)$. Then for any $s \leq t_0+1$, we find that 
\begin{align}
    \lim_{t\to t_0^-} \Lambda_Z'(t) & \geq g(t_0+1) - \frac{s-1}{s} g(s) - \frac{r}{s}\\
    & \geq g(t_0+1) - \frac{s-1}{s} g(t_0+1) - \frac{r}{s}\\
    & = \frac{1}{s} (g(t_0+1) - r) > 0,
\end{align}
where we have used that $r < g(t_0+1)$. Hence, applying the G\"{a}rtner-Ellis theorem in~\cite[Proposition 17]{hayashi2016correlation} for the random variable $Z_n$ on the interval $(0, t_0)$ and with the threshold value $0$, we have
\begin{align}
    \limsup_{n\to \infty} &-\frac{1}{n} \log \Pr[Z_n \geq 0] \notag\\
    & \leq \sup_{t \in (0, t_0)} \{-\Lambda_Z(t)\} \leq \sup_{t \in [0,t_0]} \frac{f(s,t)}{s}.
\end{align}
As this holds for any $s \in (1, t_0+1]$, we have from~\eqref{eq: optimal type I and random variable} that 
\begin{align}
    \limsup_{n\to \infty}  -\frac{1}{n} & \log (1-\alpha_{n, r}(\rho_n\| \sB_n))\notag \\
     & \leq \inf_{s \in (1, t_0+1]} \sup_{t \in [0,t_0]} \frac{f(s,t)}{s}.
\end{align}
By the convexity of $tg(1+t)$ (Remark~\ref{rem: pinch continuity and convexity}), we known that $f(s,t)$ is concave in $t$ and convex in $s$. We optimize $t$ over a compact convex set and $s$ over a convex set. Therefore, by the minimax theorem in~\cite[Proposition 21]{hayashi2016correlation}, we can exchange the order of $\inf$ and $\sup$. This gives
\begin{align}
    \inf_{s \in (1,t_0+1]} \sup_{t \in [0,t_0]} \frac{f(s,t)}{s} & = \sup_{t \in [0,t_0]} \inf_{s \in (1,t_0+1]} \frac{f(s,t)}{s}\\
    & = \sup_{t \in (0,t_0]} \inf_{s \in (1,t_0+1]} \frac{f(s,t)}{s}\\
    & \leq \sup_{t \in (0,t_0]} \frac{f(t+1, t)}{t+1}\\
    & \leq \sup_{t > 0} \frac{f(t+1, t)}{t+1},
\end{align}
where the second line follows as $f(s,0) = 0$ and the penultimate line follows by choosing $s=1+t$. 
This gives 
\begin{align}
    \limsup_{n\to \infty} & -\frac{1}{n} \log (1-\alpha_{n, r}(\rho_n\| \sB_n))\notag \\
    & \leq \sup_{t > 0} \frac{f(t+1, t)}{t+1} = {\bH}_{r}^{*,\reg}(\sA\|\sB).
\end{align}

\section{Applications}
\label{sec: applications}

Since the error exponent and strong converse exponent regimes provide a finer characterization of the trade-off between Type-I and Type-II errors than the Stein regime, we apply our results to refine and extend several existing studies. Specifically, we: (i) strengthen the generalized quantum Stein's lemma in~\cite{fang2024generalized} for hypothesis testing between sets of states; (ii) provide counterexamples to the continuity of the regularized Petz \Renyi divergence and Hoeffding divergence; (iii) derive error exponents for adversarial quantum channel discrimination; and (iv) obtain error exponents for resource detection problems in coherence theory and entanglement theory.

\subsection{Refining quantum Stein's lemma between two sets}
\label{sec: Refining the Stein lemma between two sets of quantum state}

The following generalized quantum Stein's lemma for hypothesis testing between two sets of quantum states was established in~\cite[Theorem 32]{fang2024generalized}.

\begin{theorem}[Generalized quantum Stein's lemma.]\label{thm: generalized Steins} 
Let $\sA = \{\sA_n\}_{n\in\NN}$ and $\sB = \{\sB_n\}_{n\in\NN}$ be two sequences of sets of quantum states satisfying~\cite[Assumption 24]{fang2024generalized} and $\sA_n,\sB_n \subseteq \density(\cH^{\ox n})$ and $D_{\max}(\sA_n\|\sB_n) \leq cn$, for all $n \in \NN$ and a constant $c \in \RR_{\pl}$.
For any $\ve \in (0,1)$,
\begin{align}\label{eq: adversarial quantum Steins lemma}
    \lim_{n\to \infty} - \frac{1}{n} \log \beta_\ve(\sA_n\|\sB_n) = D^\reg(\sA\|\sB).
\end{align}
\end{theorem}

The above result can be both recovered and strengthened as follows.

\begin{theorem}
    Let $\sA = \{\sA_n\}_{n\in\NN}$ and $\sB = \{\sB_n\}_{n\in\NN}$ be two sequences of sets satisfying the same assumptions in Theorem~\ref{thm: generalized Steins}. For any $0 < r < D^\infty(\sA\|\sB)$, then 
    \begin{align}
        \liminf_{n\to \infty} -\frac{1}{n} \log \alpha_{n, r}(\sA_n\| \sB_n) \geq \bH_{r}^\infty(\sA\|\sB) > 0.
    \end{align}
    For any $r > D^\infty(\sA\|\sB)$, then 
    \begin{align}
        \liminf_{n\to \infty} -\frac{1}{n} \log (1-\alpha_{n, r}(\sA_n\| \sB_n)) \geq \bH_{r}^{*,\infty}(\sA\|\sB) > 0.
    \end{align}
\end{theorem}

This result shows that any Type-II error exponent below $D^\reg(\sA\|\sB)$ is achievable, with the corresponding Type-I error decaying exponentially at a rate at least $\bH_{r}^\infty(\sA\|\sB)$. Conversely, if the Type-II error exponent exceeds $D^\reg(\sA\|\sB)$, the Type-I error inevitably converges to one exponentially, with a rate at least $\bH_{r}^{*,\infty}(\sA\|\sB)$. Thus, the regularized quantum relative entropy $D^\reg(\sA\|\sB)$ delineates a sharp threshold for the asymptotic trade-off in hypothesis testing between two sets of quantum states. 
In particular, these results apply to adversarial quantum channel discrimination, which satisfies all the required assumptions~\cite{fang2025adversarial}. See Section~\ref{sec: applications to adversarial quantum channel discrimination} for more detailed discussion in this setting.

\begin{proof}
By the assumptions on the sequences, we have
\begin{align}\label{eq: recover proof tmp1}
D^\infty(\sA\|\sB) & = \sup_{\alpha \in (0,1)} D_{\Meas,\alpha}^{\infty}(\sA\|\sB) \\
& \leq \sup_{\alpha \in (0,1)} D_{\Petz,\alpha}^{\infty}(\sA\|\sB)\\
& \leq D^\infty(\sA\|\sB),
\end{align}
where $D_{\Meas,\alpha}$ refers to the measured \Renyi divergence. The first equality follows from~\cite[Lemmas 27, 28]{fang2024generalized}, and the inequalities use that $D_{\Meas,\alpha}(\rho\|\sigma) \leq D_{\Petz,\alpha}(\rho\|\sigma) \leq D(\rho\|\sigma)$ for any $\alpha \in (0,1)$. This implies
\begin{align}
    \sup_{\alpha \in (0,1)} D_{\Petz,\alpha}^{\infty}(\sA\|\sB) = D^\infty(\sA\|\sB).
\end{align}
Note that $D_{\Petz,\alpha}^{\infty}(\sA\|\sB)$ is monotone increasing in $\alpha$. Therefore, for any $0< r < D^\infty(\sA\|\sB)$, there exists $\alpha \in (0,1)$ such that $r < D_{\Petz,\alpha}^{\infty}(\sA\|\sB)$. Then
\begin{align}
    \bH_r^\infty(\sA\|\sB) = \sup_{\alpha \in (0,1)} \frac{\alpha - 1}{\alpha} \left(r - D_{\Petz,\alpha}^\infty(\sA\|\sB)\right) > 0.
\end{align}
By Theorem~\ref{thm: Hoeffding bound sets} and Lemma~\ref{lem: Hoeffding Petz sets regularized}, we have
\begin{align}
    \liminf_{n\to \infty} & -\frac{1}{n} \log \alpha_{n, r}(\sA_n\| \sB_n) \notag \\
    &  = H_r^\infty(\sA\|\sB) \geq \bH_r^\infty(\sA\|\sB) > 0
\end{align}
which shows that the Type-I error decays exponentially, and thus $r$ is an achievable rate. This recovers the direct part of the generalized quantum Stein's lemma in Theorem~\ref{thm: generalized Steins}.

Since $\inf_{\alpha > 1} D_{\Sand,\alpha}^\infty(\sA\|\sB) = D^\infty(\sA\|\sB)$~\cite[Lemma 27]{fang2024generalized} and $D_{\Sand,\alpha}^\infty(\sA\|\sB)$ is monotone increasing in $\alpha$, for any $r > D^\infty(\sA\|\sB)$, there exists $\alpha > 1$ such that $r > D_{\Sand,\alpha}^\infty(\sA\|\sB)$. This implies that
\begin{align}
    \bH_{r}^{*,\infty}(\sA\|\sB) = \sup_{\alpha > 1} \frac{\alpha - 1}{\alpha} \left(r - D_{\Sand,\alpha}^\infty(\sA\|\sB)\right) > 0.
\end{align} 
Applying Theorem~\ref{thm: strong converse exponent for composite correlated hypotheses} and Lemma~\ref{lem: Hoeffding anti Sand sets regularized}, we obtain
\begin{align}
   \liminf_{n\to \infty}& -\frac{1}{n} \log (1-\alpha_{n, r}(\sA_n\| \sB_n))\notag \\
   & \geq H_r^{*,\infty}(\sA\|\sB) = \bH_r^{*,\infty}(\sA\|\sB) > 0.
\end{align}
This shows that the Type-I error converges to one exponentially, and thus $r$ is not an achievable rate, recovering the converse part of the generalized quantum Stein's lemma in Theorem~\ref{thm: generalized Steins}.
\end{proof}

It is worth emphasizing that the quantum Hoeffding bound in Theorem~\ref{thm: Hoeffding bound sets} and the strong converse exponent in Theorem~\ref{thm: strong converse exponent for composite correlated hypotheses} (Eq.~\eqref{eq: strong converse exponent lower bound}) hold in great generality and do not require, particularly, the polar assumption as used in~\cite{fang2024generalized}. However, to recover the Stein's setting from the error exponent regime, one needs the continuity of the regularized Petz \Renyi divergences, i.e., $\sup_{\alpha \in (0,1)} D_{\Petz,\alpha}^\infty(\sA\|\sB) = D^\infty(\sA\|\sB)$. For this, we require the polar assumption in Eq.~\eqref{eq: recover proof tmp1}. Interestingly, a similar situation also arises in the context of best-case channel discrimination, where the continuity of the regularized sandwiched \Renyi divergence between channels is sufficient to establish the quantum Stein's lemma for two quantum channels (particularly the strong converse part); see~\cite[Theorem 21]{fang2025towards} for further details.

\subsection{Discontinuity of $D_{\Petz,\alpha}^\infty(\sA\|\sB)$ at $\alpha = 1$}

As discussed above, our error exponent results can recover the Stein's exponent provided the continuity of the regularized Petz \Renyi divergence holds at $\alpha = 1$, i.e., 
\begin{align} \label{eq: regularized Petz continuity}
    \sup_{\alpha \in (0,1)} D_{\Petz,\alpha}^\infty(\sA\|\sB) = D^\infty(\sA\|\sB).
\end{align} 
Therefore, for any sequences of sets $\sA = \{\sA_n\}_{n\in \NN}$ and $\sB = \{\sB_n\}_{n\in \NN}$ that satisfy our error exponent assumptions but violate the Stein's lemma, there must be a discontinuity in $D_{\Petz,\alpha}^\infty(\sA\|\sB)$ at $\alpha = 1$, i.e., a violation of Eq.~\eqref{eq: regularized Petz continuity}. Such examples can be found in~\cite{hayashi2025general}.

In this section, we provide a more direct analysis of these counterexamples and introduce additional ones. We show that such discontinuities can occur in both nonconvex and convex settings, as well as in scenarios where a composite hypothesis is tested against a simple hypothesis (see Example~\ref{eg: counterexample 1} and Example~\ref{eg: counterexample 2}), and conversely, where a simple hypothesis is tested against a composite hypothesis (see Example~\ref{eg: counterexample 3}). Furthermore, the example presented in~\cite{lami2024asymptotic} for the nonadditivity of the Petz \Renyi divergence between two sets, makes both sides of Eq.~\eqref{eq: regularized Petz continuity} diverge, which is not a valid counterexample for discontinuity

In the following, we denote $H(p):= -p\log p - (1-p) \log (1-p)$ as the binary entropy function and $D(p\|q):= p \log \frac{p}{q} + (1-p) \log \frac{1-p}{1-q}$ as the classical relative entropy between two binary distributions $(p,1-p)$ and $(q,1-q)$. 
Denote the set of types with length $n$ by $\cP_n$, which is the set of all empirical distributions. Give $P \in \cP_n$, the type class of $P$, denoted as $\cT_{n}(P)$, is the set of bit strings whose empirical distribution is $P$.
We choose two rational numbers $p_1=\frac{m_1}{m}$ and $p_2=\frac{m_2}{m}$ such that $p_1 < p_2 \in (0, p_3) $. Let $P_j = (p_j, 1-p_j)$ be the binary distribution. Consider a two-dimensional Hilbert space ${\cal H}$ spanned by $\{|0\rangle,|1\rangle\}$ and define the qubit state 
\begin{align} \label{eq: classical state}
    \rho(p):= p|0\rangle \langle 0|+  (1-p)|1\rangle \langle 1|.
\end{align}  
We denote the uniform distribution 
over $\cT_{km}(P_j)$
in the space ${\cal H}^{\otimes km}$ by 
\begin{align} \label{eq: rho j km}
    \rho_{j,km}:= \frac{1}{|\cT_{km}(P_j)|} \sum_{x^{km} \in \cT_{km}(P_j)} \ket{x^{km}}\bra{x^{km}},
\end{align}
for $j=1,2$, where $|\cT_{km}(P_j)| = {km \choose km_j}$ is the size of the type class.
Note that $\rho_{1,km}$ and $\rho_{2,km}$ are orthogonal, as their supports are defined by different type classes. We have $\rho(p_j)=\tr_{km-1} \rho_{j,km}$, where the partial trace can be understood as tracing out any $km-1$ subsystems, as it gives the same marginal state. In the following examples, we regard the Hilbert space ${\cal H}^{\otimes m}$ as a one-copy system.

\begin{example}[Nonconvex sets, $\sA_n$ composite and $\sB_n$ simple i.i.d.]\label{eg: counterexample 1}
Let $\rho(p)$ and $\rho_{j,km}$ defined by Eqs.~\eqref{eq: classical state} and~\eqref{eq: rho j km}, respectively.
We define the convex mixiture,
\begin{align} \label{eq: rho km definition}
    \rho_{km}:= \lambda\rho_{1,km}+(1-\lambda)\rho_{2,km},
\end{align} 
with the choice of $\lambda \in (0,1)$ such that
\begin{align}\label{eq: choice of lambda}
   &\min_{i\in \{1,2\}} \frac{1}{2} \log \frac{2\pi m_i (m-m_i)}{m} \geq H(\lambda).
\end{align}
Consider the set of states
\begin{align}
{\sA}_{n}&:= \left\{
\rho_{k_1 m}\otimes \cdots \otimes \rho_{k_l m}: k_1,\ldots,k_l \geq 1, \sum_{i=1}^l k_i =n\right\},\notag\\
\sB_n&:=\bigg\{\rho(p_3)^{\ox mn}\bigg\}.\label{eq: counterexample 1}
\end{align}
Then sequences  $\sA:=\{\sA_n\}$ and $\sB:=\{\sB_n\}$ are stable under tensor product. But we have 
\begin{align}
\sup_{\alpha \in (0,1)} D_{\Petz,\alpha}^{\infty}(\sA\| \sB) < D^{\infty}(\sA\| \sB).
\end{align}
Moreover, we have 
\begin{align}
\lim_{r\to 0 }H_{r}^{\infty}(\sB\| \sA)
&< D^{\infty}(\sA\| \sB), \\
\lim_{r\to 0 }
\bH_{r}^{\infty}(\sB\| \sA)
&< D^{\infty}(\sA\| \sB),
\end{align}
indicating that the Hoeffding bound cannot recover the Stein's exponent in this case.
\end{example}
\begin{proof}
Using the standard result in type method~\cite[Theorem 11.1.2]{cover2006elements}, we have 
\begin{align}
    D(\rho_{i,km}&\|\rho(p_3)^{\otimes km })\notag \\
     & = - \log {km \choose km_i} +km (D(p_i\|p_3 )+ H(p_i) ),
\end{align}
for $i=1,2$.
As $\rho_{1,km}$ and $\rho_{2,km}$ are orthogonal, we have
\begin{align}
D(\rho_{k m}&\| \rho(p_3)^{\otimes k m } )
=
 \lambda D(\rho_{1,km}\|\rho(p_3)^{\otimes km })\notag \\
& +(1-\lambda) D(\rho_{2,km}\|\rho(p_3)^{\otimes km })  -H(\lambda). \label{eq: D estimation tmp1}
\end{align}
Note that the binomial coefficient $\binom{n}{l}$ is evaluated as~\cite{macwilliams1981theory} 
\begin{align}\label{eq: binom estimate}
    \binom{n}{l} \leq \sqrt{\frac{n}{2\pi l (n-l)}}\; 2^{n H(l/n)},
\end{align}
for all $1\leq l \leq n-1$.
Thus, we have 
\begin{align}
-  \log & {km \choose km_i}
+km H(p_i) -{H(\lambda)}\notag\\
&  \ge \frac{1}{2}\log \frac{2\pi km_i(m-m_i)}{m}
-{H(\lambda)} \ge 0,\label{eq: remaining terms tmp2}
\end{align}
for $i \in \{1,2\}$, where the last inequality follows from the choice of $\lambda$ in Eq.~\eqref{eq: choice of lambda}. 
Discarding these two positive terms, we have  
\begin{align}
D(\rho_{k m}\| \rho(p_3)^{\otimes k m } )
\ge &
km( (1-\lambda) D(p_2\|p_3 )+\lambda D(p_1\|p_3 )).
\end{align}
By the additivity of quantum relative entropy, we have 
\begin{align}
    D(\rho_{k_1 m}&\otimes \cdots \otimes \rho_{k_l m}\| \rho(p_3)^{\otimes nm } ) \notag \\
    & \geq nm ( (1-\lambda) D(p_2\|p_3 )+\lambda D(p_1\|p_3 )).
\end{align}
This implies that 
\begin{align}
D(\sA_n\| \sB_n) & \geq nm( (1-\lambda) D(p_2\|p_3 )+\lambda D(p_1\|p_3 )),\\
D^{\infty}(\sA\| \sB) & \geq m( (1-\lambda) D(p_2\|p_3 )+\lambda D(p_1\|p_3 )).\label{eq: D estimation tmp4}
\end{align}
Using $k=n$ in Eq.~\eqref{eq: D estimation tmp1},
we obtain
\begin{align}\label{eq: D estimation tmp2}
\lim_{n\to \infty} \frac{1}{n}&D(\rho_{n m}\| \rho(p_3)^{\otimes n m } )\notag\\
&=m( (1-\lambda) D(p_2\|p_3 )+\lambda D(p_1\|p_3 )),
\end{align}
where some terms vanish because of the Stirling's approximation~\cite[Example 11.1.3]{cover2006elements}
\begin{align} \label{eq: Stirling's approximation}
    \frac{1}{n+1} 2^{nH(k/n)} \leq \binom{n}{k} \leq 2^{nH(k/n)}.
\end{align} 
Note that $\rho_{nm} \in \sA_n$ by choosing $l=1$, so we have $D(\sA_n\|\sB_n) \leq D(\rho_{n m}\| \rho(p_3)^{\otimes n m })$. Combined with Eq.~\eqref{eq: D estimation tmp2}, this implies that 
\begin{align}\label{eq: D estimation tmp3}
    D^{\infty}(\sA\| \sB) \leq m( (1-\lambda) D(p_2\|p_3 )+\lambda D(p_1\|p_3 )).
\end{align}
Therefore, we have from Eqs.~\eqref{eq: D estimation tmp4} and~\eqref{eq: D estimation tmp3} that
\begin{align}
D^{\infty}(\sA\| \sB)
=m( (1-\lambda) D(p_2\|p_3 )+\lambda D(p_1\|p_3 )).
\label{NJ1B}
\end{align}

We now estimate the Petz \Renyi divergence. For $\alpha \in (0,1)$, we have
\begin{align}
    \tr  (\rho_{i,km})^{\alpha} & (\rho(p_3)^{\otimes km })^{1-\alpha}\notag\\
     & = \left[\binom{km}{km_i}2^{-  km (D(p_i\|p_3 )+ H(p_i))}\right]^{1-\alpha},\label{eq: cor tmp2}
\end{align}
for $i=1,2$, by the type method~\cite[Theorem 11.1.2]{cover2006elements}. Since $\rho_{1,km}$ and $\rho_{2,km}$ are orthogonal, we have
\begin{align}\label{eq: petz quasi entropy bound}
 \tr (\rho_{km})^{\alpha} &  (\rho(p_3)^{\otimes km })^{1-\alpha} \notag \\
& = \lambda^\alpha \tr (\rho_{1,km})^{\alpha}  (\rho(p_3)^{\otimes km })^{1-\alpha}\notag\\
& + (1-\lambda)^\alpha \tr (\rho_{2,km})^{\alpha}  (\rho(p_3)^{\otimes km })^{1-\alpha}.
\end{align}
Using Lemma~\ref{lem: log rate min} and the fact that $D(p_2\|p_3) < D(p_1\|p_3)$, we have for any $\alpha \in (0,1)$,
\begin{align}
D_{\Petz,\alpha}^{\infty}(\sA\| \sB)
&\leq 
\lim_{n\to \infty }
\frac{1}{n}
D_{\Petz,\alpha}(\rho_{n m}\| \rho(p_3)^{\otimes n m } )\\
&=m D(p_2\|p_3 ),\label{eq: NJ1B 2}
\end{align}
where some terms vanish because of the Stirling's approximation in Eq.~\eqref{eq: Stirling's approximation}.

Combining~\eqref{NJ1B} and~\eqref{eq: NJ1B 2}, we have the strict inequality,
\begin{align}
\sup_{\alpha \in (0,1)} D_{\Petz,\alpha}^{\infty}(\sA\| \sB) < D^{\infty}(\sA\| \sB).
\end{align}

Now, we further analyze the Hoeffding bound.  
For $r>0$, we have from the alternative expression of the Hoeffding bound (similar to Remark~\ref{rem: alternative expression of Hoeffding divergence}) that
\begin{align}
& H_{n, r}
(\rho(p_3)^{\otimes n m } \|\rho_{n m}) \notag\\
=& \sup_{\alpha \in (0,1)} 
\left(
-\frac{\alpha}{1-\alpha} nr +
D_{\Petz,\alpha}
(\rho_{n m}\|\rho(p_3)^{\otimes n m }) 
\right). \label{eq: hoeffding finite upper bound}
\end{align} 

By Eq.~\eqref{eq: D estimation tmp2}, we have for any $\delta > 0$, 
\begin{align}
& \frac{1}{n}D_{\Petz,\alpha}
(\rho_{n m} \| \rho(p_3)^{\otimes n m } )\notag \\
 & \leq \frac{1}{n}D
(\rho_{n m} \| \rho(p_3)^{\otimes n m } ) \\
& \le m( (1-\lambda) D(p_2\|p_3 )+\lambda D(p_1\|p_3 )) + \delta =: z,
\label{NJ2}
\end{align}
for sufficiently large $n$. Then for any $\alpha > t_r:= \frac{z}{r+z} \in (0,1)$, we can check that 
\begin{align}
-\frac{\alpha}{1-\alpha} nr +
D_{\Petz,\alpha}
(\rho_{n m} \| \rho(p_3)^{\otimes n m } ) < 0.
\end{align}
So for $r>0$, we have
\begin{align}
& \sup_{\alpha \in (0,1)} 
\left(
-\frac{\alpha}{1-\alpha} nr +
D_{\Petz,\alpha}
(\rho_{n m} \| \rho(p_3)^{\otimes n m } ) 
\right) \notag \\
&= \sup_{\alpha \in (0, t_r]} 
\left(
-\frac{\alpha}{1-\alpha} nr +
D_{\Petz,\alpha}
(\rho_{n m} \| \rho(p_3)^{\otimes n m } ) 
\right).
\end{align}
Therefore, we have
\begin{align}
& H_{n, r}
(\rho(p_3)^{\otimes n m } \|\rho_{n m}) \notag \\
& = \sup_{\alpha \in (0, t_r]} 
\left(
-\frac{\alpha}{1-\alpha} nr +
D_{\Petz,\alpha}
(\rho_{n m} \| \rho(p_3)^{\otimes n m } ) 
\right).
\end{align}
This implies that
\begin{align}
\lim_{n\to \infty}&\frac{1}{n}H_{n, r} 
(\rho(p_3)^{\otimes n m } \|\rho_{n m}) \notag\\
=&\lim_{n\to \infty}\frac{1}{n}
 \sup_{\alpha \in (0, t_r]} 
\left(
-\frac{\alpha}{1-\alpha} nr +
D_{\Petz,\alpha}
(\rho_{n m} \| \rho(p_3)^{\otimes n m } ) 
\right) \\
\leq &\lim_{n\to \infty}\frac{1}{n}
 \sup_{\alpha \in (0, t_r]} 
D_{\Petz,\alpha}
(\rho_{n m} \| \rho(p_3)^{\otimes n m } ) \\
= &\lim_{n\to \infty}\frac{1}{n}
D_{\Petz,t_r}
(\rho_{n m} \| \rho(p_3)^{\otimes n m } ) \\
 = & m D(p_2\|p_3 ),
\end{align}
where the inequality follows by throwing away a negative term, the second equality follows by the monotonicity of Petz \Renyi divergence in $\alpha$ and the last equality follows from Eq.~\eqref{eq: NJ1B 2}.

This implies that
\begin{align}
H_{r}^{\infty}
(\sB\|\sA) 
&\le 
\lim_{n\to \infty}\frac{1}{n}H_{n, r}
(\rho(p_3)^{\otimes n m } \|\rho_{n m}) \\
&\leq m D(p_2\|p_3 ).\label{XG21}
\end{align}
Similarly, we have
\begin{align}
\bH_{r}^{\infty}(\sB\|\sA) 
& =
\sup_{\alpha \in (0,1)} 
\left(
-\frac{\alpha}{1-\alpha} 
r +
D_{\Petz,\alpha}^{\infty}
(\sA\|\sB) \right) \\
& \leq
\sup_{\alpha \in (0,1)} 
\left(
-\frac{\alpha}{1-\alpha} 
nr + m D(p_2\|p_3 ) \right)\\
&= m D(p_2\|p_3 ),\label{XG11}
\end{align}
where the inequality follows from Eq.~\eqref{eq: NJ1B 2}.

Thus, combining Eqs.~\eqref{NJ1B}, \eqref{XG21}, and 
\eqref{XG11},
we have
\begin{align}
\lim_{r\to 0 }H_{r}^{\infty}(\sB\| \sA)
&< D^{\infty}(\sA\| \sB), \\
\lim_{r\to 0 }
\bH_{r}^{\infty}(\sB\| \sA)
&< D^{\infty}(\sA\| \sB).
\end{align}
This completes the proof.
\end{proof}

\begin{figure}[htbp]
\centering
\begin{minipage}{\linewidth}
    \centering
    \includegraphics[width=\linewidth]{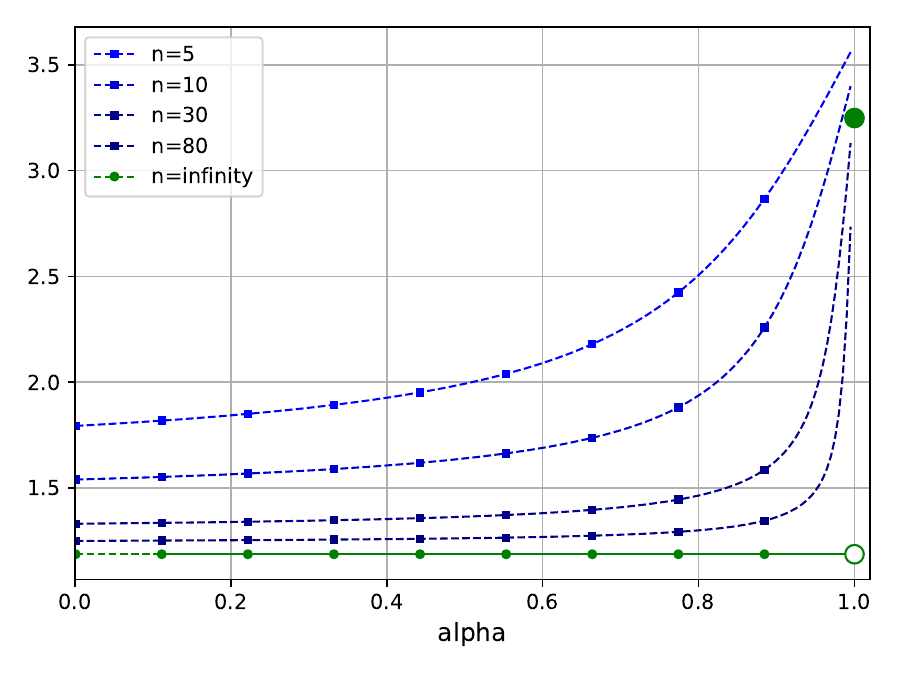}
    \caption*{(a) Petz \Renyi divergence}
\end{minipage}
\vspace{6pt}
\begin{minipage}{\linewidth}
    \centering
    \includegraphics[width=\linewidth]{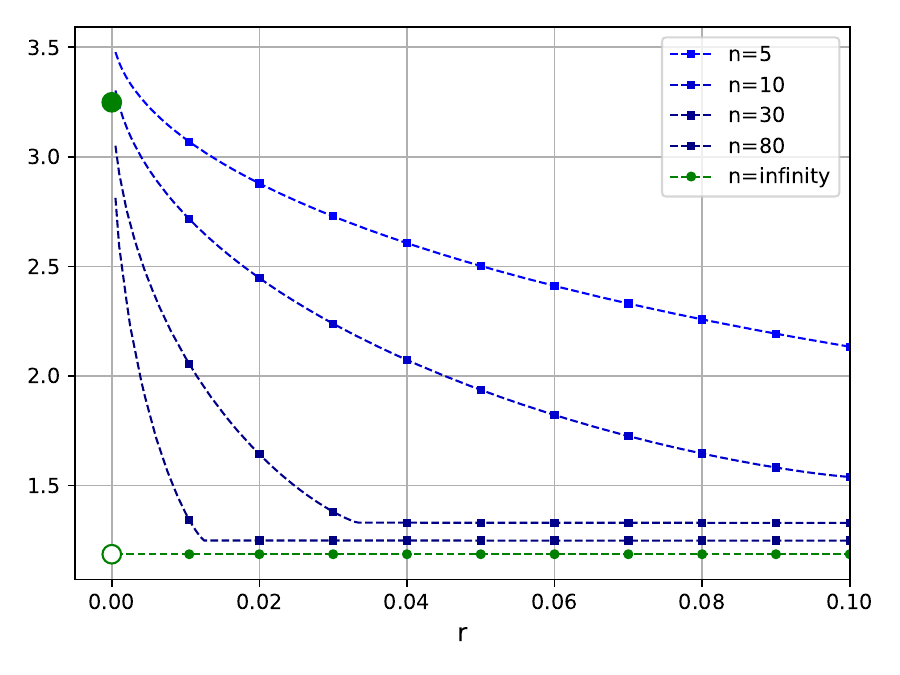}
    \caption*{(b) Hoeffding divergence}
\end{minipage}
\caption{Discontinuities in the regularized Petz \Renyi divergence and Hoeffding divergence given in Example~\ref{eg: counterexample 1}. We choose $m_1 = 1$, $m_2 = 3$, $m = 10$, $p_3 =\lambda = 0.5$. Panel (a) shows the upper bound for Petz \Renyi divergence $\frac{1}{n} D_{\Petz,\alpha}(\rho_{n m}\| \rho(p_3)^{\otimes n m})$ for different values of $n$ as a function of $\alpha$, which can be computed from Eq.~\eqref{eq: petz quasi entropy bound}. The solid line indicates tightness of the upper bound as analyzed in Eq.~\eqref{eq: petz reg tightness}, and the marked dot at $\alpha = 1$ corresponds to the regularized relative entropy in Eq.~\eqref{NJ1B}. Panel (b) displays the upper bound for Hoeffding divergence $\frac{1}{n}H_{n, r}(\rho(p_3)^{\otimes n m} \| \rho_{n m})$ from Eq.~\eqref{eq: hoeffding finite upper bound} for different $n$ as a function of $r$; for $n = \infty$, the upper bound from Eq.~\eqref{XG21} is used. The marked dot at $r = 0$ again indicates the regularized relative entropy from Eq.~\eqref{NJ1B}.}
\label{fig:petz_hoeffding_discontinuity}
\end{figure}

Here we note that, through additional analysis, we can also establish that 
$D_{\Petz,\alpha}^{\infty}(\sA\| \sB) = m D(p_2\|p_3 )$ for $\alpha\in (0,1)$ satisfying
\begin{align}
\lambda^{\alpha}  
2^{- (1-\alpha)m D(p_1\|p_3 )} 
\le
(1-(1-\lambda)^{\alpha}  )
2^{- (1-\alpha)m D(p_2\|p_3 )} .
\label{HH20}
\end{align}
At least, when $\alpha $ is sufficiently close to $1$, the above condition holds. Since 
\begin{align}
{l_1+l_2 \choose l}
=\sum_{l'=0}^l
{l_1 \choose l-l'}
{l_2 \choose l'}\ge
{l_1 \choose l-l''}
{l_2 \choose l''}
\end{align} for $l'' \in [0,l]$,
we have
\begin{align}
{km \choose km_j}^l\le {lkm \choose lkm_j}\le 2^{klm H(p_j)},
\label{HH1}
\end{align}
for $j=1,2$, where the second inequality follows from Eq.~\eqref{eq: binom estimate}.
Then, we have
\begin{align}
 \lambda^{\alpha}  &
\left[{km \choose km_1}  2^{- km (D(p_1\|p_3 )+ H(p_1))}\right]^{1-\alpha} \notag \\
& \leq
\lambda^{\alpha}  
2^{-  (1-\alpha)k m D(p_1\|p_3 )} \\
& \leq 
(1-(1-\lambda)^{\alpha}  )
2^{- (1-\alpha) m D(p_2\|p_3 )}
2^{- (1-\alpha) (k-1)  m D(p_1\|p_3 )} \\
& \le 
(1-(1-\lambda)^{\alpha}  )
2^{-(1-\alpha) k m D(p_2\|p_3 )},\label{HH30}
\end{align}
where the first inequality follows from Eq.~\eqref{HH1}
and the second inequality follows from Eq.~\eqref{HH20} and the fact that $D(p_2\|p_3) < D(p_1\|p_3)$. {Similarly, we have \begin{align}
\left[{km \choose km_2}  2^{- km (D(p_2\|p_3 )+ H(p_2))}\right]^{1-\alpha}  \leq
2^{-  (1-\alpha)k m D(p_2\|p_3 )}.\label{eq: cor tmp3} \end{align}}
Thus,
we have
\begin{align}
\tr  (\rho_{km})^{\alpha} & (\rho(p_3)^{\otimes km })^{1-\alpha} \notag \\
\leq &(1-(1-\lambda)^{\alpha}  )
2^{- (1-\alpha) km  D(p_2\|p_3 )}\notag\\
& + 
(1-\lambda)^{\alpha}  
2^{- (1-\alpha) k m D(p_2\|p_3 )}
\\
=&
2^{- (1-\alpha) k m D(p_2\|p_3 )},
\end{align}
where the inequality follows from {Eqs.~\eqref{eq: petz quasi entropy bound}, \eqref{HH30} and~\eqref{eq: cor tmp3}}. By the additivity of Petz \Renyi divergence, we have 
\begin{align}
    D_{\Petz,\alpha}(\rho_{k_1 m}\otimes \cdots \otimes \rho_{k_l m}\| \rho(p_3)^{\otimes nm } ) 
    & \geq nm D(p_2\|p_3 ).
\end{align}
This implies that 
\begin{align}\label{eq: Petz estimation tmp4}
D_{\Petz,\alpha}^{\infty}(\sA\| \sB) \geq m D(p_2\|p_3 ).
\end{align}
Combining Eqs.~\eqref{eq: NJ1B 2} and~\eqref{eq: Petz estimation tmp4}, we have
\begin{align}\label{eq: petz reg tightness}
D_{\Petz,\alpha}^{\infty}(\sA\| \sB)
= m D(p_2\|p_3 ),
\end{align}
for Example~\ref{eg: counterexample 1} when $\alpha$ satisfies Eq.~\eqref{HH20}.

An illustration for the discontinuities of the regularized Petz \Renyi divergence and the Hoeffding divergence is shown in Figure.~\ref{fig:petz_hoeffding_discontinuity}, where we can clearly see the discontinuities of both regularized quantities at $\alpha = 1$ and $r=0$, respectively.

\bigskip
The second example follows the construction in \cite[Appendix C]{hayashi2025general}, though that reference omits the technical details which we provide here.

\begin{example}[Convex sets, $\sA_n$ composite and $\sB_n$ simple i.i.d.]\label{eg: counterexample 2}
Let $\rho(p_3)$ and  $\rho_{km}$ defined in Example~\ref{eg: counterexample 1}. Consider the sets ${\sS}_{n}:=\{\tr_{km-n} \rho_{km}: k \in \NN\}$, where $\tr_{km-n}$ is the partial trace over the initial $km-n$ subsystems. Define
\begin{align}
{\sA}_{n}:= \conv \Bigg\{
\rho_{n_1 }\otimes \cdots \otimes \rho_{n_l }:n_1,\ldots,n_l \ge 1,\notag \\
\rho_{n_j }\in {\sS}_{n_j},
\sum_{j=1}^l n_j=n\Bigg\},
\end{align}
and $\sB_n=\{\rho(p_3)^{\ox n}\}$.
Then the sequences
$\sA:=\{{\sA}_{n}\}$ and $\sB:=\{\sB_{n}\}$ are convex compact and stable under tensor product. But we have
\begin{align}
\sup_{\alpha \in (0,1)} D_{\Petz,\alpha}^{\infty}(\sA\| \sB) < D^{\infty}(\sA\| \sB).
\end{align}
Moreover, we have 
\begin{align}
\lim_{r\to 0 }H_{r}^{\infty}(\sB\| \sA)
&< D^{\infty}(\sA\| \sB), \\
\lim_{r\to 0 }
\bH_{r}^{\infty}(\sB\| \sA)
&< D^{\infty}(\sA\| \sB),
\end{align}
indicating that the Hoeffding bound cannot recover the Stein's exponent in this case.
\end{example}
\begin{proof}
Let $\rho$ be any quantum state on $n$ subsystems. We can easily check the identity 
\begin{align} 
    D\left(\rho\;\bigg\| \bigotimes_{i=1}^n \sigma_i\right) = D\left(\rho\;\bigg\| \bigotimes_{i=1}^n \rho_i\right) + \sum_{i=1}^n D(\rho_i\|\sigma_i),
\end{align} 
where $\rho_i$ is the reduced state of $\rho$ on the $i$-th subsystem. Then we have for any  $\rho\in {\sA_n}$,
\begin{align}
 D(\rho\| \rho(p_3)^{\otimes n } )
& =
D(\rho\| (\tr_{n-1}\rho)^{\otimes n } )
\notag \\
& +
D( (\tr_{n-1}\rho)^{\otimes n }\| \rho(p_3)^{\otimes n } ).
\end{align}
Since $\tr_{n-1} \rho = \rho(\lambda p_1 + (1-\lambda) p_2)$, we have 
\begin{align}
D(\rho\| \rho(p_3)^{\otimes n } ) 
\ge &
n D(\rho(\lambda p_1 +  (1-\lambda) p_2) \| \rho(p_3))\\
& = n D(\lambda p_1 +  (1-\lambda) p_2\|p_3).
\end{align}
This implies that
\begin{align}
D(\sA_n\| \sB_n) 
\ge &
n D(\lambda p_1 +  (1-\lambda) p_2\|p_3).
\end{align}
On the other hand, since $\rho(\lambda p_1 +  (1-\lambda) p_2)^{\ox n} \in {\sA}_{n}$ and $\rho(p_3)^{\ox n} \in \sB_n$, the above bound is achievable at these states and therefore, we have
\begin{align}
D(\sA_n\| \sB_n)
=nD(\lambda p_1 +  (1-\lambda) p_2\|p_3).
\label{NT1}
\end{align}
Thus, we have
\begin{align}
D^\infty(\sA\| \sB)
=D(\lambda p_1 +  (1-\lambda) p_2\|p_3).
\label{NT2}
\end{align}

Since $\rho_{km} \ox \rho_{n-km} \in \sA_n$ with $0\le n-k m < m$, we have
\begin{align}
D_{\Petz,\alpha}(\sA_n\| \sB_n)
\le&
D_{\Petz,\alpha}(\rho_{k m}\|\rho(p_3)^{\otimes k m } )\notag \\
&+
D_{\Petz,\alpha}(\rho_{n-k m}\|\rho(p_3)^{\otimes n-k m } ).
\label{NT2}
\end{align}
Using the same calculation as in Eq.~\eqref{eq: NJ1B 2}, for $\alpha>0$, 
\begin{align}
D_{\Petz,\alpha}^{\infty}(\sA\| \sB)
\le D(p_2\|p_3 ).
\end{align}
Therefore, we have 
\begin{align}
 \sup_{\alpha\in (0,1)}  D_{\Petz,\alpha}^{\infty}(\sA\| \sB) &  \leq D(p_2\|p_3 )\\
 & <   D(\lambda p_1 +  (1-\lambda) p_2\|p_3) \\
 & = D^\infty(\sA\| \sB).
\label{NT3}
\end{align}
Finally, the statement for the Hoeffding bounds can be shown similarly as in Example~\ref{eg: counterexample 1}. This completes the proof.
\end{proof}

The next example demonstrates an extreme case where the regularized Petz \Renyi divergence vanishes while the regularized relative entropy diverges.
\begin{example}[Convex sets, $\sA_n$ simple i.i.d. and $\sB_n$ composite.]\label{eg: counterexample 3}
Let $\rho(p_2)$ and $\rho_{2,km}$ defined by Eqs.~\eqref{eq: classical state} and~\eqref{eq: rho j km}, respectively. Consider the sets $\sA_n:=\{\rho(p_2)^{\ox mn}\}$, and 
\begin{align}
{\sB}_{n}&:= \conv\Bigg\{
\rho_{2,k_1 m}\otimes \cdots \otimes \rho_{2,k_l m}:\notag\\
& \hspace{2cm}k_1,\ldots,k_l \geq 1, \sum_{i=1}^l k_i =n\Bigg\}.\label{eq: counterexample 3}
\end{align}
Then the sequences 
$\sA:=\{{\sA}_{n}\}$ and $\sB:=\{\sB_{n}\}$ are convex compact and stable under tensor product. But we have 
\begin{align}
\sup_{\alpha \in (0,1)} D_{\Petz,\alpha}^{\infty}(\sA\| \sB) = 0 < \infty = D^{\infty}(\sA\| \sB).
\end{align}
\end{example}
\begin{proof}
Note that the support of the state 
$\rho_{2,k_1 m}\otimes \cdots \otimes \rho_{2,k_l m}$ with $\sum_{i=1}^l k_i=n$
is included in the typical subspace ${\cal T}_{nm}(P_2)$. The support $\rho(p_2)^{\ox n m}$ is full space. Thus, we have
\begin{align}
D(\sA_n\|\sB_n) = \min_{\sigma \in \sB_n}D(\rho(p_2)^{\ox n m} \|\sigma)
=\infty,
\end{align}
and therefore,
\begin{align}
D^\infty(\sA\|\sB) 
=\infty.\label{XG1}
\end{align}
For the Petz \Renyi divergence, we have 
\begin{align}
D_{\Petz,\alpha}(\rho(p_2)^{\otimes km }\|\rho_{2,km})
=\frac{1}{\alpha-1}\log
\left[2^{- km H(p_2)}{km \choose km_2}\right]^{\alpha}, 
\end{align}
by the standard result in type method as in~\cite[Eq.~(11.15)]{cover2006elements}. By the Stirling's approximation in Eq.~\eqref{eq: Stirling's approximation}, we have
\begin{align}
\lim_{n\to \infty}\frac{1}{n}D_{\Petz,\alpha}(\rho(p_2)^{\otimes nm }\|\rho_{2,nm})
=0.
\end{align}
Thus, we have for any $\alpha \in (0,1)$,
\begin{align}
D_{\Petz,\alpha}^\infty(\sA\|\sB)
\leq \lim_{n\to \infty}\frac{1}{n}D_{\Petz,\alpha}(\rho(p_2)^{\otimes n m}\|\rho_{2,n m})
=0.\label{XG2}
\end{align}
As the divergence is always nonnegative, the above equality is attained. Finally, the combination of Eqs.~\eqref{XG1} and \eqref{XG2} implies that
\begin{align}
\sup_{\alpha \in (0,1)} D_{\Petz,\alpha}^{\infty}(\sA\| \sB) = 0 < \infty = D^{\infty}(\sA\| \sB).
\end{align}
This completes the proof.
\end{proof}

The last example is presented in~\cite{lami2024asymptotic} to demonstrate the nonadditivity of the Petz \Renyi divergence between two sets of quantum states. However, we show that this example does not give a violation of Eq.~\eqref{eq: regularized Petz continuity} as both sides diverge.
\begin{example}[Convex sets.]
Let $\sA_n := \SEP_n$ be the set of bipartite separable states on $n$ subsystems. Let $\sB_n := \{\rho^{\ox n}\}$ with $\rho := \frac{1}{d(d-1)}(I - \sum_{i,j=1}^d \ketbra{ij}{ji})$
being the antisymmetric Werner state on $\CC^d \ox \CC^d$. Then the sequences 
$\sA:=\{{\sA}_{n}\}$ and $\sB:=\{\sB_{n}\}$ are convex compact and stable under tensor product. But we have 
\begin{align}
    \sup_{\alpha \in (0,1)} D_{\Petz,\alpha}^\infty(\sA\|\sB) = D^\infty(\sA\|\sB) = \infty.
\end{align} 
\end{example}
\begin{proof}
It has been shown in~\cite[Eq. (E5)]{lami2024asymptotic} that for any $n \in \NN$ and $\alpha \in (0,1)$,
\begin{align}
    D_{\Petz,\alpha}(\sA_n\|\sB_n) = \frac{\alpha}{1-\alpha} D(\sB_n\|\sA_n).
\end{align}
This implies 
\begin{align}
   D_{\Petz,\alpha}^\infty(\sA\|\sB)= \frac{\alpha}{1-\alpha} D^\infty(\sB\|\sA),
\end{align}
by taking regularization on both sides.
Note that we have an explicit lower bound $D^\infty(\sB\|\sA) \geq \log_2 \sqrt{4/3} > 0$~\cite[Corollary 3]{christandl2012entanglement}. Taking supremum over $\alpha \in (0,1)$, we have
\begin{align}
    \sup_{\alpha \in (0,1)} D_{\Petz,\alpha}^\infty(\sA\|\sB) & = \sup_{\alpha \in (0,1)} \frac{\alpha}{1-\alpha} D^\infty(\sB\|\sA) = \infty.
\end{align}
However, as we always have the direction,
\begin{align}
    \sup_{\alpha \in (0,1)} D_{\Petz,\alpha}^\infty(\sA\|\sB) \leq D^\infty(\sA\|\sB),
\end{align}
this implies that 
\begin{align}
    \sup_{\alpha \in (0,1)} D_{\Petz,\alpha}^\infty(\sA\|\sB) = D^\infty(\sA\|\sB) = \infty.
\end{align} 
\end{proof}

\subsection{Error exponents in adversarial channel discrimination}\label{sec: applications to adversarial quantum channel discrimination}

The adversarial quantum channel discrimination was recently proposed and analyzed in~\cite{fang2025adversarial}, where a tester interacts with an untrusted quantum device that generates quantum states upon request. The device guarantees that the states are produced by either a quantum channel $\cN$ or a quantum channel $\cM$.
More formally, let $\cN_{A\to B}$ and $\cM_{A\to B}$ be the two quantum channels to be distinguished, and let $\cU_{A\to BE}$ and $\cV_{A\to BE}$ denote their respective Stinespring dilations, with $E$ representing the environmental system. Let $\CPTP(X\!:\!Y)$ denote the set of completely positive and trace-preserving maps from input system $X$ to output system $Y$. 

An \emph{adaptive strategy} for adversarial discrimination proceeds as follows. Suppose the device operates as channel $\cN$. Initially, the adversary prepares a quantum state via an operation $\cP^1 \in \CPTP(R_0E_0\!:\!A_1R_1)$, where $R_0$ and $E_0$ are trivial ($|R_0| = |E_0| = 1$), and sends system $A_1$ through the channel $\cU$, generating the output state $\cU \circ \cP^1$ and returning system $B_1$ to the tester. In the next round, the adversary performs an internal update $\cP^2 \in \CPTP(E_1R_1\!:\!A_2R_2)$, utilizing information stored in the quantum memory $R_1$ and the environmental system $E_1$ from the previous round. The adversary then sends system $A_2$ through the channel $\cU$ again, producing the output state $\cU \circ \cP^2 \circ \cU \circ \cP^1$ and returning system $B_2$ to the tester. This process can be repeated for $n$ rounds. A \emph{non-adaptive strategies} for adversarial discrimination is a subclass of adaptive strategies that disregards the environmental systems $E_i$ and performs no updates between rounds, that is, taking the operations $\cP_i, \cQ_i$ ($i \geq 2$) simply as identity maps, with the choice $R_i = A_{i+1} \cdots A_n$. 

After $n$ rounds of state generation, the tester obtains an overall state on $B_1\cdots B_n$ in their possession as:
\begin{align*}
\rho[\{\cP^i\}_{i=1}^n] := \tr_{R_nE_n} \prod_{i=1}^{n} \left[\cU_{A_i \to B_i E_i} \circ \cP^i_{R_{i-1}E_{i-1} \to A_{i}R_{i}}\right].
\end{align*}
Similarly, if the device is governed by $\cM$ and the internal operations by the adversary are given by $\cQ^i$, then the overall state is given by 
\begin{align*}
\sigma[\{\cQ^i\}_{i=1}^n] := \tr_{R_nE_n} \prod_{i=1}^{n} \left[\cV_{A_i \to B_i E_i} \circ \cQ^i_{R_{i-1}E_{i-1} \to A_{i}R_{i}}\right].  
\end{align*}
The tester needs to perform a binary quantum measurement $\{M_n, I-M_n\}$ on systems $B_1 \cdots B_n$ to determine which channel was used inside the black box.

Due to limited knowledge of the device's internal workings and the adversary's strategies, the tester only gets partial information, knowing that the state belongs to one of two sets:
\begin{align*} 
    \sA_n & := \{\rho[\{\cP^i\}_{i=1}^n] : \cP^i \in \CPTP(R_{i-1}E_{i-1}\!\!:\!\!A_iR_i), \forall R_i, \forall i\},\\
    \sB_n & := \{\sigma[\{\cQ^i\}_{i=1}^n] : \cQ^i \in \CPTP(R_{i-1}E_{i-1}\!\!:\!\!A_iR_i), \forall R_i, \forall i\},
\end{align*}
where the adversary's internal memory $R_i$ may have arbitrarily large dimension. In particular, for non-adaptive strategies, where the adversary ignores the environmental systems $E_i$ and performs no updates between rounds, the sets $\sA_n$ and $\sB_n$ reduce to
\begin{align} 
    \sA_n' & := \{\cN^{\ox n}(\rho_n): \rho_n \in \density(A^{\ox n})\},\\
    \sB_n' & := \{\cM^{\ox n}(\sigma_n): \sigma_n \in \density(A^{\ox n})\}.
\end{align}

It has been shown in~\cite{fang2025adversarial} that the optimal Stein exponent for both adaptive and non-adaptive strategies is given by the regularized quantum relative entropy between the two channels:
\begin{align}
    D^{\inf,\reg}(\cN\|\cM) := \lim_{n\to \infty} \frac{1}{n} D^{\inf}(\cN^{\ox n}\|\cM^{\ox n}),
\end{align}
where the quantum relative entropy between two channels is defined by
\begin{align}
    D^{\inf}(\cN\|\cM) := \inf \left\{ D(\cN(\rho)\| \cM(\sigma)): \rho, \sigma \in \density(A)\right\}.
\end{align}

\paragraph{Error exponents} We can check that $\{\sA_n\}, \{\sA'_n\}, \{\sB_n\}, \{\sB_n'\}$ are all stable sequences of convex and compact sets of quantum states (the convexity of $\{\sA_n\}, \{\sB_n\}$ is proved in~\cite{fang2025adversarial}). Therefore, our result directly applies to adversarial quantum channel discrimination. In particular, for any $0 < r < D^{\inf,\reg}(\cN\|\cM)$, the optimal Type-I error exponent under an exponential constraint on the Type-II error is given by
\begin{align}
    \lim_{n\to \infty} -\frac{1}{n} \log \alpha_{n, r}(\sA_n\| \sB_n) & = H_r^\infty(\sA\|\sB),\\
    \lim_{n\to \infty} -\frac{1}{n} \log \alpha_{n, r}(\sA_n'\| \sB_n') & = H_r^\infty(\sA'\|\sB') > 0,
\end{align}
where we can ensure the strict positivity as the continuity of the regularized Petz \Renyi divergence holds in this case~\cite{fang2025adversarial}. 
It would be interesting to further investigate whether adaptive strategies can outperform non-adaptive strategies in terms of the error exponent, i.e., whether $H_r^\infty(\sA\|\sB) < H_r^\infty(\sA'\|\sB')$ for some channels $\cN, \cM$ and rate $r$.

The authors in~\cite{hayashi2016correlation} consider the error exponent and strong converse exponents for distinguishing the null hypothesis
\(\sA_n''=\{\rho_{AB}^{\otimes n}\}\) from the composite alternative
\(\sB_n''=\{\rho_A^{\otimes n}\otimes\sigma_{B^n}:\ \sigma\in\density(B^n)\}\), which reflects the scenario of correlation detection. This can be modelled as an adversarial channel discrimination problem as well. In particular, let $\cR_\rho$ be a replacer channel that always outputs $\rho$ regardless of the input. Then the above $\sA_n''$ can be regarded as the image set of the replacer channel $(\cR_{\rho_{AB}})^{\ox n}$ and $\sB_n''$ is the image set of the channel $(\cR_{\rho_A} \ox I_B)^{\ox n}$. All assumptions in Theorem~\ref{thm: Hoeffding bound sets} and Theorem~\ref{thm: strong converse exponent for composite correlated hypotheses} are satisfied in this case. Therefore, we can recover the results in~\cite{hayashi2016correlation} as:
\begin{align}
    \lim_{n\to \infty} & -\frac{1}{n} \log \alpha_{n, r}(\sA_n''\| \sB_n'')\notag\\
     & = \sup_{\alpha \in (0,1)} \frac{\alpha - 1}{\alpha} \left(r - I_{\Petz,\alpha}(A\!:\!B)_\rho\right),
\end{align} 
{for $0 < r < I(A\!:\!B)_\rho$}, where \(I(A\!:\!B)_\rho := D(\sA_1''\| \sB_1'')\) and \(I_{\Petz,\alpha}(A\!:\!B)_\rho := D_{\Petz,\alpha}(\sA_1''\| \sB_1'')\) denote, respectively, the (Umegaki) mutual information and the Petz \Renyi mutual information (which is additive), and
\begin{align}
    \lim_{n\to \infty} & -\frac{1}{n} \log (1-\alpha_{n, r}(\sA_n''\| \sB_n''))\notag\\
     & = \sup_{\alpha > 1} \frac{\alpha - 1}{\alpha} \left(r - I_{\Sand,\alpha}(A\!:\!B)_\rho\right),
\end{align}
{for $I(A\!:\!B)_\rho < r < \sup_{\alpha > 1} I_{\Sand,\alpha}(A\!:\!B)_\rho$,}
where $I_{\Sand,\alpha}(A\!:\!B)_\rho := D_{\Sand,\alpha}(\sA_1''\| \sB_1'')$ denotes the sandwiched \Renyi mutual information (which is also additive).

\paragraph{Examples for nonadditivity} The adversarial channel discrimination also provides a scenario where we can find nonadditivity examples for the Petz \Renyi divergence and Hoeffding divergence between two sets of quantum state. For this, we consider two qutrit quantum channels. Let $\cN(\cdot)=\tr[\cdot] \rho$ to be the replacer channel with 
\begin{align}
    \rho = 0.9 \cdot \ket{\psi}\bra{\psi} + 0.1 \cdot \frac{I}{3},
\end{align}
where $\ket{\psi} = \frac{1}{\sqrt{2}}(\ket{0}+\ket{2})$.
Let $\cM$ be the platypus channel~\cite{Leditzky_2023}, $\cM(X) = M_0 X M_0^\dagger + M_1 X M_1^\dagger$ with Kraus operators
\begin{align}
    M_0 = \begin{bmatrix}
        \sqrt{p} & 0 & 0\\
        0 & 0 & 0\\
        0 & 1 & 0
    \end{bmatrix},
    \qquad 
    M_1 = \begin{bmatrix}
        0 & 0 & 0\\
        \sqrt{1-p} & 0 & 0\\
        0 & 0 & 1
    \end{bmatrix}.
\end{align}
In this case, we have the image sets of these channels as
\begin{align}
    \sA_n' & = \{\rho^{\ox n}\}, \\
    \sB_n' & = \{\cM^{\ox n}(\sigma_n): \sigma_n \in \density((\CC^3)^{\ox n})\}.
\end{align}
As these sets are given by semidefinite constraints, we can efficiently evaluate the Petz \Renyi divergence and Hoeffding divergence between them via semidefinite programming (see Remark~\ref{rem: computability of Hoeffding sets}). More explicitly, for fixed $\alpha \in (0,1)$, we can efficiently evaluate 
\begin{align}
D_{\Petz, \alpha}(\sA_n'\|\sB_n') & = \inf_{\sigma_n \in \density} D_{\Petz,\alpha}(\rho^{\ox n}\|\cM^{\ox n}(\sigma_n)),
\end{align}
by the QICS package~\cite{he2024qics}.
Moreover, we can also evaluate the Hoeffding divergence
\begin{align}
H_{n,r}&(\sA_n'\|\sB_n')\notag\\
 & = \bH_{n, r}(\sA_n'\|\sB_n')\\
& = \sup_{\alpha \in (0,1)} \frac{\alpha - 1}{\alpha} \left(nr - D_{\Petz,\alpha}(\sA_n'\|\sB_n')\right),
\end{align}
by scanning the parameter $\alpha \in (0,1)$ with fine grid.

\begin{figure}[htbp]
\centering
\begin{minipage}{\linewidth}
    \centering
    \includegraphics[width=\linewidth]{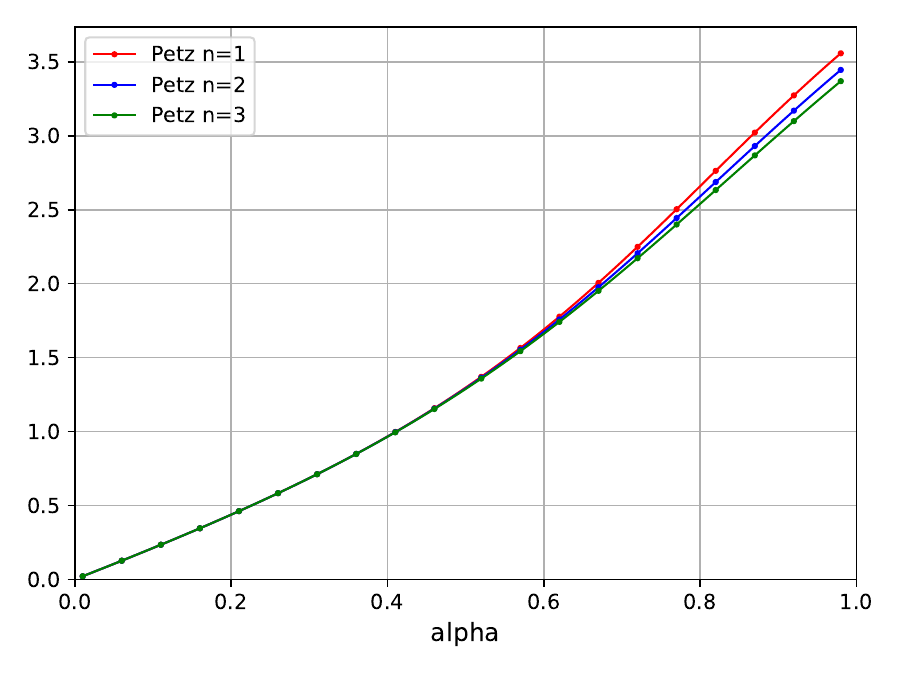}
    \caption*{(a) Petz \Renyi divergence}
\end{minipage}
\vspace{6pt}
\begin{minipage}{\linewidth}
    \centering
    \includegraphics[width=\linewidth]{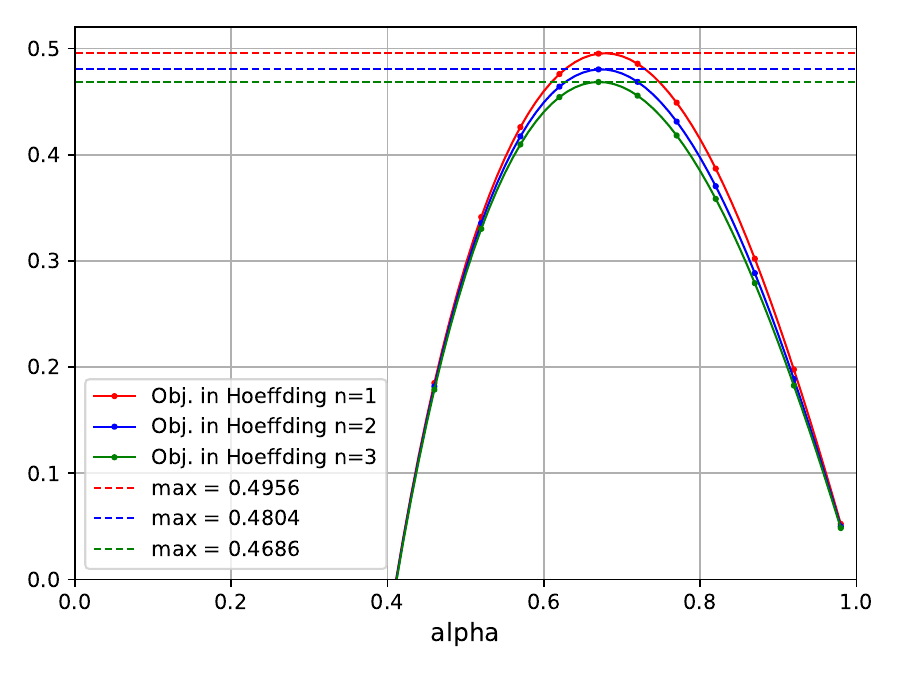}
    \caption*{(b) Hoeffding divergence}
\end{minipage}
\caption{Non-additivity for the Petz \Renyi divergence and the Hoeffding divergence between the image set of two channels. Here, we consider the replacer channel $\cN$ with output state $\rho = 0.9 \cdot \ket{\psi}\bra{\psi} + 0.1 \cdot I/3$, where $\ket{\psi} = (\ket{0}+\ket{2})/\sqrt{2}$, and the platypus channel $\cM$ with channel parameter ${p}=0.01$. We plot (a) the Petz \Renyi divergence $D_{\Petz,\alpha}(\sA_n'\|\sB_n')/n$ and (b) the objective function in Hoeffding divergence for $n=1,2,3$ as functions of $\alpha \in (0,1)$, respectively. The Hoeffding divergence $H_{n,1}(\sA_n'\|\sB_n')/n$ is indicated by the maximum value in dashed lines.}
\label{fig:petz_hoeffding_nonadd}
\end{figure}

The numerical results are shown in Figure~\ref{fig:petz_hoeffding_nonadd} where we use channel parameter ${p}=0.01$ and scan $\alpha\in(0,1)$ with step size $0.01$. Panel (a) displays the Petz \Renyi divergence $D_{\Petz,\alpha}(\sA_n'\|\sB_n')/n$ for $n=1,2,3$ as a function of $\alpha$, while panel (b) shows the objective function of the Hoeffding divergence versus $\alpha$, together with its maximum (i.e., the Hoeffding divergence $H_{n,r}(\sA_n'\|\sB_n')/n$) for $n=1,2,3$. Here we choose the rate $r=1$. The plots exhibit a clear separation between different numbers of copies for both the Petz \Renyi and Hoeffding divergences, illustrating non-additivity in this example and justifying the necessity of regularization in our main results.

\subsection{Error exponents in quantum resource detection}
\label{sec: applications to resource theories}

As our results only require minimal assumptions, they can be readily applied to the detection problem in various quantum resource theories (see e.g.~\cite{hayashi2025entanglement}), where the stability of the set of free states is naturally satisfied~\cite{brandao2015reversible,chitambar2019quantum}, including entanglement~\cite{brandao2010generalization,fang2019non,regula2019one,theurer2025single}, coherence~\cite{Winter_2016,diaz2018using,fang2018probabilistic,regula2018one,hayashi2021finite}, asymmetry~\cite{Marvian_2013}, thermodynamics~\cite{Brandao2015thermodynamics} and purity~\cite{horodecki2003reversible} and magic~\cite{Bravyi_2005,fang2020no,fang2022no,fang2025oneshotdistillationconstantoverhead}. Our refined analysis of error exponents can thus provide a better understanding of the trade-offs between Type-I and Type-II errors in hypothesis testing within these resource theories.

We showcase coherence theory and entanglement theory as two illustrative examples with explicit expressions. These examples have been previously studied in~\cite{hayashi2025entanglement}, but the error exponents obtained therein are not tight. Here we provide tight characterizations using our results.

\paragraph{Quantum coherence detection} In the resource theory of coherence, the set of free states is given by the set of incoherent states $\sI_n$, which is convex, compact, and stable under tensor products. Moreover, the Petz \Renyi divergence between a quantum state and the set of incoherent states is known to be additive~\cite[Theorem 3]{Zhu_2017}. Therefore, our results yield explicit expressions for the error exponent in hypothesis testing between a quantum state and the set of incoherent states
\begin{align}
    \lim_{n\to \infty} -\frac{1}{n} \log \alpha_{n, r}(\rho^{\ox n}\| \sI_n) & = \sup_{\alpha \in (0,1)} \frac{\alpha - 1}{\alpha} \left(r - C_\alpha(\rho)\right),
\end{align}
{for $0 < r < C(\rho):= D(\rho\|\sI_1)$} where $C_\alpha(\rho) := D_{\Petz,\alpha}(\rho\|\sI_1)$ has closed form expression~\cite{Chitambar_2016comparison} as
\begin{align}
    C_{\alpha}(\rho) = \frac{\alpha}{\alpha-1} \log \tr\left[(\Delta(\rho^\alpha))^{1/\alpha}\right].
\end{align}
where $\Delta(\cdot)$ is the completely dephasing operation in the reference basis. 

As for the strong converse exponent, the sandwiched \Renyi divergence between these sets is also additive~\cite[Theorem 3]{Zhu_2017} and is given by
\begin{align}
    C^*_\alpha(\rho) := D_{\Sand,\alpha}(\rho\|\sI_1) = \inf_{\sigma_B \in \density(B)} D_{\Sand,\alpha}(\rho_{\text{mc}}\|I_A \ox \sigma_B),
\end{align}
where $\rho_{\text{mc}} := \sum_{i,j} \rho_{ij} \ket{ii}\bra{jj}$ is the associated maximally correlated state if $\rho=\sum_{i,j} \rho_{ij}\ket{i}\bra{j}$. The differentiability of $C^*_\alpha(\rho)$ with respect to $\alpha$ has been established in~\cite[Proposition 11]{hayashi2016correlation}. Therefore, our results yield the strong converse exponent in this case as well:
\begin{align}
    \lim_{n\to \infty} -\frac{1}{n} \log & (1-\alpha_{n, r}(\rho^{\ox n}\| \sI_n)) \notag \\
    & = \sup_{\alpha > 1} \frac{\alpha - 1}{\alpha} \left(r - C^*_\alpha(\rho)\right),
\end{align}
{for $C(\rho) < r < \sup_{\alpha > 1} C^*_{\alpha}(\rho)$.}

\paragraph{Quantum entanglement detection} In the resource theory of entanglement, the set of free states is the set of separable states $\SEP_n$, which is convex, compact, and stable under tensor products. For a maximally correlated state $\rho_{\text{mc}} := \sum_{i,j} \rho_{ij} \ket{ii}\bra{jj}$—which includes many important cases of interest, e.g., pure states and mixtures of Bell states—the Petz and sandwiched \Renyi divergences relative to the set of separable states are known to be additive. The error exponent and strong converse exponent are then analogous to those in the resource theory of coherence. In particular, the error exponent is given by
\begin{align}
    \lim_{n\to \infty} -\frac{1}{n} & \log \alpha_{n, r}(\rho_{\text{mc}}^{\ox n}\| \SEP_n)\notag \\
     & = \sup_{\alpha \in (0,1)} \frac{\alpha - 1}{\alpha} \left(r - E_{\alpha}(\rho_{\text{mc}})\right),
\end{align}
{for $0<r<E(\rho_{\text{mc}}):= D(\rho_{\text{mc}}\|\SEP_1)$},
where $E_{\alpha}(\rho_{\text{mc}}) := D_{\Petz,\alpha}(\rho_{\text{mc}}\|\SEP_1) = D_{\Petz,\alpha}(\rho\|\sI_1)$ is the \Renyi relative entropy of entanglement, which has the closed-form expression~\cite[Corollary 1]{Zhu_2017}:
\begin{align}
    E_{\alpha}(\rho_{\text{mc}}) = \frac{\alpha}{\alpha-1} \log \tr\left[(\Delta(\rho_{\text{mc}}^\alpha))^{1/\alpha}\right].
\end{align}
Similarly, the strong-converse exponent is given by
\begin{align}
    \lim_{n\to \infty} & -\frac{1}{n} \log (1-\alpha_{n, r}(\rho_{\text{mc}}^{\ox n}\| \SEP_n)) \notag \\
    & = \sup_{\alpha > 1} \frac{\alpha - 1}{\alpha} \left(r - E^*_\alpha(\rho_{\text{mc}})\right),
\end{align}
{for $E(\rho_{\text{mc}}) < r < \sup_{\alpha > 1} E_{\alpha}^*(\rho_{\text{mc}})$},
where $E^*_\alpha(\rho_{\text{mc}}) := D_{\Sand,\alpha}(\rho_{\text{mc}}\|\SEP_1) = D_{\Sand,\alpha}(\rho\|\sI_1)$ is the sandwiched \Renyi relative entropy of entanglement and we can choose a universal state as the tensor product of the universal state on $A^n$ and $B^n$, respectively. Finally, the same result works if we consider the set of PPT states as the free states, as the Petz and sandwiched \Renyi divergences are the same as those relative to the set of separable states for maximally correlated states~\cite{Zhu_2017}.

\section{Discussion}
\label{sec: discussion}

We have established a framework for analyzing error exponents in quantum hypothesis testing between two sets of quantum states, extending beyond the conventional i.i.d. setting to encompass composite correlated hypotheses. Our main contributions include a generalization of the quantum Hoeffding bound and strong converse exponent to stable sequences of convex, compact sets of quantum states. These results provide a finer characterization than the Stein regime, yielding deeper insights into the fundamental trade-offs between Type-I and Type-II errors in quantum state discrimination with composite correlated hypotheses.
The broad applicability of our framework is demonstrated through concrete applications to adversarial quantum channel discrimination and various quantum resource theories, highlighting its potential for further advances across quantum information theory.

Several open questions remain. While we showed $H_r^\infty(\sA\|\sB) \ge \bH_r^\infty(\sA\|\sB)$, proving equality in full generality remains open and would likely require a minimax theorem suited to the regularized setting. 
{It would also be interesting, as a first step, to establish equality under some physically natural assumptions.}
Extending the strong converse upper bound beyond the current technical restrictions is another important challenge. Resolving these issues would complete the asymptotic characterization of quantum state discrimination for composite, correlated hypotheses. Finally, since the generalized quantum Stein's lemma implies asymptotic reversibility in the corresponding resource theory~\cite{fang2024generalized,hayashi2024generalized,lami2024solutiongeneralisedquantumsteins}, our refined error exponent analysis may shed new light on conversion rates and convergence properties of resource interconversion, enabling finer control of asymptotic reversibility and efficiency. We leave these questions for future work.


\appendix

\section{Useful lemmas}

The following lemma is a minimax theorem that accounts for the infinity values of the function.
Let $X$ be a convex set in a linear space. A
 function $f: X \to (-\infty,-\infty]$ said to be convex,
  if $f(px+(1-p)y) \leq pf(x)+(1-p)f(y)$, the 
  multiplication $0\cdot f(x)$ is interpreted as $0$ 
  and $p\cdot \infty=\infty$ for  $p\neq 0$. Similar 
  definition holds for concave functions.

\begin{lemma}\cite[Theorem 5.2]{farkas2006potential}\label{lem: minimax infinity value}
    Let $X$ be a compact, convex subset of a Hausdorff topological vector space and $Y$ be a convex subset of the linear space. Let $f : X \times Y \to (-\infty,\infty]$ be lower semicontinuous on $X$ for fixed $y \in Y$, and assume that $f$ is convex in the first and concave in the second variable. Then we have
\begin{align}
\sup_{y \in Y} \inf_{x \in X} f(x,y) 
= \inf_{x \in X} \sup_{y \in Y} f(x,y).
\end{align}
\end{lemma}

The following lemmas are standard results in mathematical analysis and will be used frequently in our proofs. For detailed proofs, see, e.g.,~\cite[Lemma 2.8, 2.9]{ben2023new}.

\begin{lemma}
\label{lem: compact lsc}
Let $X$ be a nonempty compact topological space, and let $f : X \to \overline{\mathbb{R}}$ be a function. Then if $f$ is upper semicontinuous, it attains its maximum, meaning there is some $x \in X$ such that for all $x' \in X$, $f(x') \leq f(x)$. Similarly, if $f$ is lower semicontinuous, it attains its minimum.
\end{lemma}

\begin{lemma} \label{lem: inf usc}
    Let $X$ be a topological space, let $I$ be a set, and let $\{f_i\}_{i \in I}$ be a collection of functions $f_i : X \to \overline{\mathbb{R}}$. Then if each $f_i$ is upper semicontinuous, the function $f(x) = \inf_{i \in I} f_i(x)$ is also upper semicontinuous. Similarly, if each $f_i$ is lower semicontinuous, the pointwise supremum is lower semicontinuous.
\end{lemma}

\begin{lemma}\label{lem: log rate min}
    For two scalar sequences $x_n,y_n>0$ with $\limsup_{n\to \infty}  -\frac{\log x_n}{n} = x_R$ and $\lim_{n\to \infty} - \frac{\log y_n}{n} = y_R$, then $\limsup_{n\to \infty} -\frac{1}{n}\log (x_n+y_n) = \min\{x_R, y_R\}$. 
\end{lemma}
\begin{proof}
    For any scalars $x,y>0$, we can verify that~\cite{audenaert2008asymptotic}, 
    \begin{align}
        \max\{\log x, \log y\} \leq \log (x+y) \leq \max\{\log x, \log y\} + 1.
    \end{align}
    Therefore, we have 
    \begin{align}
    \limsup_{n\to \infty} -\frac{1}{n}\log (x_n+y_n) & \leq \limsup_{n\to \infty} -\frac{1}{n} \log x_n = x_R,\\
    \limsup_{n\to \infty} -\frac{1}{n}\log (x_n+y_n) & \leq \limsup_{n\to \infty} -\frac{1}{n} \log y_n = y_R.
    \end{align}
    This gives $\limsup_{n\to \infty} -\frac{1}{n}\log (x_n+y_n) \leq \min\{x_R, y_R\}$. For the other direction, by the definition of $x_R$, for any $\ve > 0$, there exists an infinite subsequence $n_m$ such that $-\frac{1}{n_m} \log x_{n_m} \geq x_R - \ve$. As the limit of $y_n$ exists, we know that for this subsequence $-\frac{1}{n_m} \log y_{n_m} \geq y_R - \ve$. This implies that 
    \begin{align}
        \min\left\{-\frac{1}{n_m} \log x_{n_m}, -\frac{1}{n_m} \log y_{n_m}\right\} \geq \min\{x_R, y_R\} - \ve
    \end{align} 
    Therefore, we have
    \begin{align}
    & \limsup_{n\to \infty} -\frac{1}{n}\log (x_n+y_n)\notag\\
     & \geq \limsup_{n\to \infty} -\frac{1}{n} \max\{\log x_n, \log y_n\} \\
    & = \limsup_{n\to \infty} \min\left\{-\frac{1}{n} \log x_n, -\frac{1}{n} \log y_n\right\} \\
    & \geq \limsup_{m\to \infty} \min\left\{-\frac{1}{n_m} \log x_{n_m}, -\frac{1}{n_m} \log y_{n_m}\right\} \\
    & \geq \min\{x_R, y_R\} - \ve.
    \end{align}
    As this holds for any $\ve > 0$, we have
    \begin{align}
    \limsup_{n\to \infty} -\frac{1}{n}\log (x_n+y_n) & \geq \min\{x_R, y_R\}.
    \end{align}
    This completes the proof.
\end{proof}

Note that the existence of the limit for one sequence is necessary in Lemma~\ref{lem: log rate min}.
Consider the case where $x_n = 1$ if $n$ is odd and $x_n = 2^n$ if $n$ is even; and $y_n = 2^n$ if $n$ is odd and $y_n = 1$ if $n$ is even. Then we have $\limsup_{n\to \infty} -\frac{1}{n} \log x_n = \limsup_{n\to \infty} -\frac{1}{n} \log y_n = 0$, but $\limsup_{n\to \infty} -\frac{1}{n} \log (x_n + y_n) = -1$. This gives a counterexample to Lemma~\ref{lem: log rate min} if the limit does not exist for either sequence.

\section{Mathematical properties of $\phi(s)$}
\label{sec: math properties of phi}

In this section, we discuss some useful properties of the function~\footnote{Note that this definition differs slightly from the one in the main text Lemma~\ref{LH1}, but they are equivalent up to a constant factor. Thus, all subsequent results work as the same.} $\phi(s|\rho\|\sigma) := \log \tr  \rho^{1-s} \sigma^{s}$. These results can be found in the book~\cite[Chapter 3]{hayashi2017quantum} and we list them here for completeness. Let $P_{\rho,\sigma}$ and $Q_{\rho,\sigma}$ denote the Nussbaum-Szko\l{}a distributions for $\rho$ and $\sigma$. These classical distributions preserve many important information of the original states, including
\begin{align}
    D(\rho\|\sigma) & = D(P_{\rho,\sigma}\|Q_{\rho,\sigma}),\\
    \phi(s|\rho\|\sigma) & = \phi(s|P_{\rho,\sigma}\|Q_{\rho,\sigma}).
\end{align}
Due to the second equality, we may simplify the notation as $\phi(s)$ when there is no confusion.
Moreover, the Hoeffding divergence can also be expressed as
\begin{align}
H_{1,r}(\rho\|\sigma) = \max_{s\in (0,1)} \frac{-\phi(s)-s r}{1-s}.
\end{align}

\begin{lemma}\label{lem: first and second derivative of phi}
Let $\rho,\sigma \in \PSD$. Then the first and second derivatives of $\phi(s|\rho\|\sigma)$ with respect to $s$ are given by~\cite[Exercise 3.5]{hayashi2017quantum},
    \begin{align}
    \phi'(s|\rho\|\sigma) & = \frac{\tr \rho^{1-s} \sigma^s(\log \sigma - \log \rho)}{\tr \rho^{1-s} \sigma^s},\\
    \phi''(s|\rho\|\sigma) & = \frac{\tr \rho^{1-s}(\log \sigma - \log \rho) \sigma^s (\log \sigma - \log \rho)}{\tr \rho^{1-s} \sigma^s}\notag \\
    &\quad  - \frac{(\tr \rho^{1-s} \sigma^s (\log \rho - \log \sigma) )^2}{(\tr \rho^{1-s} \sigma^s)^2}.
\end{align}
In particular, $\phi'(0\|\rho\|\sigma) = - D(\rho\|\sigma)$ and $\phi'(1|\rho\|\sigma) = D(\sigma\|\rho)$. Moreover, using Schwarz inequality, we know that $\phi''(s|\rho\|\sigma) \geq 0$ and therefore $\phi(s|\rho\|\sigma)$ is convex in $s \in \RR$. {Moreover, $\phi(s|\rho\|\sigma)$ is strictly convex in $s \in \RR$ if $\rho$ and $\sigma$ are not proportional.}
\end{lemma}

\begin{proof}
For the first derivative, we have
    \begin{align}
        \phi'(s|\rho\|\sigma) & = \frac{(\tr  \rho^{1-s} \sigma^{s})'}{\tr  \rho^{1-s} \sigma^{s}}\\
        & = \frac{\tr [ (\rho^{1-s} \sigma^{s})']}{\tr \rho^{1-s} \sigma^s}\\
        & = \frac{\tr [(\rho^{1-s})' \sigma^s + \rho^{1-s} (\sigma^s)']}{\tr \rho^{1-s} \sigma^s}\\
        & = \frac{\tr [-\rho^{1-s} (\log \rho) \sigma^s + \rho^{1-s}\sigma^s (\log \sigma)]}{\tr \rho^{1-s} \sigma^s}\\
        & = \frac{\tr [-(\log \rho)\rho^{1-s}  \sigma^s + \rho^{1-s} \sigma^s (\log \sigma)]}{\tr \rho^{1-s} \sigma^s}\\
        & = \frac{\tr \rho^{1-s} \sigma^s(\log \sigma - \log \rho)}{\tr \rho^{1-s} \sigma^s}
    \end{align}

For the second derivative, we have first that
\begin{align}
    & (\tr \rho^{1-s} \sigma^s(\log \sigma - \log \rho))'\notag \\
     & = \tr [(\rho^{1-s} \sigma^s(\log \sigma - \log \rho))']\\
    & = \tr [(\rho^{1-s} \sigma^s)'(\log \sigma - \log \rho)]\\
    & = \tr [-\rho^{1-s} (\log \rho)  \sigma^s + \rho^{1-s} \sigma^s (\log \sigma)](\log \sigma - \log \rho)\\
    & = \tr [-\rho^{1-s} (\log \rho)  \sigma^s + \rho^{1-s}  (\log \sigma)\sigma^s](\log \sigma - \log \rho)\\
    & = \tr [\rho^{1-s}  (\log \sigma - \log \rho) \sigma^s  (\log \sigma - \log \rho)].
\end{align}
Then we have the second derivative as
\begin{align}
    & \phi''(s|\rho\|\sigma) \notag \\
    & = \frac{X}{(\tr \rho^{1-s} \sigma^s)^2}\\
    & = \frac{\tr [\rho^{1-s}  (\log \sigma - \log \rho) \sigma^s  (\log \sigma - \log \rho)]}{\tr \rho^{1-s} \sigma^s} \notag \\
    & \quad - \frac{(\tr \rho^{1-s} \sigma^s (\log \rho - \log \sigma) )^2}{(\tr \rho^{1-s} \sigma^s)^2},
\end{align}
where $X = (\tr \rho^{1-s} \sigma^s(\log \sigma - \log \rho))' (\tr \rho^{1-s} \sigma^s) - (\tr \rho^{1-s} \sigma^s(\log \sigma - \log \rho)) (\tr \rho^{1-s} \sigma^s)'$.
By the Schwarz inequality $|\tr(XY^\dagger)|^2 \leq |\tr(XX^\dagger)| |\tr(YY^\dagger)|$, we have 
\begin{align}
& (\tr \rho^{1-s} \sigma^s (\log \rho - \log \sigma) )^2 \\
& \leq \tr [\rho^{1-s}  (\log \sigma - \log \rho) \sigma^s  (\log \sigma - \log \rho)] \tr [\rho^{1-s} \sigma^s],\notag
\end{align}
by considering $X = \rho^{(1-s)/2}(\log \sigma - \log \rho)\sigma^{s/2}$ and $Y = \rho^{(1-s)/2}\sigma^{s/2}$. This implies that $\phi''(s|\rho\|\sigma) \geq 0$ and therefore $\phi(s|\rho\|\sigma)$ is convex in $s \in \RR$. {Note that the Schwarz inequality is saturated if and only if $X$ and $Y$ are linearly dependent, which is equivalent to the condition that $\rho^{(1-s)/2}(\log \sigma - \log \rho)\sigma^{s/2}$ and $\rho^{(1-s)/2}\sigma^{s/2}$ are linearly dependent. This is then equivalent to $\rho = \lambda \sigma$ for some $\lambda \in \RR$. So in case $\rho$ and $\sigma$ are not proportional, we have $\phi''(s|\rho\|\sigma) > 0$ for all $s \in \RR$, which means that $\phi(s|\rho\|\sigma)$ is strictly convex in $s \in \RR$.}
\end{proof}

The following result can be found in~\cite[Exercise 3.45]{hayashi2017quantum}.
\begin{lemma}\label{lem: exponent optimization}
Let $r < D(\rho\|\sigma)$. There exists a unique maximum achieved at $s_r \in (0,1)$ for 
\begin{align}
  \sup_{0 \leq s \leq 1} f(s),\quad \text{with} \quad f(s) := \frac{-sr - \phi(s|\rho\|\sigma)}{1-s}.
\end{align}
Moreover, the optimal solution $s_r$ satisfies
\begin{align}
    r = (s_r - 1) \phi'(s_r) - \phi(s_r).
\end{align}
\end{lemma}
\begin{proof}
Let $P,Q$ be the Nussbaum-Szko\l{}a distributions and let $P_s(x):= P(x)^{1-s}Q(x)^s e^{-\phi(s)}$.
Then the derivative of $f(s)$ is given by
\begin{align}
    f'(s) = \frac{-r + (s-1) \phi'(s) - \phi(s)}{(1-s)^2}.
\end{align}
The sign of $f'(s)$ depends on the sign of its nominator $g(s):=-r + (s-1) \phi'(s) - \phi(s)$. By taking the derivative of $g(s)$, we have
\begin{align}
    g'(s) = (s-1) \phi''(s) < 0, \quad \text{for } s \in (0,1).
\end{align}
This means that $g(s)$ is strictly decreasing in $s \in (0,1)$. We also have~\cite[Exercise 3.45]{hayashi2017quantum} that,
\begin{itemize}
\item $D(P_s\|P_1) = (s-1)\phi'(s) - \phi(s)$ and $D(P_s\|P_0) = s\phi'(s) - \phi(s)$.
\item $\frac{d }{ds} D(P_s\|P_1) = (s-1) \phi''(s) < 0$  and $\frac{d }{ds} D(P_s\|P_0) = s \phi''(s) >0$ for $s \in (0,1)$.
\item Let $r < D(P\|Q)$. Then there uniquely exists $s_r \in (0,1)$ such that $D(P_{s_r}\|P_1) = r$.
\end{itemize}
The last item implies that there exists a unique critical point $s_r \in (0,1)$ such that $g(s_r) = -r + D(P_{s_r}\|P_1) = 0$. Together with the monotonicity of $g(s)$, we know that $g(s) > 0$ for $s < s_r$ and $g(s) < 0$ for $s > s_r$. This implies that $f'(s) > 0$ for $s < s_r$ and $f'(s) < 0$ for $s > s_r$. So $f(s)$ is increasing for $s < s_r$ and decreasing for $s > s_r$. This shows that $f(s)$ attains its maximum at $s_r$. Moreover, we can show that $f(s_r) = D(P_{s_r}\|P)$.
\end{proof}

The following shows that the optimal solution $s_r$ is differentiable in $r$.
\begin{lemma}\label{lem: sr differentiable}
    The optimal solution $s_r$ is differentiable in $r$ and ~\cite[Exercise 3.51]{hayashi2017quantum},
    \begin{align}
        \frac{d s_r}{dr} = \frac{1}{(s_r - 1) \phi''(s_r)}.
    \end{align}
\end{lemma}
\begin{proof}
    We have the following
    \begin{align}
        1 & = \frac{dr}{dr}\\
        & = \frac{d}{dr} ((s_r - 1) \phi'(s_r) - \phi(s_r))\\
        & = \frac{d s_r}{dr} \cdot \frac{d}{ds}((s-1) \phi'(s) - \phi(s))|_{s=s_r}\\
        & = \frac{d s_r}{dr} \cdot ((s_r - 1)\phi''(s_r)),
    \end{align}
    where the third line uses the chain rule of differentiation. This implies the desired result.
\end{proof}

\bibliographystyle{IEEEtran}
\bibliography{Bib}

\vspace{-1cm}
\begin{IEEEbiographynophoto}{Kun Fang}
received the BS degree in Mathematics from Wuhan University in 2015 and PhD degree in Quantum Information from the University of Technology Sydney in 2018. After that, he worked as a postdoctoral researcher at the University of Cambridge and the University of Waterloo from 2018 to 2020. He served as a senior researcher and tech lead at the Institute for Quantum Computing, Baidu, from 2020 to 2023. Currently, he is an assistant professor in the School of Data Science at The Chinese University of Hong Kong, Shenzhen. His research interests lie in understanding the capabilities and limitations of quantum resources, as well as their usage in quantum computation and quantum communication. 
\end{IEEEbiographynophoto}

\begin{IEEEbiographynophoto}{Masahito Hayashi} (Fellow, IEEE) was born in Japan, in 1971. He received the B.S. degree from the Faculty of Sciences, Kyoto University, Japan, in 1994, and the M.S. and Ph.D. degrees in mathematics from Kyoto University in 1996 and 1999, respectively. 
He was with Kyoto University as a Research Fellow of Japan Society of the Promotion of Science (JSPS) from 1998 to 2000 and the Laboratory for Mathematical Neuroscience, Brain Science Institute, RIKEN, as a Researcher, from 2000 to 2003. He was with ERATO Quantum Computation and Information Project, Japan Science and Technology Agency (JST), as the Research Head, from 2003 to 2006, and ERATO-SORST Quantum Computation and Information Project, JST, as the Group Leader, from 2006 to 2007. He was also with the Superrobust Computation Project Information Science and Technology Strategic Core (21st Century COE by MEXT) Graduate School of Information Science and Technology, The University of Tokyo, as an Adjunct Associate Professor, from 2004 to 2007. He was with the Graduate School of Information Sciences, Tohoku University, as an Associate Professor, from 2007 to 2012. In 2012, he joined the Graduate School of Mathematics, Nagoya University, as a Full Professor. He was with Shenzhen Institute for Quantum Science and Engineering, Southern University of Science and Technology, Shenzhen, China, as the Chief Research Scientist, from 2020 to 2023. In 2023, he joined the School of Data Science, The Chinese University of Hong Kong, Shenzhen (CUHK-Shenzhen), as a Full Professor, and joined International Quantum Academy (SIQA) as the Chief Research Scientist. He also has the title of Presidential Chair Professor in CUHK-Shenzhen from 2023. Also, he was with the Centre for Quantum Technologies, National University of Singapore, as a Visiting Research Associate Professor, from 2009 to 2012, and a Visiting Research Professor from 2012 to 2024. He was with the Center for Advanced Intelligence Project, RIKEN, as a Visiting Scientist, from 2017 to 2020. He was with Shenzhen Institute for Quantum Science and Engineering, Southern University of Science and Technology, as a Visiting Professor, from 2018 to 2020, and Center for Quantum Computing, Peng Cheng Laboratory, Shenzhen, as a Visiting Professor, from 2019 to 2020. He is currently with the Department of Information Engineering, The Chinese University of Hong Kong as an Adjunct Professor since 2024. In 2006, he published the book Quantum Information: An Introduction (Springer), whose revised version was published as Quantum Information Theory: Mathematical Foundation from Graduate Texts in Physics (Springer) in 2016. In 2016, he published other two books Group Representation for Quantum Theory and A Group Theoretic Approach to Quantum Information (Springer). He is on the editorial board of New Journal of Physics and International Journal of Quantum Information. His research interests include classical and quantum information theory and classical and quantum statistical inference. He received the Information Theory Society Paper Award for Information-Spectrum Approach to SecondOrder Coding Rate in Channel Coding in 2011. In 2016, he received the Japan Academy Medal from Japan Academy and the JSPS Prize from Japan Society for the Promotion of Science. In 2022, he was elected as an IMS Fellow and an Asia–Pacific Artificial Intelligence Association (AAIA) Fellow.
\end{IEEEbiographynophoto}




\end{document}